\documentclass[fleqn, usenatbib]{mnras}

\usepackage{newtxtext,newtxmath}

\usepackage[T1]{fontenc}

\DeclareRobustCommand{\VAN}[3]{#2}
\let\VANthebibliography\thebibliography
\def\thebibliography{\DeclareRobustCommand{\VAN}[3]{##3}\VANthebibliography}

\usepackage{graphicx}	
\usepackage{amsmath}	
\usepackage{subcaption} 
\usepackage{multicol}
\usepackage{multirow}
\usepackage{booktabs}
\usepackage{titlesec}
\usepackage{xspace}
\usepackage[version=4]{mhchem}

\newcommand{\change}{ }

\newcommand{\dsct}{$\delta$\,Sct\xspace}
\newcommand{\mesa}{\textsc{mesa}\xspace}
\newcommand{\gyre}{\textsc{gyre}\xspace}

\titleclass{\subsubsubsection}{straight}[\subsubsection]
\newcounter{subsubsubsection}[subsubsection]
\renewcommand\thesubsubsubsection{\thesubsubsection.\arabic{subsubsubsection}}
\titleformat{\subsubsubsection}
    {\itshape\normalsize}{\thesubsubsubsection}{1em}{} 
\titlespacing*{\subsubsubsection}
  {0pt}{1.25ex plus 1ex minus .2ex}{0.75ex plus .2ex}

\title[$\delta$\,Scuti model grid]{Modelling $\delta$\,Scuti pulsations: A new grid of p, g, and f modes across pre-main-sequence to post-main-sequence evolution}

\author[Gautam et al.]{
Anuj Gautam,$^{1}$\thanks{E-mail: anuj.gautam@unisq.edu.au (AG)}, Simon J. Murphy,$^{1}$\thanks{E-mail: simon.murphy@usq.edu.au (SJM)}, Timothy R. Bedding,$^{2}$\\
$^{1}$ Centre for Astrophysics, University of Southern Queensland, Toowoomba, QLD 4350, Australia\\
$^{2}$ Sydney Institute for Astronomy, School of Physics, University of Sydney, Sydney, NSW 2006, Australia\\
}

\date{Accepted XXX. Received YYY; in original form ZZZ}
\pubyear{YYYY}

\begin{document}
\label{firstpage}
\pagerange{\pageref{firstpage}--\pageref{lastpage}}
\maketitle

\begin{abstract}
    Space-based photometry reveals regular high-frequency patterns in many young $\delta$\,Scuti stars. These pulsations provide a powerful means of inferring stellar properties, particularly ages, for young $\delta$\,Scuti stars for which traditional age-dating methods are poorly constrained. Realising this potential requires theoretical models that capture the complexities of stellar structure and evolution. We present a comprehensive grid of 25 million stellar pulsation models, computed using the \mesa stellar evolution code and the \gyre stellar oscillation code, tailored to $\delta$\,Scuti stars. The grid spans a wide range of masses, metallicities and rotation velocities, and covers evolutionary phases from the early pre-main-sequence through the main sequence and into the post-main sequence contraction phase. For each model, we computed adiabatic pulsation frequencies for degrees $\ell = 0$--$3$, capturing p\:modes, g\:modes, f\:modes and their interactions through avoided crossings. \change{We find that f and low-order g\:modes have mode inertias comparable to or lower than the fundamental radial mode during the late pre-MS and early MS, implying that these modes should be observable.} We revisit $\delta$\,Scuti scaling relations and map asteroseismic observables, including the large frequency separation ($\Delta\nu$) and phase offset parameter ($\varepsilon$), across age, mass, metallicity, and rotation. This new model grid, which is publicly available, improves upon previous such model grids by facilitating interpretation of $\delta$\,Scuti pulsations, allowing for more reliable age estimates and tighter constraints on stellar evolutionary pathways, and planet formation in A- and F-type stars.
\end{abstract}




\section{Introduction}
\label{sec:intro}

Accurately determining stellar ages is essential for advancing our understanding of Galactic chemical evolution and the dynamics of planet formation. Precise age-dating is particularly significant for young stars, whose nascent planetary systems evolve quickly within short-lived circumstellar discs. Without reliable age estimates, it is difficult to place robust temporal constraints on protoplanetary disc observations from facilities such as ALMA, VLT/SPHERE, and \textit{JWST}, thereby limiting our understanding of the timescales of planet formation.

Current age-dating methods often rely on isochrone fitting of star clusters, which is highly sensitive to the properties of high-mass stars ($M > 2.5\,\textrm{M}_\odot$) near the main sequence turn-off. Other techniques, such as lithium depletion or kinematics, are more effective for low-mass stars ($M < 1.5\,\textrm{M}_\odot$). In contrast, age determination for intermediate-mass stars remains particularly challenging because traditional age-estimation techniques such as isochrone fitting and gyrochronology often suffer from significant uncertainties in this mass regime, owing to sparse empirical calibrators and the absence of strong magnetic braking \citep{soderblom_ages_2010}.

Asteroseismology, the study of stellar oscillations, offers a promising alternative for age determination, particularly for intermediate-mass stars \citep{aerts_book_2010, kurtz_asteroseismology_2022}. Among these, \dsct stars, typically with masses between 1.5 and 2.5~$\textrm{M}_\odot$, are of particular interest. These stars are found in associations spanning ages from a few million years (e.g. Upper Centaurus; \citealt{murphy_precise_2021}) to early- or mid-main-sequence clusters (e.g. Pleiades; \citealt{FoxMachado_pleiades_2006, Bedding_pleiades_2023}; Praesepe; \citealt{Hernandez_Praesepe_1998, frandsen__2001, dall__2002, breger_delta_2012}). At older ages, \dsct stars are also observed near the terminal-age main sequence and beyond, though they are typically found as field stars, since most clusters disperse before reaching such ages.

Oscillations have been detected in a large number of \dsct stars, revealing pulsations dominated by low-overtone radial and non-radial pressure (p) modes with periods ranging from 18 minutes to 8 hours \citep{Handler_dsctgdor_2002, Goupil_dsct_2005}. These pulsations modulate the stellar luminosity with photometric amplitudes ranging from a few $\mu$mag to over 300 mmag \citep{rodriguez_catalogue_2000}. High-precision photometry from space-based missions such as \textit{WIRE}, \textit{MOST}, \textit{BRITE}, \textit{CoRoT}, \textit{Kepler}, and \textit{TESS} has played a central role in detecting and characterizing these pulsations across a wide population of \dsct stars \citep{Handler_dsctgdor_2002, Goupil_dsct_2005, buzasi_altair_2005, gilliland_kepler_2010, Uytterhoeven_AFkepler_2011, breger_delta_2012, paparo_corot_2013, balona_pulsation_2015, michel_corot_2017, forteza_envelope_2018, bowman_characterizing_2018, antoci_first_2019, guzik_highlights_2021, hasanzadeh_relations_2021, barac_revisiting_2022, chen_TESS_2022, zwintz_catalogue_2024}. 

Pulsations in \dsct stars are generally attributed to the $\kappa$-mechanism acting in the He II ionization zone, which causes a local opacity increase that drives pulsational instability \citep{pamyatnykh_pulsational_2000}. However, not all observed features, particularly the lowest-amplitude, high-frequency modes, are readily explained by this mechanism alone. In addition to $\kappa$-mechanism, turbulent pressure in the outer convective layers, especially near the hydrogen partial ionization zone, may also contribute to the excitation of specific high-order p\:modes ($7 \leq n \leq 10$) \citep{Houdek_convective_2000, antoci_turbulentp_2014, smalley_pulsation_2017}. Such high-frequency p\:modes often exhibit very small amplitudes, which may indicate that turbulent pressure, rather than the standard $\kappa$-mechanism, contributes to their excitation \citep{bedding_highf_2020}. Turbulent pressure may be especially relevant in Am stars, where helium depletion and chemical stratification alter the opacity profile and suppress classical $\kappa$-driving \citep{preston_chemically_1974}. In addition to these excitation mechanisms, the edge-bump mechanism has been identified as an alternative opacity source capable of driving pulsations, even in stars with significantly depleted helium envelopes \citep{Stellingwerf_edgebump_1979, TheadoCunha_2004, Murphy_Ap_2020}, and appears to explain the confinement of Am stars to the red half of the instability strip \citep{Murphy_Ap_2020}. 

Regular frequency patterns in \dsct stars have been both theoretically predicted and observationally confirmed. Theoretical studies have shown that high-radial-order p\:modes can retain near-regular frequency spacings even in the presence of moderate rotation, owing to their confinement in the outer stellar envelope \citep{reese_regular_2008, Ouazzani_Roxburgh_Dupret_2015, reese_frequency_2017, mirouh_mode_2019}. Observationally, such regularities have been detected in many \dsct stars \citep{garcia_hernandez_asteroseismic_2009, breger_regularities_2011, garcia_hernandez_-depth_2013, paparo_unexpected_2016, bedding_highf_2020, singh_2025}. In particular, the high-frequency p\:mode spectra of young \dsct stars often display regularly spaced overtone series that can be lined up in \'echelle diagrams, allowing for identification of radial and non-radial modes \citep{bedding_highf_2020, murphy_precise_2021,scutt_asteroseismology_2023, Bedding_pleiades_2023}. The ability to identify pulsation modes means that fundamental stellar properties including mass, metallicity, and age can be determined via forward asteroseismic modelling, without requiring cluster membership or empirical calibrations. Additionally, this approach allows measurement of the internal angular rotation frequency ($\Omega$) independent of the inclination angle, eliminating the geometric projection uncertainty that affects spectroscopic measurements of rotational velocities ($v\sin i$). Moreover, older \dsct stars, particularly those exhibiting hybrid characteristics with $\gamma$\,Doradus pulsations, can also show gravity (g) modes that provide further insight into their deeper interior regions \citep{kurtz_asteroseismic_2014, saio_asteroseismic_2015, li_gravity-mode_2020, aerts_asteroseismic_2024, aerts_evolution_2025}.

The large frequency separation, $\Delta\nu$, represents the regular spacing between consecutive radial orders of pressure modes and scales approximately with the square root of the mean stellar density \citep{ulrich_determination_1986}, making it a powerful diagnostic of global stellar properties \citep{Garcia_loggDnu_2017, murphy_grid_2023}. This scaling has been empirically validated in studies of \dsct stars in eclipsing binary systems, where independent mass and radius measurements enable direct calibration of the $\Delta\nu$-density relation \citep{garcia_hernandez_observational_2015}. 

Even in the presence of stellar rotation, the relation remains robust. For moderate rotators, the scaling holds well in the low-radial-order regime, with some dependence on stellar mass and evolutionary state \citep{rodriguez-martin_study_2020}. At more rapid rotation rates, where centrifugal distortion and mode coupling become significant, $\Delta\nu$ continues to scale approximately with the square root of the mean density, although with reduced precision \citep{reese_regular_2008, garcia_hernandez_observational_2015}. This persistence highlights the fundamental connection between stellar structure and pulsation properties. 

Beyond mean density, pulsation modelling has shown that $\Delta\nu$, when combined with luminosity measurements, can also constrain stellar inclinations and rotation rates \citep{murphy_cepher_2024}. While individual frequency modelling yields the most precise stellar parameters, $\Delta\nu$ serves as a reliable and measurable quantity for population studies and for stars with less distinct pulsation patterns where mode identification remains challenging.

Grid-based asteroseismic modelling has already yielded tight age constraints for young \dsct stars. For example, HD\,139614, a pre-main-sequence \dsct star in Upper Centaurus--Lupus, was modelled with $<8\%$ uncertainty on its age \citep{murphy_precise_2021}. Similar analyses have been applied to \dsct stars in the Pleiades \citep{murphy_five_2022}, and in the SPYGLASS sample \citep{kerr_spyglassII_2022, Kerr_spuglassIII_2022}, using \'echelle diagrams to identify radial and dipole modes and fit them with 1D evolutionary models. Moreover, asteroseismology can vet cluster or moving group membership by corroborating seismic and group ages. For HIP\,99770, a \dsct pulsator, $\Delta\nu=4.86\pm0.03~\mathrm{d^{-1}}$ implies an age of $\sim115$--$200$\,Myr from rotating \mesa models, favoring Ursa Majoris moving group over the $\sim 40$\,Myr Argus association \citep[][and supplementary material]{currie_direct_2023}. For the youngest pre–MS \dsct stars, we note that differing accretion histories can leave subtle imprints on internal structure, and hence on pulsation frequencies \citet{steindl_pms_2021, steindl_imprint_2022}, although these signatures may be difficult to detect observationally.

Applying asteroseismic modelling to ensembles of \dsct stars within the same cluster or association enables cross-validation of stellar parameters and yields high-precision estimates of cluster ages. Because all members of a cluster are expected to share the same age and initial composition, differences in the inferred parameters from individual stars directly reflect the precision and limitations of the modelling framework. In the Pleiades, for example, asteroseismic fits to five \dsct stars revealed that non-rotating models systematically overestimated stellar densities, emphasizing the necessity of incorporating rotational effects in stellar models \citep{murphy_five_2022}. Similarly, large-sample analyses in the Cepheus-Hercules complex found that pulsator fractions are higher in younger populations and that rapid rotators, while more likely to pulsate, exhibit less regular mode patterns \citep{murphy_cepher_2024}. 

Despite these advances, existing model grids have significant limitations. Previous grids either cover only a fraction of the main sequence lifetime \citep{murphy_grid_2023}, neglect rotation entirely, or include a limited range of pulsation modes. To address these shortcomings and provide a comprehensive resource for the asteroseismic community, we have developed an extensive grid of 25 million stellar pulsation models computed using the stellar evolution code \mesa \citep{Paxton2011, Paxton2013, Paxton2015, Paxton2018, Paxton2019, Jermyn2023_MESA} and the stellar oscillation code \gyre \citep{gyre1, gyre2, gyre3}. For this grid, we simulated the evolution of approximately 20,000 stars with varying masses, initial metallicities, and surface rotation velocities, across thousands of ages each (described below). Our grid extends beyond previous work by:

\begin{enumerate}
    \item \textit{Incorporation of rotation}:

    Stars exhibiting \dsct pulsations typically lie above the Kraft break, a well-defined transition in stellar structure and rotational behaviour that separates low-mass, slowly rotating stars from higher-mass, rapidly rotating ones \citep{Kraft_break_1967}. The break occurs near $T_{\rm eff} \sim 6550$\,K, and $M \sim 1.37$\,M$_\odot$, with a characteristic width of only $\sim$200\,K and 0.11\,M$_\odot$ in mass \citep{beyer_kraft_2024}. This boundary separates lower-mass stars with deep convective envelopes, capable of sustaining magnetic dynamos and undergoing significant angular momentum loss via magnetized winds, from higher-mass stars with predominantly radiative envelopes. In the latter, magnetic braking is inefficient or absent. The Kraft break coincides with a structural transition to thin, ionization-driven surface convection zones \citep{cantiello_envelope_2019, jermyn_atlas_2022}.

    As a result of lying above the Kraft break, \dsct stars are able to retain much of their initial angular momentum throughout the main sequence. Their observed surface rotation velocities span a broad range, with some reaching up to 300\,km\,s$^{-1}$ \citep{uesugi_vizier_1982, peterson_resolving_2006}, while others rotate much more slowly. Early studies by \citet{AbtMorrell_rot_1995} suggested a bimodal distribution: one population of fast-rotating, chemically normal stars and another of slower-rotating, chemically peculiar (Am) stars. More detailed analyses by \citet{royer_rotational_2007} and \citet{Zorec_rot_2012} revealed that the rotational velocity distribution among A-type stars, which includes \dsct stars, is more complex. Specifically, they described the observed distribution as a mass-dependent, lagged Maxwellian with a distinct low-velocity component probably associated with stars in binary or multiple systems, and a high-velocity component representing single, rapidly rotating stars. 

    Asteroseismic studies of A/F stars, including \dsct variables, reveal very little radial differential rotation across their interiors for slow to moderate rotators \citep{kurtz_asteroseismic_2014, saio_asteroseismic_2015, murphy_near-uniform_2016}, suggesting the presence of efficient internal angular momentum transport mechanisms in these stars. However, the internal rotation profiles of the most rapidly rotating \dsct stars remain largely unconstrained by asteroseismology \citep{aerts_combined_2019, aerts_asteroseismic_2024}, mainly because mode identification becomes significantly more challenging in these cases due to the effects of strong centrifugal distortion and complex mode coupling. The empirical evidence for slow to moderate rotators aligns with theoretical predictions \citep{aerts_combined_2019}, which highlight the role of internal gravity waves and hydrodynamical instabilities in enforcing near-rigid rotation. 

    Beyond angular momentum transport, rotation also induces large-scale meridional circulation and shear instabilities that drive chemical mixing between the convective core and radiative envelope \citep{Maeder_rot_2000}. This mixing transports fresh hydrogen to the core, extending the main-sequence lifetime of the star and altering the internal chemical gradients that affect pulsation properties.
    
    In our computations, we adopt uniform rotation profiles, initializing rotation at the onset of the pre-main-sequence phase (see Section~\ref{sec:methodology}). This approach is consistent with observations and enables a tractable treatment of rotational effects across a large model grid.
    
    \item \textit{Broader pulsation modelling}: 

    We extended our analysis by including higher-degree modes up to $\ell = 3$ in our models. Non-radial pressure mode pulsations propagate along paths not directed through the stellar center. As these waves travel inward, their lower portions of their wavefront encounter regions of higher temperature and sound speed than their upper portions, causing them to refract back toward the surface where they are reflected inward again. The degree $\ell$ of a mode corresponds to the number of nodal lines on the stellar surface, with higher-degree modes having more surface reflection points and being confined to shallower propagation cavities. Because each mode samples the sound speed along its ray path, its frequency reflects conditions within a specific region of the star. As a result, modes of varying degree carry information about different depths within the stellar interior \citep{kurtz_stellar_2006, aerts_book_2010}. While higher-degree modes are more difficult to detect observationally due to geometric cancellation effects, their inclusion in our models provide sensitivity to a wider range of depths within the stellar structure.

    In addition to p\:modes for radial orders $n = 1$ to $11$, we also compute g\:modes for radial orders $n = 1$ to $6$ (i.e., $n_\mathrm{pg} = -1$ to $-5$) for $\ell = 1$, $2$, and $3$, along with the f\:modes for $\ell = 2$, and $3$ (see Sec. \ref{sec:avoided_crossings_GYRE}). The low-order g\:modes considered here are distinct from the high-order g\:modes that are typically a feature of $\gamma$\,Doradus ($\gamma$\,Dor) stars and hybrid pulsators. Together with f\:modes, which occupy frequencies slightly higher than the lowest frequency p\:mode, they must be included for a complete representation of the non-radial mode spectra in these stars.
    
    Identifying and matching the rotational splittings of $\ell = 1$--3 modes, for p, g and f modes, allows for a better chance of successfully measuring the rotation rate of the star. Although our models assume solid-body rotation, the differing depth sensitivities of these modes allows us to compare observed splitting ratios to a solid-body baseline, revealing departures from uniform rotation and offering clues about angular-momentum transport. With well identified splittings of each mode type, we can estimate first-order core versus envelope rotation rates.
    
    \item \textit{Extended age range}: 
    
    The pulsation spectra of \dsct stars change significantly as they evolve \citep[for a review, see][]{JCD_introdsct_2000}. The expansion of the stellar envelope lengthens the acoustic travel time, leading to a decrease in p-mode frequencies. Meanwhile, the buoyancy frequency in the core increases, leading to an increase in the frequencies of the g\:modes. Eventually, these p and g\:modes overlap in frequency and couple to form mixed modes, resulting in mode bumping and \emph{avoided crossings}. 
    
    These avoided crossings are a resonance-like phenomenon caused by changes in the propagation region, where modes of similar frequencies but differing spatial characteristics (p and g\:modes) interact as they traverse regions with sharp density gradients, such as at the convective core boundary. Instead of crossing, the modes couple and exchange characteristics. This interaction effectively increases the restoring force for the p\:mode-dominated oscillation, shifting its frequency upward to resemble that of a higher radial order mode, while also altering the spatial structure of the coupled mode \citep{JCD_introdsct_2000, aerts_book_2010}. Avoided crossings are the reason that our previous work only modelled stars up to one-third of their main sequence lifetime \citep{murphy_grid_2023}.

    To capture these complex pulsation behaviours -- the gradual decrease of p-mode frequencies, the initial increase followed by the decrease in g-mode frequencies, and the effects of avoided crossings -- we computed our evolutionary tracks from the pre-main-sequence, across the full main sequence, and through to the end of the post–main-sequence contraction phase.

\end{enumerate}

This paper is the first in a series aimed at integrating our pulsation model grid into a neural network based inference framework that will also account for systematic modelling uncertainties. The paper is organized as follows: In Section~\ref{sec:methodology}, we describe the computational methods and the physical parameters used in constructing the model grid. Section~\ref{sec:gridanalysis} presents an analysis of our model grid, including the effects of rotation and age on pulsation frequencies. Finally, Section~\ref{sec:conclusions} discusses the implications of our findings for asteroseismology with \dsct stars, summarizes our conclusions, and outlines prospects for future work.

\section{Methodology}
\label{sec:methodology}
    
    \subsection{Stellar evolutionary models}
    \label{sec:stellar_evol_models}
        
        Stellar evolutionary models were calculated using \mesa (Modules for Experiments in Stellar Astrophysics; r24.03.1; \citealt{Paxton2011, Paxton2013, Paxton2015, Paxton2018, Paxton2019, Jermyn2023_MESA}). This code computes one-dimensional stellar models by solving the fully coupled structure and composition equations, providing equilibrium models at discrete time-steps based on our temporal sampling (see Sec. \ref{sec:sampling}).
    
        \subsubsection{Grid input parameters}
        \label{sec:grid_input_parameters}
            The model grid spans initial stellar masses ($M$) from 1.4 to 2.5\,M$_\odot$ in increments of 0.02\,M$_\odot$, metal mass fractions ($Z$) from 0.001 to 0.026 in steps of 0.001, and initial surface rotational velocities ($v_{\rm eq,0}$) from 0 to 15\,km\,s$^{-1}$ in 1 km\,s$^{-1}$ increments. These velocities increase markedly as the pre-main-sequence models contract, with an initial surface rotational velocity 10\,km\,s$^{-1}$ corresponding to a zero-age main sequence surface rotation velocity of 142.62\,km\,s$^{-1}$ for a solar-metallicity model ($Z=0.015$) of 1.7\,M$_\odot$. Each evolutionary track is initialized as described in Section\,\ref{sec:evolutionary_phases} and Section\,\ref{sec:dsct_evolution}.
            
        \subsubsection{Evolutionary phases}
        \label{sec:evolutionary_phases}
\begin{figure}
                \begin{center}
                \includegraphics[width=0.48\textwidth]{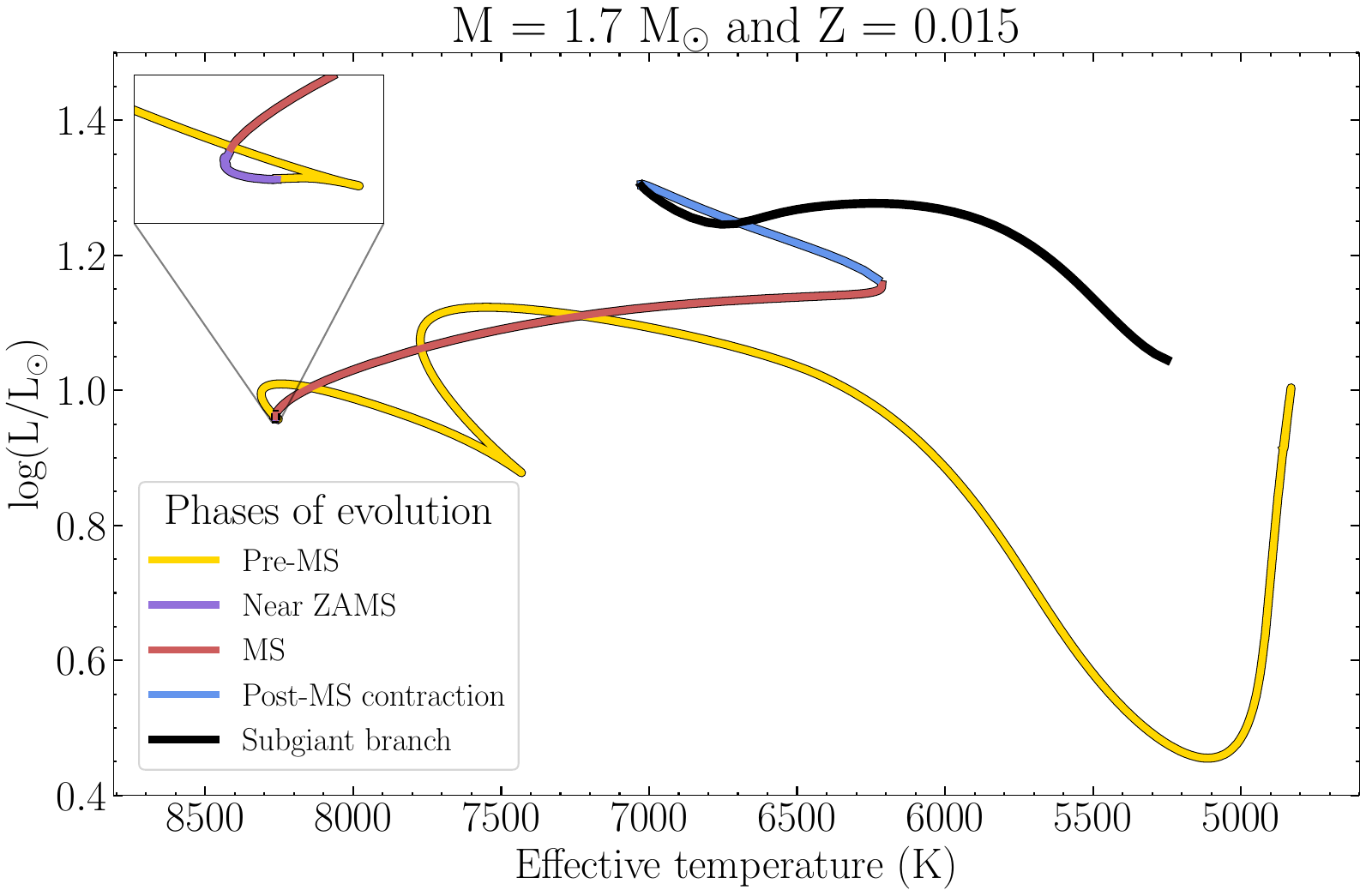}
                \caption{Phases of evolution of a typical 1.7-M$_\odot$, solar-metallicity \dsct star on the Hertzsprung-Russell diagram.}
                \label{fig:phases_basic}
                \end{center}
            \end{figure}

            In our models, we choose a temporal sampling that varies between different evolutionary phases (refer to Figure\:\ref{fig:phases_basic} and Table~\ref{tab:temporal_resolution}), defining each phase based on structural and evolutionary criteria. The distinct phases considered are the pre-main-sequence (pre-MS), near zero-age main sequence (near-ZAMS), main sequence (MS), and post-main-sequence (post-MS) contraction phases. The specific criteria for distinguishing each phase are described in detail below.
    
            We begin by generating a pre-MS model with a specified mass and metallicity following the standard \mesa routine. This routine initializes a chemically homogeneous star with a core temperature below $9 \times 10^5$ K, no nuclear burning, and uniformly contracts the model under gravity to generate sufficient luminosity to make the star fully convective.
    
            The \textit{pre-MS phase} spans the evolution from the initial pre-MS model to the ZAMS and is characterized by gravitational contraction and structural reconfiguration as the star evolves toward sustained hydrogen fusion. During this phase, the star transitions from a fully convective, contracting configuration to one with a radiative envelope and a convective core, where nuclear burning becomes the dominant energy source. 

\begin{figure}
                \begin{center}
                \includegraphics[width=0.48\textwidth]{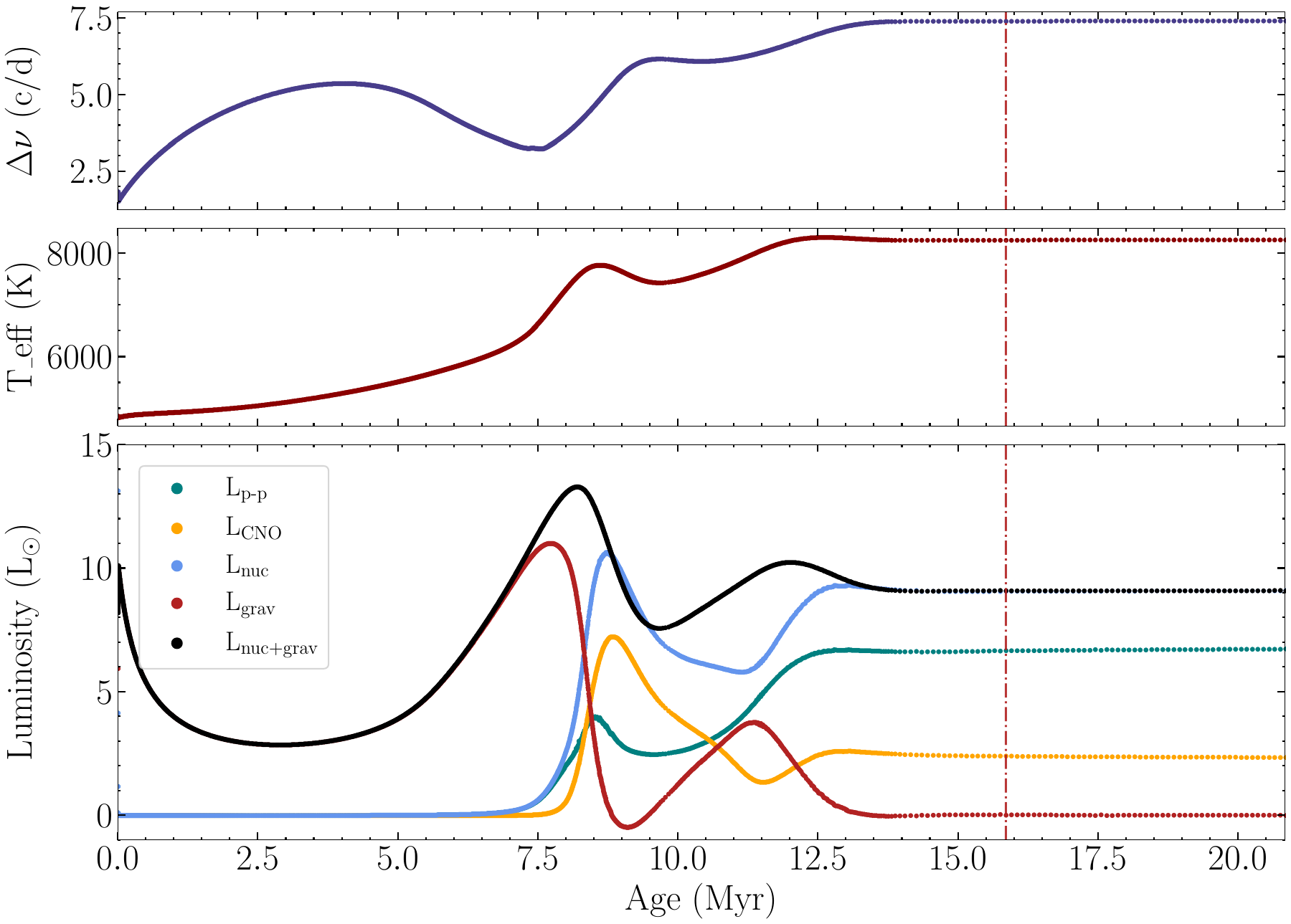}
                \caption{Evolution of the large frequency separation ($\Delta\nu$) for a typical 1.7 solar mass star with solar metallicity. The dashed line shows the ZAMS stabilization point, where both the stellar density (hence $\Delta\nu$) and nuclear burning have reached equilibrium.}
                \label{fig:zams_stabilization}
                \end{center}
            \end{figure}

            We define the \textit{zero-age main sequence} (ZAMS) as the point at which the second derivative of the large frequency separation, $\ddot{\Delta\nu}$, becomes stable. Specifically, we identify the earliest age prior to 50 Myr where the rolling standard deviation of $\ddot{\Delta\nu}$ remains below 0.01 d$^{-1}$ Myr$^{-2}$ over a sliding window of 100 models (spanning $\sim$10 Myr), as illustrated in Figure~\ref{fig:zams_stabilization}. 

            We also note that defining ZAMS as the point of highest $\Delta\nu$ (and therefore highest density) is not appropriate in general. In many tracks, both $\Delta\nu$ and $T_\mathrm{eff}$ can exhibit a bump followed by a dip during the final stages of pre-MS contraction, before true equilibrium is established. For many stars, this bump represents the global maxima of these quantities. Such features can lead to premature ZAMS identification if stability is not explicitly assessed.
            
            To ensure this pulsational stability corresponds to genuine structural equilibrium, we additionally require that the rolling standard deviations of the second derivatives of the effective temperature and mean stellar density remain below 0.5\,K\,Myr$^{-2}$ and $10^{-4}$\,g\,cm$^{-3}$\,Myr$^{-2}$, respectively, over the same window. These quantities are not used to define the ZAMS independently, but act as consistency checks to avoid falsely identifying transient dips in $\ddot{\Delta\nu}$ as stable phases. 

            This physically motivated approach corresponds to the end of gravitational contraction, where the star reaches hydrostatic and thermal equilibrium, and sustained hydrogen burning dominates its energy output. Compared to the $\Delta\nu$-based ZAMS definition used in \citet{murphy_grid_2023}, our method offers more consistent ZAMS tagging across a wide range of masses and metallicities. It also avoids the ambiguity inherent in abundance-based definitions such as that used in \citet{steindl_imprint_2022}, which mark ZAMS by a fixed decrement in central hydrogen content ($\Delta X_c = -0.010$), but which does not consistently correspond to structural equilibrium across the range of masses, metallicities, and initial rotation rates in our grid. Since $\Delta\nu \propto \sqrt{\bar{\rho}}$, where $\bar{\rho}$ is the mean stellar density, a similar density-based criterion to ours can also be applied to models that do not include pulsation calculations.

            The \textit{near-ZAMS phase} begins at the ZAMS and extends until an age of 50\,Myr. Note that the age at the ZAMS is not known a priori and varies significantly across the parameter space of mass, metallicity and rotation. Maintaining dense sampling through this near-ZAMS phase ensures that we capture the full range of ZAMS onset times in our grid. By 50\,Myr, all stars within our model range have stably entered core hydrogen burning \footnote[1]{disequilibrium H burning, including via the CNO cycle, occurs prior to this on the pre-MS (see Sec.~\ref{sec:dsct_evolution})}, making this a conservative upper limit that guarantees inclusion of the early main-sequence across all tracks.
    
            The \textit{main sequence (MS) phase} is defined as the period of sustained core hydrogen burning following the near-ZAMS phase. During this time, the star evolves gradually as hydrogen is depleted in the core.

            We define the \textit{terminal-age main sequence} (TAMS) as the point that corresponds to the local minimum in $T_\mathrm{eff}$ towards the end of late-MS. Formally, we compute this as the first time step where $T_\mathrm{eff}$ shows a positive time derivative after the point where the star's central hydrogen fraction ($X_\mathrm{c}$) drops below 30\% of its initial value. Here, the $X_\mathrm{c}\le0.3X_\mathrm{c,0}$ condition allows us to select the late-MS evolution and makes sure we get the correct $T_\mathrm{eff}$ turning point that corresponds to the TAMS. The TAMS marks the end of stable core hydrogen burning and the onset of post-MS evolution.
            
            The \textit{post-MS contraction phase}, often identified as the `hook' or `blue loop' in the HR diagram, extends from the TAMS to the point where the stellar luminosity, $L$, reaches a local maximum. This point corresponds to the ignition of hydrogen burning in a shell around the core. We terminate our models at this stage rather than continuing into the subgiant phase, as $\delta$\,Scuti pulsators become increasingly rare beyond this point due to stars rapidly evolving out of the instability strip (see Sec. \ref{sec:dsct_evolution}).
            
        \subsubsection{Sampling and numerical convergence}
        \label{sec:sampling}
            We controlled the maximum time-step during different evolutionary phases by setting the \texttt{max\_years\_for\_timestep} parameter in \mesa. The limits were set as follows: 0.01\,Myr for creating the pre-MS starting model, 0.1\,Myr for pre-MS evolution and near-ZAMS evolution (up to 50\,Myr), and 2\,Myr for evolution along the MS to TAMS, until the end of the post-MS contraction. High temporal resolution was required during the pre-MS and early MS phases due to the rapid changes in stellar structure, energy transport, and nuclear burning during these phases. Having larger time steps during this interval can lead to numerical inaccuracies that propagate through later evolution (a detailed analysis of which is being prepared for future work). However, dense profile output is not strictly necessary to maintain this accuracy. To ensure adequate sampling for \gyre pulsation frequency calculations such that key structural transitions are adequately captured, the \texttt{profile\_interval} parameter in \mesa was configured to save stellar structure profiles at every time-step during early phases and every five steps during later phases (refer to Table~\ref{tab:temporal_resolution}). These adjustments provided a balance between computational efficiency and model accuracy across all evolutionary phases. Note that the temporal sampling used in Figures \ref{fig:zams_stabilization} and \ref{fig:phases} is 5$\times$ finer during the MS and post-MS phases than that of the model grid.

            \begin{table}
                \centering
                \caption{Temporal resolution settings used in the model grid. The second column specifies the maximum allowed evolution time-step, and the third column shows the effective time interval between saved profiles for pulsation frequency calculations.}
                \resizebox{\columnwidth}{!}{
                \begin{tabular}{lcc}
                    \hline
                    \textbf{Evolutionary Phase} & \textbf{Evolution Time-step} & \textbf{Output Interval} \\
                    \hline
                    Pre-MS evolution      & 0.1 Myr  & 0.1 Myr  \\
                    Near-ZAMS evolution    & 0.1 Myr  & 0.1 Myr  \\
                    MS evolution to TAMS     & 2 Myr  & 10 Myr \\
                    Post-MS evolution     & 2 Myr & 10 Myr \\
                    \hline
                \end{tabular}
                }
                \label{tab:temporal_resolution}
            \end{table}

            The spatial resolution of stellar models was controlled using the \texttt{mesh\_delta\_coeff} parameter in \mesa, which governs the maximum permissible variation in properties between adjacent grid points. Testing \texttt{mesh\_delta\_coeff} values from 0.2 (high resolution) to 1.2 (low resolution) revealed fractional frequency differences of less than 0.5\%, with computational time reduced by 80\% for the coarser models. Fractional differences in resolution effects diminished near the ZAMS; beyond the ZAMS, during the MS, fractional differences in p-mode frequencies remained consistently below 0.1\%. We adopted the default \mesa value of \texttt{mesh\_delta\_coeff} = 1.0 to balance precision and computational efficiency.

            \change{To ensure a complete grid of models, we developed a systematic approach for the small fraction ($\sim$6.7\%) of tracks that failed to converge at the beginning of the pre-MS phase. If a model failed to converge and \mesa exited with a terminated evolution error, we recomputed from the last saved model with the option to restore the structural mesh to its previous configuration using \texttt{restore\_mesh\_on\_retry} in \mesa. In a minority of cases, two or three fresh attempts (with identical inlists and \texttt{restore\_mesh\_on\_retry} enabled) were required, consistent with floating-point-level sensitivity of the first pre-MS timestep and mesh placement near steep gradients. Reverting the mesh at each retried step likely removes the numerical trigger that causes this issue with the \mesa Newton-iteration solver. We validated this method on successfully converging evolutionary tracks and found that the resulting structural histories and adiabatic pulsation spectra agree to within $0.1\%$ in fractional frequency. For completeness, additional contingency options (e.g., tolerance relaxation) were prepared but not used. Non-converging models that benefit by relaxing tolerances are usually very rapid rotators, especially at high metallicity. These were not included in this work because we restrict to $\Omega/\Omega_{\rm crit}\le 0.3$. We archive our full \mesa run logs, which are available upon request.}
    
    \subsection{Input physics}
    \label{sec:input_physics}
        We provide our \mesa inlist template as a supplementary file, should experienced \mesa users find this a preferable way to read through the input physics and numerical setup. Any parameters not explicitly discussed in the following sections remained at their default \mesa settings.
        \subsubsection{Rotation}
            \mesa includes a one-dimensional treatment of stellar rotation based on the shellular approximation, where rotation frequency is assumed to be constant on isobaric surfaces \citep{zahn_circulation_1992, meynet_stellar_1997}. This allows rotational effects to be incorporated into 1D models by modifying the stellar structure equations to account for centrifugal acceleration and rotational deformation \citep{kippenhahn_simple_1970, Paxton2013, Paxton2019}. 

            \change{In our grid, we adopted {\sc mesa}’s standard rotation physics and made only minimal modifications. Rotation was initiated from the beginning of the pre-MS phase and each model was assigned a solid-body rotation profile throughout the evolution, with an initial surface velocity drawn from the range mentioned in Section \ref{sec:grid_input_parameters}. To allow for convergence, we relaxed each model to its target surface velocity over 100 steps, using a maximum timestep of $10^{-5}$ years.}

            The treatment of rotation in \mesa includes centrifugal corrections to pressure and temperature gradients through modification factors $f_P$ and $f_T$, which account for the deviation of equipotential surfaces from spherical symmetry \citep{Paxton2019}. These corrections are computed using analytical fits to the Roche potential of a point mass, allowing for stable calculations up to approximately $\Omega/\Omega_{\rm crit} \sim 0.9$. The implementation accounts for the reduced mean stellar density resulting from centrifugal distortion, which affects both the stellar structure and pulsation properties.

            \mesa approximates the effects of rotationally induced instabilities using diffusive prescriptions for angular momentum transport and chemical mixing. These approximations use diffusive prescriptions derived from multidimensional hydrodynamic theory but implemented within a one-dimensional stellar structure framework \citep{heger_presupernova_2000, Paxton2013}. While this approach captures the first-order effects of rotation on stellar evolution and structure, including centrifugal deformation effects on the mean density, it does not include full two-dimensional stellar structure, latitudinal differential rotation, or angular momentum transport by waves or internal gravity modes.

            To remain within the regime where first-order perturbative pulsation theory is valid, we limited our grid to models with rotation frequencies $\Omega / \Omega_{\mathrm{crit}} \leq 0.3$. This threshold avoids complications arising from centrifugal distortion and mode coupling that would otherwise affect the interpretation of pulsation frequencies. Since our goal is to construct linear adiabatic pulsation models that are representative of \dsct stars and use them to constrain stellar ages, this cutoff ensures consistency between the stellar models and the assumptions of the \gyre's pulsation framework (see Section~\ref{sec:rotation_GYRE}).
            
        \subsubsection{Convective mixing}
            \change{Convection in our models is treated using the mixing-length theory (MLT) formalism of \citet{Henyey1965}. The mixing-length parameter was set to $\alpha_{\text{MLT}} = 1.9$, which is a reasonable and commonly adopted choice for \dsct stellar models; several studies either used values near $1.8$--$1.92$ or fixed $1.9$ across grids for these stars \citep{templeton_asteroseismology_2001,Joyce_MLT_2023,murphy_grid_2023,guo_oscillation_2024}. Our chosen value lies squarely within the $\alpha_{\rm MLT}\sim1.7$--$2.1$ range supported by Time Dependent Convection (TDC) and 3D radiation--hydrodynamic calibrations across the $T_{\rm eff}$--$\log g$ space that overlaps \dsct stars \citep{dupret_theoretical_2004,trampedach_mlt_2014b,magic_stagger-grid_2015}. These calibrations further indicate trends of slight increase in $\alpha_{\rm MLT}$ toward higher $\log g$ and lower $T_{\rm eff}$, implying that $\alpha_{\rm MLT}\approx1.8$--$2.0$ is broadly applicable to \dsct stars. On a star-by-star basis, asteroseismic results have been found to be insensitive to the value of $\alpha_{\rm MLT}$ \citep{murphy_precise_2021}, albeit over a narrow region of the HR diagram. We extend this test in this work.}
            
            Convective boundaries in our models are determined using the Ledoux criterion. MLT, in general, assumes that convective energy transport occurs via fluid parcels that travel a mean free path, or mixing length ($ \lambda $), before dissolving and releasing their energy. The efficiency of this process is governed by the parameter $ \alpha_{\text{MLT}} $, defined as the ratio of the mixing length to the pressure scale height ($ H_P $), which quantifies the distance over which pressure decreases by a factor of $ e $:
            \begin{eqnarray} 
                \alpha_{\text{MLT}} \equiv \frac{\lambda}{H_P}.
            \end{eqnarray}

            The mixing-length parameter represents the efficiency of convective energy transport; larger values imply a greater distance a convective parcel travels, indicating more efficient transport of flux by convection. However, MLT simplifies convection by assuming it to be vertically bidirectional, overlooking the complex, asymmetrical and dynamic nature of actual convective flows that are characterised by the continuous deformation and restructuring of flow channels, complex flow dynamics, turbulent mixing, and multiscale dynamics. Despite these limitations in capturing the full complexity of convection, its simplicity and efficiency make it valuable for large datasets. While not perfect, it provides reasonable estimates of overall energy transport, especially for dominant large-scale features \citep{Joyce_MLT_2023}. 

            To assess the sensitivity of our pulsation models to the treatment of convection, we varied $\alpha_{\text{MLT}}$ by $\pm0.2$ around the fiducial value. We found that fractional differences in pulsation frequencies induced by this variation can reach up to 10\% during the pre-MS phase, particularly for lower-mass stars in our sample. Importantly, varying $\alpha_{\text{MLT}}$ had a negligible effect on the timing of ZAMS arrival, with differences of less than 1 Myr across all tested evolutionary tracks. On the MS, the fractional frequency differences decrease to below 5\% for lower-mass stars and below 1\% for the rest of the sample, indicating a reduced sensitivity to convective efficiency once stars settle onto the MS. The associated systematic uncertainties arising from the choice of $\alpha_{\text{MLT}}$ and the treatment of convection will be studied in detail in future work.

        \subsubsection{Convective core overshooting}
            Core overshooting was modelled using the exponential diffusive overshooting scheme, which allows for partial mixing beyond the convective boundary defined by the Ledoux criterion. This mixing occurs due to the inertia of convective cells, which penetrate into the radiative zone and extend mixing beyond the core. Such overshooting significantly impacts the evolution of stars by altering core hydrogen burning, stellar lifespans, and internal structure.

            The exponential overshooting formulation is characterized by two parameters: $ f $, which determines the extent of the overshooting layer as a fraction of the pressure scale height ($ H_P $), and $ f_0 $, which sets the width of the region near the convective boundary where the diffusion coefficient begins to decay exponentially. 

            \begin{figure*}
                \centering
                \includegraphics[width=\textwidth]{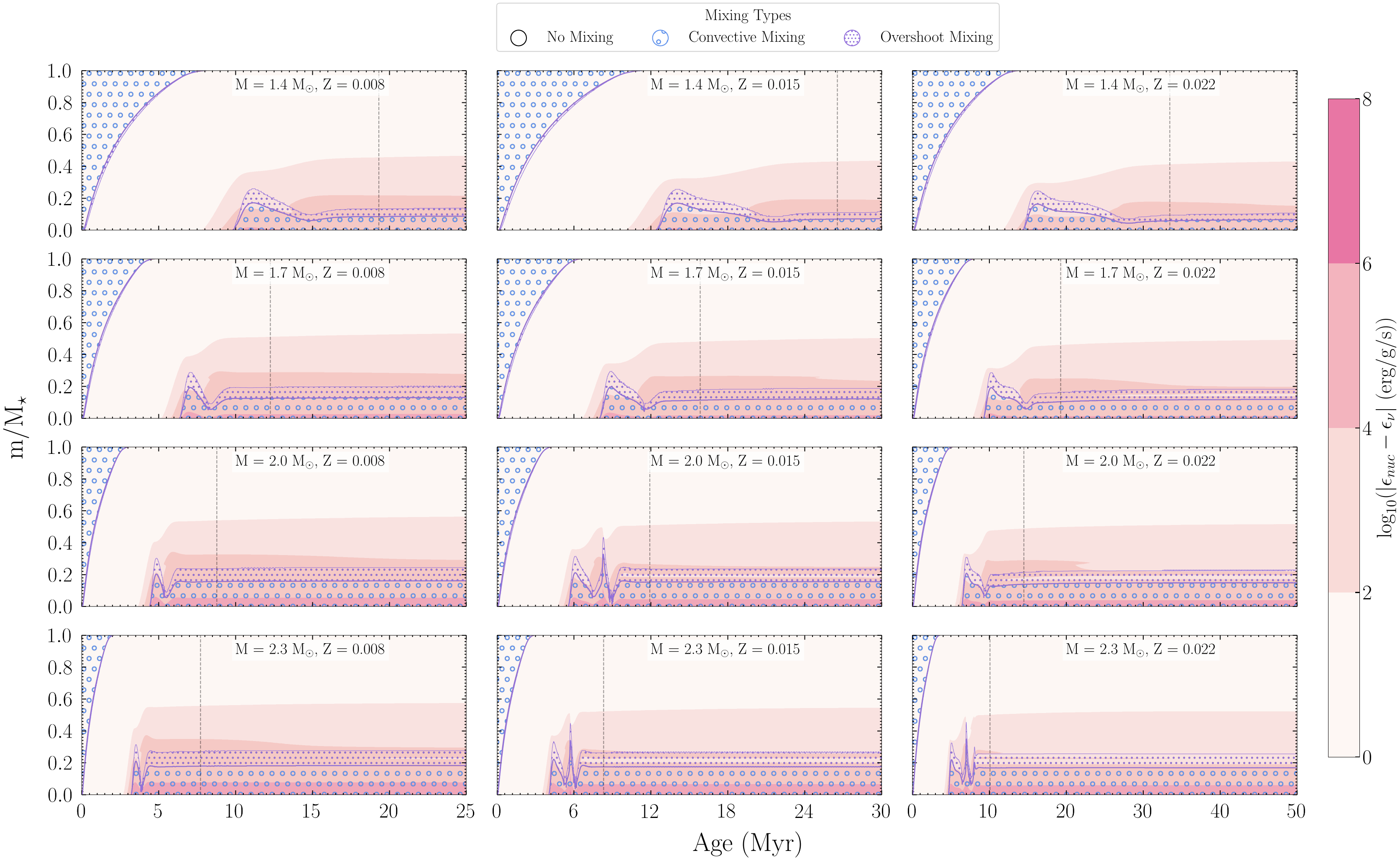}
                \caption{Kippenhahn diagrams for 12 stars from our model grid, illustrating how variations in stellar mass and metallicity influence core structure and evolution. These factors affect the size of the convective core, the extent of near-core overshooting, and the timing of the transition to stable hydrogen burning at the ZAMS (indicated by black dashed lines). Shading represents $\log_{10}$ of the net energy generation rate, defined as nuclear energy production minus neutrino losses, in units of \text{erg g}$^{-1}$\text{s}$^{-1}$. Mixing occurs within convective zones and the adjacent overshoot regions, as indicated by the shaded areas and legend. Stellar mass increases from top to bottom, and metallicity increases from left to right.}
                \label{fig:kippenhahn}
            \end{figure*} 
            
            The diffusion coefficient ($ D_{\rm ov} $) in the overshooting region is given by:

            \begin{eqnarray}
                D_{\rm ov} = D_0 e^{-\frac{2z}{H_\nu}},
            \end{eqnarray}
            where $ D_0 $ is the diffusion coefficient at the convective boundary, $ z $ is the distance from the boundary, and $ H_\nu = f H_P $ is the velocity scale height.

            In our grid, overshooting parameters were set as:
            \begin{itemize}
                \item $ f = 0.017 $ and $ f_0 = 0.002 $ at the top of convective zones, and
                \item $ f = 0.006 $ and $ f_0 = 0.001 $ at the bottom of convective zones.
            \end{itemize}
            This prescription for overshooting and the chosen parameter values are consistent with convective boundary mixing and overshooting studies \citep{Claret_Mfov_2019, pedersen_internal_2021, anders_convective_2023}. The `top of convective zones' refers to regions where a convective zone interacts with an overlying radiative zone (e.g., at the outer boundary of a convective core), while the `bottom of convective zones' corresponds to cases where a convective layer extends into the radiative interior (e.g., at the base of an envelope convection zone). 

            In intermediate-mass stars like \dsct stars, large convective cores and steep temperature gradients increase the sensitivity of stellar structure to overshooting, making the resulting variations pronounced. The size of the convective core and the extent of the overshooting region both increase with stellar mass. At fixed mass, meanwhile, increasing metallicity leads to a slight reduction in the size of the convective core. For instance, across models with masses of 1.4, 1.7, 2.0, and 2.3\,M$_\odot$, increasing $Z$ from 0.008 to 0.022 reduces the fractional enclosed core mass by approximately 0.02\,M$_\odot$. This trend is consistent across the mass range studied, indicating a weak dependence on stellar mass. This sensitivity to metallicity arises because higher metallicity increases the mean molecular weight due to the greater abundance of heavy elements, and tends to reduce the pressure scale height near the core by lowering the core temperature and increasing opacity. Despite these effects on the core, the mass encompassed by the overshooting region remains largely unaffected by metallicity. Figure~\ref{fig:kippenhahn} presents Kippenhahn diagrams for various stars across our model grid, illustrating the evolution of convective boundaries and the extent of overshooting during the pre-MS and near-ZAMS phases.

            We adopted a single overshooting value throughout our grid, based on the upper limit derived by \citet{Claret_Mfov_2019}, who examined the mass dependence of overshooting. Their study found that $ f_{\text{ov}} $ peaks at around 1.8 M$_\odot$, with lower values at lower masses and a roughly constant value at higher masses. By selecting an overshooting parameter near this maximum, we ensure consistency with empirical constraints while avoiding unphysically large convective cores in lower-mass stars. We acknowledge that overshooting is inherently mass-dependent and should ideally be treated as such to better reflect its variation with stellar mass and metallicity \citep{Claret_Mfov_2019, claret_dependence_2018, claret_dependence_2017, claret_dependence_2016, lovekin_convection_2017, dornan_effects_2022}. This will be addressed in future work, where we will investigate the impact of overshooting variations on pulsation frequencies as part of a broader study on systematic uncertainties.

        \subsubsection{Nuclear reaction network}
            In this work, we adopted a custom nuclear reaction network (\texttt{pp\_extras+hot\_cno.net}) constructed by combining the \texttt{pp\_extras} and \texttt{hot\_cno} extensions. This network incorporates the enhanced proton--proton (p--p) chain reactions and the hot CNO cycles, which are relevant for \dsct stars, while omitting unnecessary reactions like those from the cold CNO-IV cycle, which only become significant in massive stars. 

            The \texttt{pp\_extras+hot\_cno.net} network closely matches the accuracy of the more complete \texttt{pp\_and\_cno\_extras.net} network, with fractional frequency errors ($\delta f/f$) of less than 0.5\%, while reducing computation time by approximately 10\%. Compared to simpler networks like \texttt{basic.net}, which can introduce systematic errors up to 5\%, this configuration ensures efficient and reliable modelling of \dsct stars across their evolutionary phases \citep{murphy_grid_2023}.

        \subsubsection{Opacity}
            Opacity values in our models were computed using {\sc mesa}’s \texttt{kap} module, which dynamically selects and blends opacity tables according to the local temperature, density, and chemical composition throughout the stellar structure \citep{Paxton2011, Paxton2015}. The total opacity, $\kappa$, was calculated as a combination of radiative opacity ($\kappa_{\text{rad}}$) and conductive opacity ($\kappa_{\text{cond}}$):
            \begin{eqnarray}
                \frac{1}{\kappa} = \frac{1}{\kappa_{\text{rad}}} + \frac{1}{\kappa_{\text{cond}}}.
            \end{eqnarray}
            
            For higher temperatures (T $\ge 10^4$K) and hydrogen-rich conditions (Type 1 tables), the models primarily utilized the \texttt{a09} opacity tables, which are part of the OPAL \citep{Iglesias1993, Iglesias1996} dataset and include high-resolution metallicity ($Z$) values across a wide range of hydrogen fractions ($X$). Seismic modelling comparisons indicate that OPAL tables yield better agreement with observed \dsct temperatures and luminosities than OP or OPLIB alternatives, making them the preferred choice for modelling intermediate-mass pulsators \citep{jadwiga_opacity_2023}. 
            In metal-rich or hydrogen-poor conditions, the \texttt{a09\_co} (Type 2) opacity tables are employed, which account for opacity variations in chemically evolved layers.

            At lower temperatures, our models used the \texttt{lowT\_fa05\_a09p} opacity tables \citep{Ferguson2005}, which provide coverage of molecular and grain opacities relevant to cooler stellar layers. While \dsct stars have effective temperatures between approximately 6,000 and 10,000 K---where molecular species are largely dissociated and contribute negligibly to the stellar spectra---the inclusion of low-temperature opacity tables is done solely for completeness. This ensures consistency should future work extend these models to include metal-rich or chemically peculiar stars, to incorporate very low-mass stellar companions, or to investigate pulsating stars undergoing high-amplitude, shock-driven variations that momentarily cool their outer layers \citep{niu_hads_2024}.

            Recent expansions to OPLIB tables (1194 Type-1 tables) show $\le15$\% differences in solar convection zone opacities compared to OPAL \citep{farag_expanded_2024}, but \dsct modelling still favors OPAL's temperature-luminosity consistency \citep{jadwiga_opacity_2023, lenz__2010, adassuriya_mode_2022}. Our future work will explore the impact of the expanded OPLIB tables and other opacity table sources on \dsct pulsations to assess their influence on stellar structure and pulsation properties (Gautam et al. in preparation).

            \mesa blends opacity tables to allow for smooth transitions across temperature and composition conditions within each stellar model. In zones where the local temperature or composition falls within the overlap of different opacity tables, \mesa blends the tabulated values using smooth weighting functions, effectively a form of interpolation, to prevent discontinuities. This includes blending between high- and low-temperature opacity tables based on the local temperature, and between Type 1 (hydrogen-rich) and Type 2 (hydrogen-poor or C/O-enhanced) tables based on local hydrogen abundance and metallicity. In our models, all opacity blending settings were left at their default \mesa values.

        \subsubsection{Atmospheres}
            In this work, we adopted {\sc mesa}’s default atmosphere setup with a fixed Eddington $T$--$\tau$ relation \citep{eddington_internal_1926}. This approach determines the surface temperature ($T_{\text{surf}}$) and pressure ($P_{\text{surf}}$) by solving for the atmospheric structure based on the optical depth ($\tau$). 
    
            For \dsct stars during the pre-MS, near-ZAMS, MS and post-MS contraction phases, the default atmospheric treatment provides a sufficient solution. \citet{murphy_precise_2021} tested four different atmospheric prescriptions for the pre-MS star HD 139614 and found no significant effect on asteroseismic age determination. \citet{steindl_pms_2021} noted that the Eddington $T$--$\tau$ relation successfully recreates the instability region for pre-MS \dsct stars and thus preferred its use in their models. Our choice of the default Eddington $T$--$\tau$ relation is consistent with these previous studies.

            While atmospheric boundary conditions may influence specific aspects of the instability strip location, their overall impact on \dsct stars remains secondary. The pulsations in these stars are driven by the $\kappa$ mechanism in the second helium ionization zone, which is located deep within the stellar envelope, far below the atmospheric layers. Consequently, the choice of atmospheric prescription has minimal influence on global pulsation frequencies or asteroseismic properties, making the default $T$--$\tau$ treatment sufficient for this work. An assessment of systematic effects of atmosphere prescription will be presented in future work on modelling uncertainties.
    
        \subsubsection{Chemical composition}
        \label{sec:chemical_composition}
            Initial chemical abundances were determined using metallicity ($Z_{\text{init}}$) and the helium enrichment relation:
            \begin{eqnarray}
                Y_{\text{init}} = Y_0 + \left(Z_{\text{init}} - Z_{\odot,\ \text{bulk}}\right) \frac{dY}{dZ},
            \end{eqnarray}
            where $Y_0 = 0.28$ is the recommended helium abundance based on typical Galactic evolution values, and $\frac{dY}{dZ} = 1.4$ accounts for the helium enrichment ratio \citep{Verma_heliumastero_2019, Lyttle_Kepdwarfssubgiants_2021, Brogaard_helium_2012}. We used solar reference values from \citet{Asplund_solchemcomp_2009}: $Z_{\odot,\ \text{bulk}} = 0.0142$, and $Y_{\odot,\ \text{bulk}} = 0.2703$. The hydrogen ($X_{\text{init}}$) fraction was calculated as:
            \begin{eqnarray}
                X_{\text{init}} = 1 - Y_{\text{init}} - Z_{\text{init}}.
            \end{eqnarray}
    
            Isotopic ratios were used to refine the initial hydrogen and helium abundances. The deuterium-to-hydrogen ratio was set to $^{2}\text{H}/^{1}\text{H} = 2.0 \times 10^{-5}$ \citep{stahler_evolution_1980, linsky_deuterium_1998}, and the helium-3 to helium-4 ratio was set to $^{3}\text{He}/^{4}\text{He} = 1.66 \times 10^{-4}$. The individual isotopic abundances were calculated as:
            \begin{eqnarray}
            \text{Initial } ^{2}\text{H} =\ ^{2}\text{H}/^{1}\text{H} \cdot X_{\text{init}}, \quad \text{Initial } ^{1}\text{H} = (1\ -\ ^{2}\text{H}/^{1}\text{H}) \cdot X_{\text{init}},
            \end{eqnarray}
            \begin{eqnarray}
            \text{Initial } ^{3}\text{He} =\ ^{3}\text{He}/^{4}\text{He} \cdot Y_{\text{init}}, \quad \text{Initial } ^{4}\text{He} = (1\ -\ ^{3}\text{He}/^{4}\text{He}) \cdot Y_{\text{init}}.
            \end{eqnarray}
            This calculation ensures consistency with observed solar abundances while accounting for galactic helium enrichment since the Sun's formation. It provides initial conditions for hydrogen and helium abundances that match the metallicity input for our grid.

    \subsection{Stellar pulsation calculations}
    \label{sec:pulsation_calculations}
        Stellar pulsation frequencies were computed using \gyre (v7.2.1; \citealt{gyre1, gyre2, gyre3}), which numerically solves the linearized pulsation equations for stellar equilibrium models. The calculations included adiabatic frequencies for p\:modes, g\:modes, and mixed modes, up to a radial order of $n \sim 11$ and harmonic degrees $\ell = 0, 1, 2, 3$. 
    
        In our \gyre computations, we used the second-order Gauss-Legendre Magnus difference scheme (\texttt{MAGNUS\_GL2}) for models younger than 50 Myr and the second-order collocation scheme (\texttt{COLLOC\_GL2}) for older models. This selection was informed by computational efficiency tests, which showed that \texttt{MAGNUS\_GL2} delivers accurate results while being computationally faster than higher order schemes, while \texttt{COLLOC\_GL2} offers comparable accuracy for MS models with significantly reduced computation time \citep{murphy_grid_2023}.
    
        We performed frequency scans over the ranges of 0.8--95 c/d (0.8--150 c/d for models with metallicity less than 0.003) with a frequency grid of 500 points. To match the expected mode distribution, we used a linear grid (\texttt{grid\_type = `LINEAR'}) for p\:modes, which exhibit near-uniform frequency spacing at high radial orders, and an inverse frequency grid (\texttt{grid\_type = `INVERSE'}) for g\:modes, which tend toward uniform period spacing in the asymptotic limit. Since we were modelling rotating stars, we set \texttt{grid\_frame = `COROT\_I'}, following the recommendation that asymptotic behaviours apply more accurately in the co-rotating frame than in the inertial one.
        
        Spatial grid points followed the underlying \mesa model structure. We explicitly set the inner boundary coordinate to \mbox{\texttt{x\_i} = $10^{-5}$} to resolve g\:mode and mixed-mode eigenfunctions near the core while minimizing truncation errors. This also ensured consistency across models, avoiding systematic differences. We set \mbox{\texttt{w\_osc} = 10} to give at least 10 grid points per wavelength in oscillation cavities, providing sufficient resolution for p\:modes, which have short radial wavelengths. The exponential weighting parameter was set to \mbox{\texttt{w\_exp} = 2} to maintain at least two grid points per scale length in evanescent regions, preventing under-sampling in steep temperature and density gradients. Finally, \mbox{\texttt{w\_ctr} = 10} was used to increase resolution near the stellar core, allowing us to compute g\:mode eigenfunctions and mixed modes under the traditional approximation of rotation. 
        
        \change{Although \gyre\ offers both adiabatic and non-adiabatic pulsation computations, we adopted the adiabatic approximation throughout this work. This choice follows standard asteroseismic practice for large grids of models, where determining the frequencies of all possible pulsation modes is the primary objective. Non-adiabatic calculations, by contrast, are primarily used to evaluate mode stability through work integrals and growth rates ($\eta$), or to predict flux--displacement phase relations via the complex $f$-parameter. These quantities are highly sensitive to the adopted convection and mixing prescriptions.}
        
        \change{To quantify the magnitude of non-adiabatic corrections, we computed non-adiabatic pulsation frequencies using the same equilibrium models for a subset of stars from our grid. The mean fractional frequency difference between adiabatic and non-adiabatic computations, averaged over all computed modes, was $<0.4\%$ during the pre-MS, and $<0.2\%$ on the MS across our test subset. Frequencies of the highest-order p modes showed the largest absolute shifts (as expected, since they have low inertia and sample the near-surface layers where non-adiabatic effects are strongest, which means that a given perturbation shifts their frequencies more). However, these shifts remained small in fractional terms and are much smaller than modelling uncertainties arising from rotation, convection, and mixing treatments. Moreover, non-adiabatic calculations are typically $5$-$7\times$ more computationally expensive than adiabatic ones, which is prohibitive at the scale of our full grid.}

        These choices, and the others detailed in this section, follow best practices recommended in the \gyre documentation. We provide a \texttt{gyre\_template} file with all input parameters used in our computations.
        
        \subsubsection{Rotation treatment in \gyre}
        \label{sec:rotation_GYRE}
            For a spherical, non-rotating star, pulsation modes of a given spherical degree $\ell$ and azimuthal order $m$ are degenerate in frequency. Stellar rotation lifts this degeneracy through a hierarchy of effects. Among these, \gyre computes the lowest-order rotational corrections: a kinematic Doppler shift arising from the transformation between inertial and co-rotating frames, and first-order dynamical perturbations due to the Coriolis force. Together, these constitute the first-order rotational corrections and are typically sufficient for modelling frequency splittings in slowly to moderately rotating stars. 
    
            In \gyre, mode frequencies are expressed as angular frequencies $\sigma$ (in radians per second) in the inertial frame. The transformation between inertial and co-rotating frames is given by:
            \begin{eqnarray}
                \sigma_c = \sigma - m \Omega,
            \end{eqnarray}
            where $\sigma_c$ is the angular frequency in the co-rotating frame, $\Omega$ is the star's angular rotation frequency, and $m$ is the azimuthal order. This Doppler shift causes prograde modes ($m > 0$) to appear at higher frequencies and retrograde modes ($m < 0$) at lower frequencies in the observer's frame. The corresponding cyclic frequency is given by $\nu = \sigma / 2\pi$.
            
            Beyond this frame transformation, the Coriolis force introduces additional frequency perturbations, leading to the splitting of non-radial modes into multiplets. For a mode of degree $\ell$, the number of components in the multiplet is $2\ell + 1$, each corresponding to a distinct azimuthal order $m$. The observed frequency $\nu_{n \ell m}$ of a non-radial mode is given by:
            \begin{eqnarray} \label{eq:rot_splitting}
                \nu_{n \ell m} = \nu_{n \ell 0} + \frac{m\Omega}{2\pi}(1-C_{n\ell}) + D_L \frac{m^2}{\nu_{n\ell 0}} \bigg(\frac{\Omega}{2\pi}\bigg)^2,
            \end{eqnarray}
            where $\nu_{n \ell 0}$ is the unperturbed frequency (for $m = 0$), $C_{n\ell}$ is the Ledoux constant \citep{Ledoux_nonradial_1951}, and $D_L$ is a coefficient dependent on the mode geometry and the star's structural gradients \citep{Ledoux_nonradial_1951, Saio_rot_1981, Dziembowski_rot_1992}.
    
            The Ledoux constant differs significantly between g\:modes and p\:modes, owing to the differing role of the Coriolis force in each case \citep{ouazzani_rotational_2012, townsend_pulsation-rotation_2013, kurtz_asteroseismic_2014, reed_analysis_2014, guo_oscillation_2024}. With increasing radial order, the Ledoux constant $C_{n\ell}$ for g\:modes asymptotically approaches $\frac{1}{\ell(\ell+1)}$, whereas for p\:modes it tends toward zero. This contrast reflects the differing geometry of the mode eigenfunctions: high-order g\:modes are dominated by horizontal displacements, with the spherical degree $\ell$ determining the horizontal wavelength, while high-order p\:modes exhibit increasingly short radial wavelengths and more radial nodes, resulting in predominantly radial motion. \change{Because the Coriolis force acts primarily on horizontal displacements, its influence remains important for high-order g\:modes but diminishes for high-order p\:modes, leading to a correspondingly smaller $C_{n\ell}$ and larger rotational splittings} \citep{Ledoux_nonradial_1951, unno_nonradial_1989, Dziembowski_rot_1992, lopes_new_2000, aerts_book_2010}.
            
            In \gyre, the first-order effects of the Coriolis force on non-radial modes can be incorporated using either a perturbative or a non-perturbative treatment. The choice of method depends on the type of pulsation mode. In this work, we employed these methods as follows:
    
            \begin{enumerate}
                \item \textit{Perturbative treatment} \\
                In the perturbative approach, the effects of the Coriolis force are applied as a post-calculation correction to non-rotating eigenfrequencies. \gyre implements this method for p\:mode calculations, since their frequencies are much higher than the stellar rotation frequency. Consequently, the Coriolis force acts as a minor perturbation, and first-order corrections are sufficient to capture its effects. The rotational shift is expressed as:
                \begin{eqnarray}
                    \Delta \sigma \approx m (1 - C_{nl}) \Omega.
                \end{eqnarray} 
                This shift is calculated using the \texttt{domega\_rot} term output by \gyre as:
                \begin{eqnarray} 
                    \Delta \sigma = m \int_{0}^{R} \Omega \frac{\partial \beta}{\partial r} \, \mathrm{d}r,
                \end{eqnarray}
                where $ \frac{\partial \beta}{\partial r} $ is the rotational splitting kernel, and $ \beta $ depends on the mode eigenfunctions.
            
                \item \textit{Non-perturbative treatment: Traditional approximation of rotation (TAR)} \\
                In the non-perturbative approach, the rotational effects are incorporated directly into the computed frequencies using the Traditional Approximation of Rotation (TAR). The TAR simplifies the oscillation equations by assuming that horizontal displacements dominate over radial displacements, which is valid for g\:modes but not for p\:modes. \gyre does this by replacing spherical harmonics, which describe the angular dependence of oscillation modes, with Hough functions. These are solutions to Laplace's tidal equations, which explicitly account for the Coriolis force's influence on the angular structure of g\:modes. The TAR is commonly used for computing g-mode frequencies in rotating stars \citep{lee_acoustic_1993, bildsten_ocean_1996, lee_low-frequency_1997, townsend_asymptotic_2003, townsend_semi-analytical_2003}, and recent studies found that it remains applicable for rotation rates up to $\Omega = 0.2$--$0.4\,\Omega_\text{crit}$ \citep{prat_period_2017, mathis_traditional_2019, dhouib_traditional_2021}.

                \change{We applied this treatment to g\:modes, for which Coriolis effects are strong because their low pulsation frequencies are comparable to their rotation frequencies, and because their eigenfunctions are dominated by horizontal displacement, upon which the Coriolis acceleration acts most directly. The strength of rotation relative to the mode frequency is quantified by the spin parameter, $q$, defined as:
                \begin{eqnarray}
                q = \frac{2\Omega}{\sigma_\mathrm{c}},
                \end{eqnarray}
                where $\Omega$ is the supplied rotation rate and $\sigma_\mathrm{c}$ is the co-rotating eigenfrequency. Because $\sigma_\mathrm{c}$ is unknown, \gyre iteratively solves for it until it agrees with the eigenfrequency, and then maps the solution to the inertial frame by adding the standard $m\Omega$ Doppler term. Larger $q$ implies a stronger Coriolis imprint and correspondingly stronger rotation signatures in g\:mode period spacings. As stated in the \gyre documentation, this treatment is intended for low-frequency modes; it neglects centrifugal deformation, and results in a rotational shift that is consistent with the perturbative result but derived using a non-perturbative framework that incorporates the full effects of the Coriolis force on g\:modes.}
            \end{enumerate}
            
            For slowly rotating stars, p\:mode splitting is generally uniform and well described by the first-order term, $\frac{m\Omega}{2\pi}(1 - C_{n\ell})$. However, in rapidly rotating stars, second-order ($\Omega^2$) effects introduce additional perturbations \citep{Saio_rot_1981, Dziembowski_rot_1992, Reese_TOP_2021, ouazzani_rotational_2012}, which are not accounted for in \gyre's computations. These effects can introduce asymmetries in the splitting and cause deviations from the simple symmetrical multiplet patterns expected in the slow-rotation regime \citep{guo_oscillation_2024, keen_kic_2015}. Additionally, rapid rotation induces centrifugal deformation, causing the star to become oblate rather than spherically symmetric, further complicating pulsation mode calculations \citep{Reese_TOP_2021, reese_2d_2022, mirouh_forward_2022}. The Coriolis force also modifies pulsation modes by inducing circular motions, which alter the frequency structure \citep{aerts_book_2010}. The influence of these higher-order rotational effects on pulsations will be investigated in future work.
    
        \subsubsection{Addressing avoided crossings and f modes}
        \label{sec:avoided_crossings_GYRE}
            In \gyre, pulsation modes are classified using the Eckart-Scuflaire-Osaki-Takata (ESO) scheme \citep{Eckert_TAR_1963, Scuflaire_oscillations_1974, Osaki_oscillations_1975, Takata_npg_2006a, Takata_npg_2006b}. For dipole modes ($\ell = 1$), the \citet{Takata_npg_2006b} formalism was used to address limitations in the ESO scheme. Here, the generalized radial order is given by $n_{pg} = n_p - n_g + 1$ for $n_p - n_g \ge 0$ and $n_{pg} = n_p - n_g$ for $n_p - n_g < 0$. For higher-degree modes ($\ell > 1$), ESO scheme is followed and the generalized radial order is defined as $n_{pg} = n_p - n_g$. Modes with $n_{pg} > 0$ are classified as p\:modes, modes with $n_{pg} = 0$ are classified as f\:modes, while those with $n_{pg} < 0$ are classified as g\:modes. Notably, f\:modes exist only for $\ell\ge2$, since an $\ell=1$ f\:mode would physically correspond to a displacement of the stellar centre of mass, and is thus prohibited under the ESO scheme \citep{unno_nonradial_1989, Takata_npg_2006a}. These classifications are important to distinguish between p, g and f modes, particularly in more evolved \dsct stars with mixed modes where avoided crossings can occur.
    
            During avoided crossing interactions, the p and g\:modes exchange characteristics, with p\:modes gaining sensitivity to the stellar core and g\:modes becoming more influenced by the outer layers. These avoided crossings could potentially be a valuable diagnostic for probing the stellar core and envelope properties, although to our knowledge no modes that have undergone avoided crossings have yet been identified in any \dsct star. Modelling these mixed modes accurately may help them to be identified from observations.

            In addition to p and g\:modes, our models also include non-radial f\:modes for degrees $\ell = 2$ and $\ell = 3$, consistent with theoretical expectations from linear pulsation theory. These f\:modes are node-less modes with radial order $n_{pg} = 0$, distinct from both p and g\:modes. Traditionally, f\:modes have been described as surface gravity modes confined to the outer layers of the star \citep[][Sec. 7.1.4.3]{aerts_book_2010}, where the dominant restoring force is gravity rather than pressure (as in p\:modes) or buoyancy (as in g\:modes), with predominantly horizontal displacement patterns. Recent studies have revealed that f\:modes can exhibit two distinct behaviours depending on their degree: low-degree f\:modes ($\ell \lesssim 15$) behave as Lamb-like waves that propagate in the bulk of the star, while high-degree f\:modes ($\ell \gtrsim 15$) remain confined to the surface as classical surface gravity waves \citep{perrot_topological_2019, leclerc_topological_2022, saux_core-sensitive_2025}. The Lamb-like f\:modes propagate through the stellar interior and are trapped near a specific radius which is dictated by the acoustic--buoyant frequency \citep{leclerc_topological_2022}. \change{In our \dsct models ($\ell=2,3$), the f\:mode eigenfunctions show significant amplitude both in the outer layers and in the stellar interior, consistent with Lamb-like bulk propagation together with near-surface trapping. We illustrate this behaviour with eigenfunction profiles in Sec.~\ref{sec:eigenfunctions_mode_inertia}.}
            
    \subsection{Asteroseismic parameters}
    \label{sec:asteroseismic_parameters}
        The large frequency separation ($\Delta\nu$) and offset parameter ($\varepsilon$) are defined by the asymptotic relation for high-order p\:modes:
        \begin{eqnarray}
            \nu = \Delta\nu \left(n + \frac{\ell}{2} + \varepsilon\right),
        \end{eqnarray}
        where $\Delta\nu$ is inversely proportional to the acoustic wave travel time across the star's diameter and directly proportional to the square root of the mean density ($\bar{\rho}$) \citep{ulrich_determination_1986}. To determine $\Delta\nu$ and $\varepsilon$, we performed a linear fit to the frequencies of radial modes ($\ell = 0$) in the range $n = 5$ to 9, where modes are approximately equally spaced and form a vertical ridge in the \'echelle diagram \citep{bedding_highf_2020}. At lower orders, however, the spacing becomes non-uniform, causing the ridge to curve; consequently, it is not appropriate to measure $\Delta\nu$ at $n < 5$. The slope of the fit yields $\Delta\nu$, while the y-intercept divided by the slope gives $\varepsilon$ \citep{white_calculating_2011}. 
    
        The $\varepsilon$ parameter depends on the upper and lower turning points of the standing waves, and is therefore sensitive to the surface layers \citep[][Sec. 3.4.2.1]{aerts_book_2010}. While $\Delta\nu$ and $\varepsilon$ have been shown to constrain mass and age on the main sequence (MS) \citep{bedding_highf_2020, murphy_grid_2023}, their behaviour in the pre-main-sequence (pre-MS) phase is more complex. On the MS, $\Delta\nu$ primarily reflects metallicity and age, whereas $\varepsilon$ is more sensitive to mass, making them useful for disentangling fundamental stellar properties \citep{murphy_grid_2023}. However, on the pre-MS, the dependence of these parameters on age is non-monotonic, and metallicity plays an important role in shaping $\Delta\nu$ \citep{murphy_grid_2023}. This complicates the distinction between pre-MS and MS stars using only these parameters, which becomes further complicated by rotation. We analyse the behaviour of $\Delta\nu$ and $\varepsilon$ in our rotating models in Sec.\,\ref{sec:gridanalysis}.
        
\section{Grid of \dsct pulsation models}
\label{sec:gridanalysis}
    To understand the complex behaviour of asteroseismic parameters such as $\Delta\nu$ and the evolutionary progression of pulsation properties across our model grid, it is helpful to first examine the stellar evolution of a typical \dsct star. In this section, we trace the evolutionary path of a representative model to establish the physical context for the pulsation behaviours discussed throughout this work. We then examine how key stellar parameters --- mass, metallicity, and rotation --- systematically affect the distribution of models across the Hertzsprung-Russell diagram and explore the temporal evolution of individual pulsation modes. Finally, we investigate asteroseismic scaling relations and their dependencies on fundamental stellar properties, providing both theoretical insights and practical tools for interpreting observations of \dsct stars.

    \subsection{Typical \dsct Evolution}
    \label{sec:dsct_evolution}

        \begin{figure*}
            \centering
            \begin{subfigure}[t]{1.9\columnwidth} 
                \centering
                \includegraphics[width=1\linewidth]{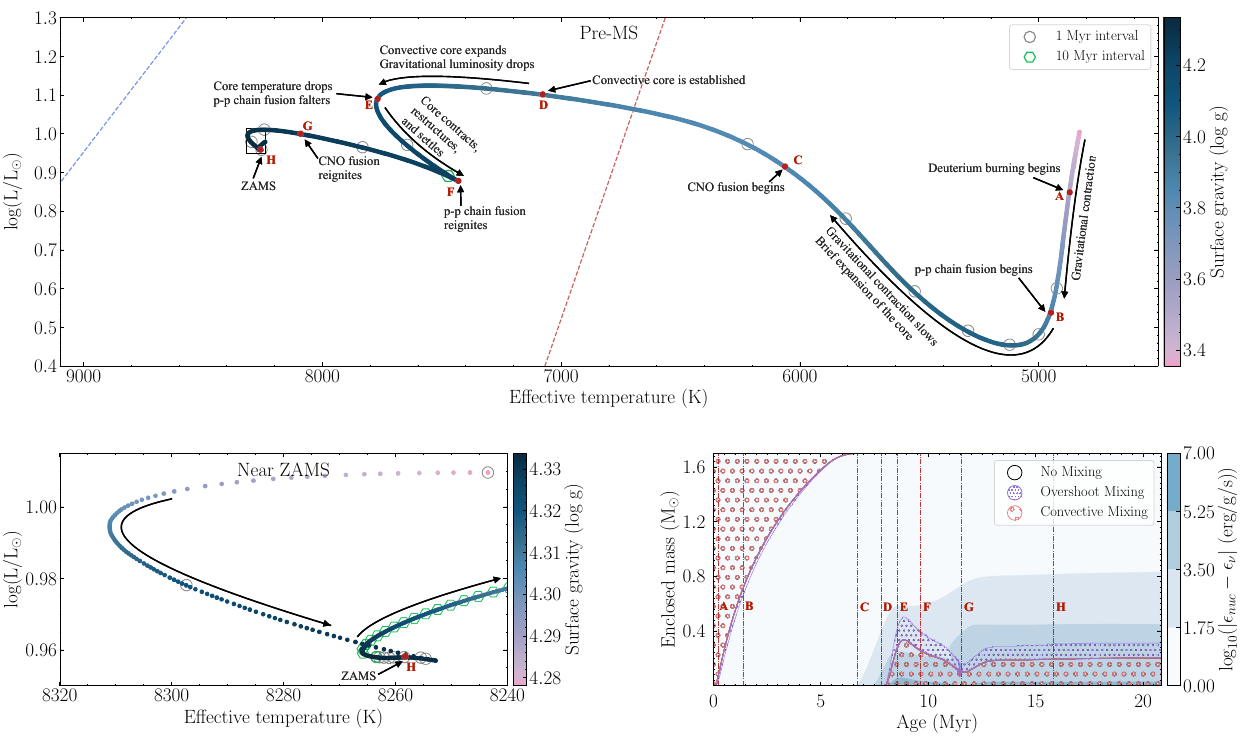}
                \caption{Pre-MS evolution up to ZAMS}
                \label{fig:phases1}
            \end{subfigure}
            
            \vspace{0.6cm}

            \begin{subfigure}[t]{1.9\columnwidth}
                \centering
                \includegraphics[width=1\linewidth]{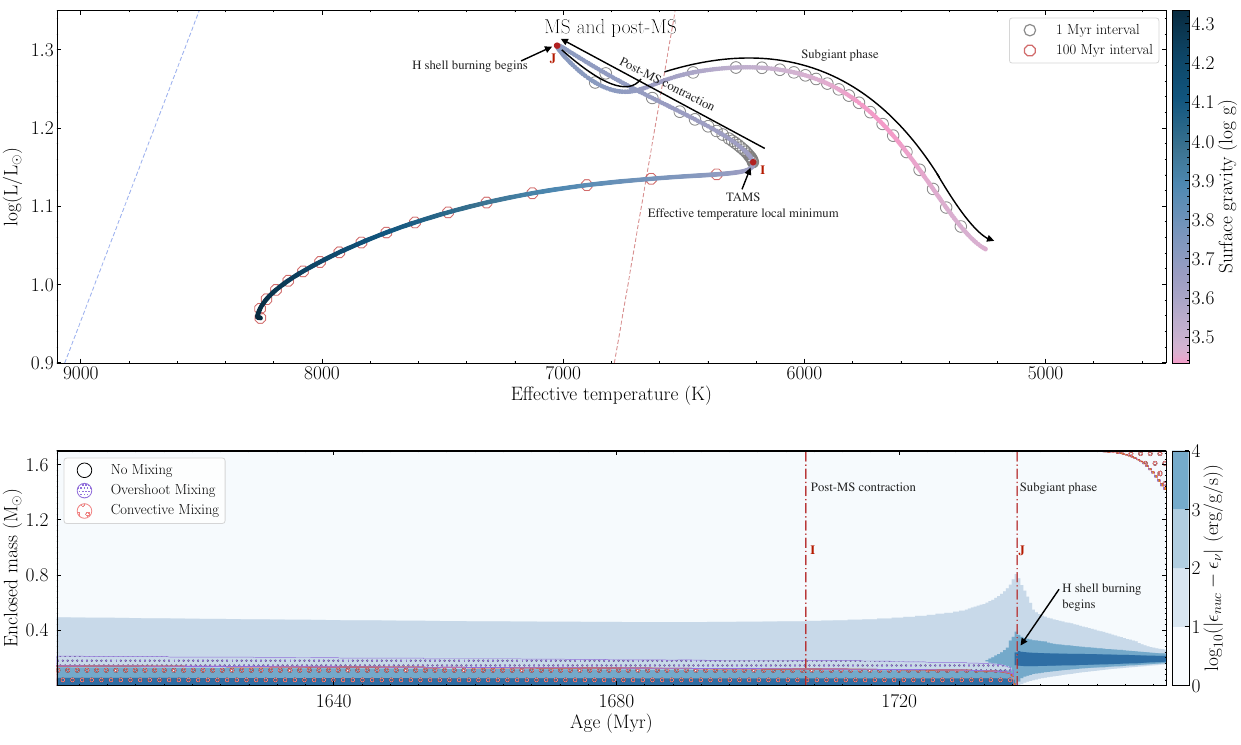}
                \caption{MS and post-MS evolution}
                \label{fig:phases2}
            \end{subfigure}
        
            \caption{Evolution of a 1.7-M$_\odot$ star with solar metallicity. Figure~\ref{fig:phases1} shows the pre-MS evolution up to the ZAMS with three panels: the top panel displays the evolutionary track on the HR diagram coloured by $\log_{10}$ of the surface gravity (cm s$^{-2}$), the bottom left panel zooms in to show the ZAMS, while the bottom right panel presents Kippenhahn diagram corresponding to the top panel. Figure~\ref{fig:phases2} shows the MS and post-MS evolution and contains two panels. The top panel shows the HR diagram evolution from the ZAMS through TAMS up to the end of the subgiant phase, while the bottom panel presents the same evolution as a Kippenhahn diagram, in which shading represents $\log_{10}$ of the net energy generation rate, defined as nuclear energy production minus neutrino losses, in units of erg g$^{-1}$s$^{-1}$. Shaded areas show mixing regions as indicated in the legend. The stellar track shown here was computed at 5$\times$ higher temporal resolution during the MS and post-MS than the grid models to more clearly illustrate phase transitions.}
            \label{fig:phases}
        \end{figure*}

        Figures~\ref{fig:phases1} and \ref{fig:phases2} illustrate the evolutionary path of a typical 1.7-M$_\odot$ star with solar metallicity, representative of \dsct pulsators in our model grid.
        During the pre-MS phase, the star begins its evolution on the Hayashi track as a fully convective protostar powered by gravitational contraction. As the core temperature rises to approximately 5~million~K \citep{stahler_formation_2004, prialnik_introduction_2010}, deuterium burning ignites via $\ce{^2_1H + ^1_1H -> ^3_2He + \gamma}$ (point A in Figure\:\ref{fig:phases1}), temporarily slowing contraction and increasing luminosity. 

        Continued contraction raises the central temperature to $\sim$6--8~million~K, initiating hydrogen fusion via the proton--proton (p--p) chain (point B). Initially, energy generation is dominated by the p--p\,I branch, which converts hydrogen to helium through a sequence beginning with $\ce{^1_1H + ^1_1H -> ^2_1H + e^+ + \nu_e}$ and culminating in $\ce{^3_2He + ^3_2He -> ^4_2He + 2 ^1_1H}$. This pathway operates efficiently at relatively low temperatures without requiring seed nuclei heavier than hydrogen.

        As the temperature increases further, the p--p\,II branch begins contributing significantly through the $\ce{^3_2He + ^4_2He -> ^7_4Be + \gamma}$ reaction sequence. This transition occurs just before the onset of CNO burning and reflects the accumulation of $\ce{^3_2He}$ from earlier p--p reactions. However, nuclear energy generation remains insufficient to fully balance gravitational contraction, and the core continues evolving toward higher temperatures and densities.

        At a core temperature of $\sim$11~million~K, the CNO cycle begins (point C). This cycle is significantly more temperature-sensitive than the p--p chains, with energy generation rates scaling approximately as $T^{16-20}$. Shortly thereafter ($\sim$30~million~K), the p--p\,III branch becomes active through $\ce{^7_4Be + ^1_1H -> ^8_5B + \gamma}$, marking the onset of $\ce{^8_5B}$ production. Here, both $\ce{^7_4Be}$ and $\ce{^8_5B}$ are short-lived intermediates rather than accumulating species. $\ce{^7_4Be}$ is destroyed either by electron capture ($\ce{^7_4Be + e^- -> ^7_3Li + \nu_e}$) followed by rapid proton capture ($\ce{^7_3Li + p -> \alpha\ + \alpha}$), or by proton capture in the p--p\,III branch. $\ce{^8_5B}$ then undergoes $\beta^+$ decay ($\ce{^8_5B ->^8_4Be^* + e^+ + \nu_e}$) to unbound $\ce{^8_4Be^*}$, which is highly unstable and promptly splits into two $\alpha$ particles.
        
        The steep temperature dependence of both the CNO cycle and p--p\,III branch drives the formation of a convective core (point D), as the corresponding rise in nuclear luminosity and central energy flux creates the conditions necessary for core convection.
        
        The growth of the convective core (from point D to E) causes a transient expansion of the stellar interior, leading to a temporary decrease in core temperature and density. This results in a brief dip in both p--p and CNO luminosities (point F). As contraction resumes, the core temperature climbs to $\sim$17--19~million~K, allowing for full operation of the p--p chain and complete activation of the CNO cycle (point G). The star soon settles into equilibrium, with nuclear burning balancing contraction, marking the ZAMS (point H).

        The main sequence (MS) phase follows and is characterized by gradual structural evolution as hydrogen is depleted in the core. The post-main-sequence (post-MS) contraction phase begins at the TAMS (point I). During this phase, as core hydrogen becomes exhausted, the lack of central energy generation causes the core to contract under its own gravity. This core contraction releases gravitational potential energy, which initially heats the overlying layers and causes the envelope to expand. The envelope expansion leads to a brief reduction in surface gravity and $T_{\rm eff}$. However, as the core continues to contract rapidly, the star as a whole begins to shrink. The resulting increase in core temperature and density drives a rise in $T_{\rm eff}$ and surface gravity, which offsets the decreasing radius and produces an increase in total luminosity. This phase concludes when the luminosity reaches its local maximum, corresponding to the ignition of hydrogen burning in a shell around the core and marking the transition to the subgiant branch. We did not follow the full subgiant evolution in our models. Instead, evolutionary tracks were terminated at the end of the post-MS contraction phase, whereupon hydrogen shell burning develops (point J).

        \begin{figure*}
            \begin{center}
            \includegraphics[width=1.9\columnwidth]{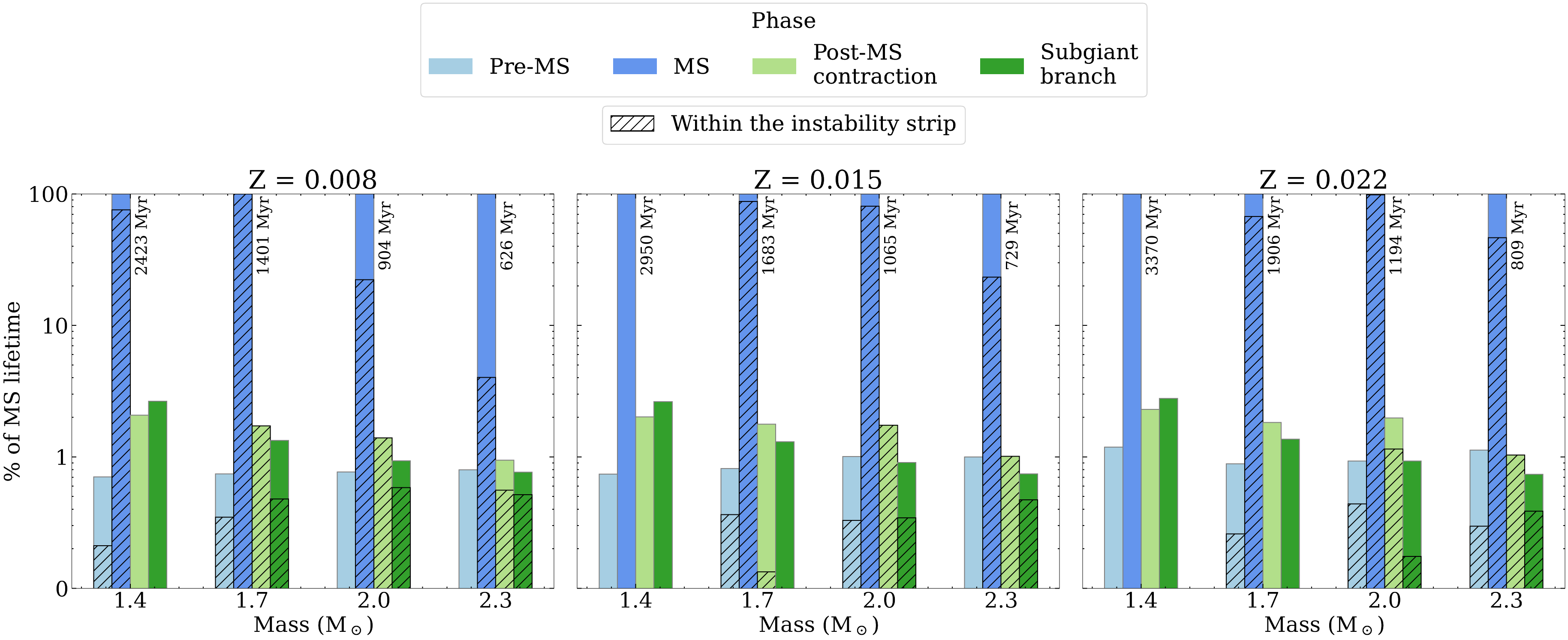}
            \caption{Percentage of the main-sequence lifetime spent in each evolutionary phase for stellar models across a range of four masses and three metallicities. Each bar represents a star with a given mass (x-axis) and metallicity (panel title), subdivided into pre-main-sequence (Pre-MS), main-sequence (MS), post-MS contraction, and subgiant branch phases. The hatched portion of each bar indicates the time spent within the classical instability strip \citep{dupret_theoretical_2004} during that phase.}
            \label{fig:barplot_phases}
            \end{center}
        \end{figure*} 

        To give an idea of timeframes of these phases, note that our 1.7-M$_\odot$ model with solar metallicity spends approximately 13.7\,Myr on the pre-MS, 1683\,Myr on the MS, and 29.8\,Myr in the post-MS contraction phase. Figure\:\ref{fig:barplot_phases} summarizes the fractional time spent in each phase for models across a range of masses and metallicities. Table\,\ref{tab:rot_impact} presents the corresponding absolute times and peak surface rotation velocities for models initiated with different rotational velocities. We find that moderate rotation (up to $\Omega/\Omega_{\rm crit} \sim 0.3$) has little effect on the durations of each evolutionary phase, with differences remaining below 1\% for the pre-MS and MS, and under 20\% for the post-MS contraction phase.

        \begin{table*}
            \centering
            \caption{Evolutionary phase durations and peak rotation velocities for stellar models of varying initial surface rotation. Each row corresponds to a unique combination of mass, metallicity, and initial surface rotation velocity.}
            \label{tab:rot_impact}
            \begin{tabular}{cc c cc cc cc}
                \toprule
                \multirow{2}{*}{Mass (M$_\odot$)} & \multirow{2}{*}{Z} & \multirow{2}{*}{$v_{\rm eq,0}$ (km/s)} &
                \multicolumn{2}{c}{Pre-MS} &
                \multicolumn{2}{c}{MS} &
                \multicolumn{2}{c}{Post-MS Contraction} \\
                \cmidrule(lr){4-5} \cmidrule(lr){6-7} \cmidrule(lr){8-9}
                 & & & Duration (Myr) & Max $v_{\rm eq}$ (km/s) & Duration (Myr) & Max $v_{\rm eq}$ (km/s) & Duration (Myr) & Max $v_{\rm eq}$ (km/s) \\
                \midrule
                \multirow{2}{*}{1.4} & \multirow{2}{*}{0.008} & 4 & 17.10 & 51.8 & 2423.3 & 53.1 & 50.3 & 54.9 \\
                \cmidrule{3-9}
                 &  & 8 & 16.53 & 104.5 & 2433.6 & 107.4 & 48.2 & 112.2 \\
                \cmidrule{1-9}
                \multirow{2}{*}{1.4} & \multirow{2}{*}{0.015} & 4 & 21.84 & 46.5 & 2950.2 & 47.8 & 59.4 & 46.4 \\
                \cmidrule{3-9}
                 &  & 8 & 22.17 & 93.7 & 2967.5 & 96.4 & 60.4 & 93.5 \\
                \cmidrule{1-9}
                \multirow{2}{*}{1.4} & \multirow{2}{*}{0.022} & 4 & 40.05 & 41.8 & 3369.9 & 42.8 & 77.4 & 39.1 \\
                \cmidrule{3-9}
                 &  & 8 & 40.08 & 83.8 & 3378.3 & 85.9 & 75.8 & 78.7 \\
                \cmidrule{1-9}
                \multirow{2}{*}{1.7} & \multirow{2}{*}{0.008} & 4 & 10.40 & 62.7 & 1401.1 & 61.9 & 24.2 & 61.4 \\
                \cmidrule{3-9}
                 &  & 8 & 10.71 & 126.8 & 1407.8 & 125.3 & 26.4 & 126.2 \\
                \cmidrule{1-9}
                \multirow{2}{*}{1.7} & \multirow{2}{*}{0.015} & 4 & 13.75 & 55.8 & 1683.0 & 55.3 & 29.8 & 53.0 \\
                \cmidrule{3-9}
                 &  & 8 & 13.94 & 112.8 & 1692.6 & 111.7 & 24.7 & 108.3 \\
                \cmidrule{1-9}
                \multirow{2}{*}{1.7} & \multirow{2}{*}{0.022} & 4 & 16.88 & 52.1 & 1905.9 & 51.7 & 34.9 & 48.5 \\
                \cmidrule{3-9}
                 &  & 8 & 17.02 & 105.2 & 1917.8 & 104.3 & 32.9 & 98.7 \\
                \cmidrule{1-9}
                \multirow{2}{*}{2.0} & \multirow{2}{*}{0.008} & 4 & 6.93 & 72.8 & 903.8 & 71.2 & 12.6 & 68.4 \\
                \cmidrule{3-9}
                 &  & 8 & 7.09 & 148.0 & 907.4 & 144.3 & 16.3 & 141.6 \\
                \cmidrule{1-9}
                \multirow{2}{*}{2.0} & \multirow{2}{*}{0.015} & 4 & 10.74 & 65.3 & 1065.0 & 63.7 & 18.6 & 59.4 \\
                \cmidrule{3-9}
                 &  & 8 & 10.63 & 132.4 & 1073.3 & 128.9 & 18.8 & 122.0 \\
                \cmidrule{1-9}
                \multirow{2}{*}{2.0} & \multirow{2}{*}{0.022} & 4 & 11.10 & 60.1 & 1194.4 & 59.0 & 23.6 & 54.3 \\
                \cmidrule{3-9}
                 &  & 8 & 11.02 & 121.7 & 1203.3 & 119.3 & 20.3 & 111.1 \\
                \cmidrule{1-9}
                \multirow{2}{*}{2.3} & \multirow{2}{*}{0.008} & 4 & 4.99 & 83.4 & 625.7 & 80.9 & 5.9 & 76.6 \\
                \cmidrule{3-9}
                 &  & 8 & 5.41 & 170.1 & 627.4 & 164.6 & 9.2 & 160.6 \\
                \cmidrule{1-9}
                \multirow{2}{*}{2.3} & \multirow{2}{*}{0.015} & 4 & 7.29 & 73.9 & 728.7 & 71.9 & 7.4 & 66.2 \\
                \cmidrule{3-9}
                 &  & 8 & 7.70 & 150.5 & 726.8 & 145.9 & 13.2 & 137.3 \\
                \cmidrule{1-9}
                \multirow{2}{*}{2.3} & \multirow{2}{*}{0.022} & 4 & 9.11 & 68.3 & 808.7 & 66.4 & 8.3 & 60.5 \\
                \cmidrule{3-9}
                 &  & 8 & 9.19 & 138.9 & 807.9 & 134.6 & 13.1 & 124.7 \\
                \bottomrule
            \end{tabular}
        \end{table*}

        Although subgiant \dsct stars may offer insights into the late stages of pulsational evolution, several factors justify our decision to exclude this phase from our model grid. While the total time spent in the subgiant phase is comparable to that of the post-MS contraction phase, the time spent within the classical instability strip is significantly shorter. As a result, the expected fraction of observable pulsating subgiant \dsct stars is low.
        Moreover, mode identification in this regime is especially challenging due to the emergence of complex pulsation spectra involving mixed modes, avoided crossings, and potential mode coupling. These features emerge due to sharp sound-speed gradients at the edge of the receding convective core, caused by chemical composition gradients left by core hydrogen burning \citep{reese_frequency_2017, dornan_effects_2022, winther_K444_2023}. \gyre, does not account for mode coupling and so computed frequencies in this regime, where such coupling is expected to be significant, may be inaccurate. The traditional approximation of rotation, which we adopted for computing g-mode frequencies, also becomes increasingly inaccurate in post-MS models \citep{mathis_traditional_2019, dhouib_traditional_2021}. Accurate treatment of pulsation--rotation coupling in this regime requires full two-dimensional modelling, which is computationally infeasible for grids of this scale.
        
        The onset of hydrogen shell burning during the subgiant phase further amplifies this complexity by introducing additional discontinuities and steep gradients in the Brunt-V\"ais\"al\"a  frequency. These structural changes reshape the p and g mode propagation cavities, strengthen mode trapping, and disrupt the regular frequency spacings required for reliable mode identification. The stellar structure during this phase also becomes highly sensitive to microphysical inputs such as convective overshooting, opacity profiles, and chemical mixing prescriptions, increasing model dependence and limiting the reliability of asteroseismic inference.

    \subsection{HR diagram trends across mass, metallicity, and rotation}
    \label{sec:hr_trends}
        \begin{figure*}
            \centering
            \begin{subfigure}[t]{1.97\columnwidth} 
                \centering
                \includegraphics[width=1\linewidth]{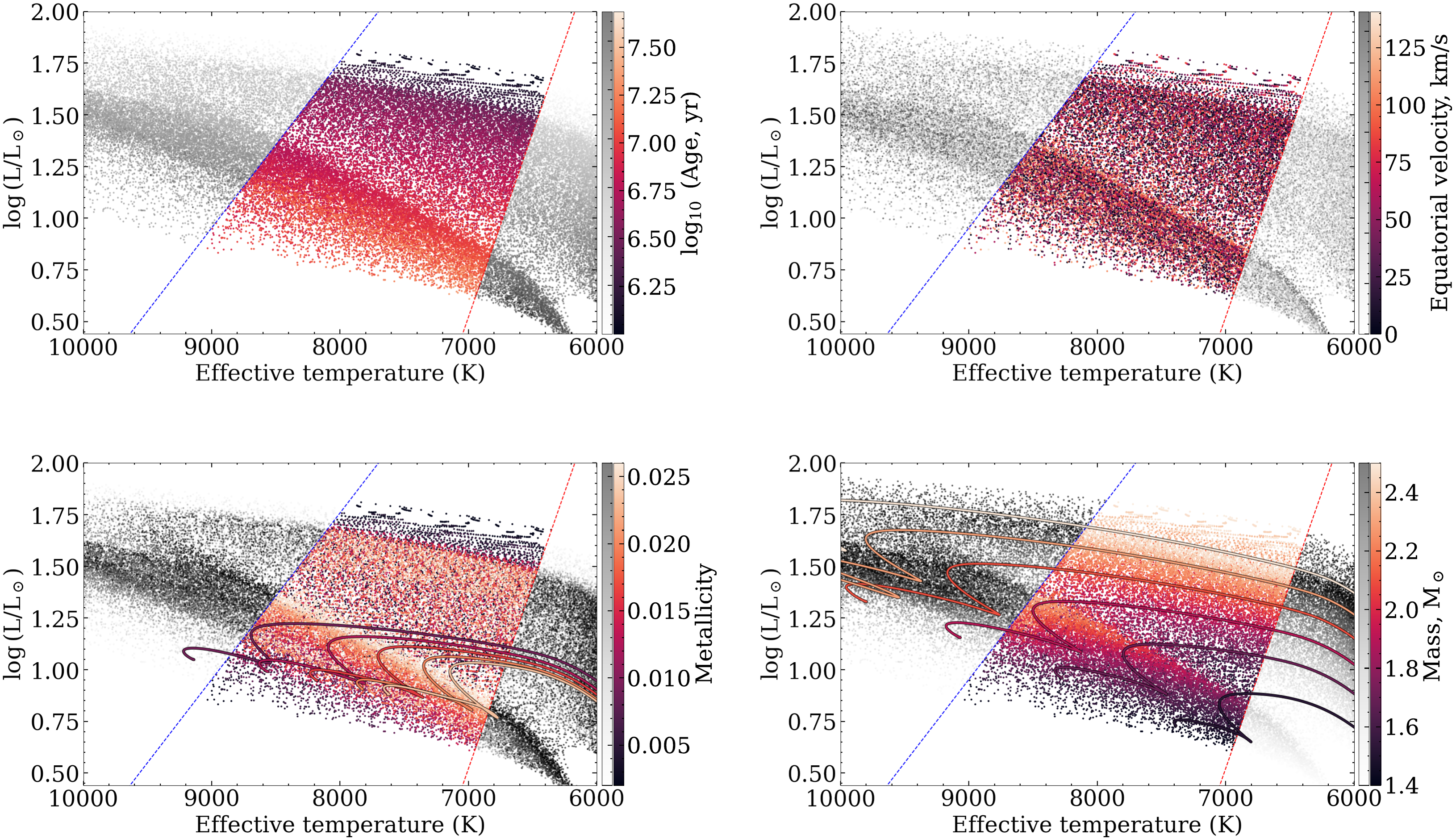}
                \caption{Pre-MS evolution up to ZAMS}
                \label{fig:HR_preMS}
            \end{subfigure}
            
            \vspace{0.6cm}

            \begin{subfigure}[t]{1.97\columnwidth}
                \centering
                \includegraphics[width=1\linewidth]{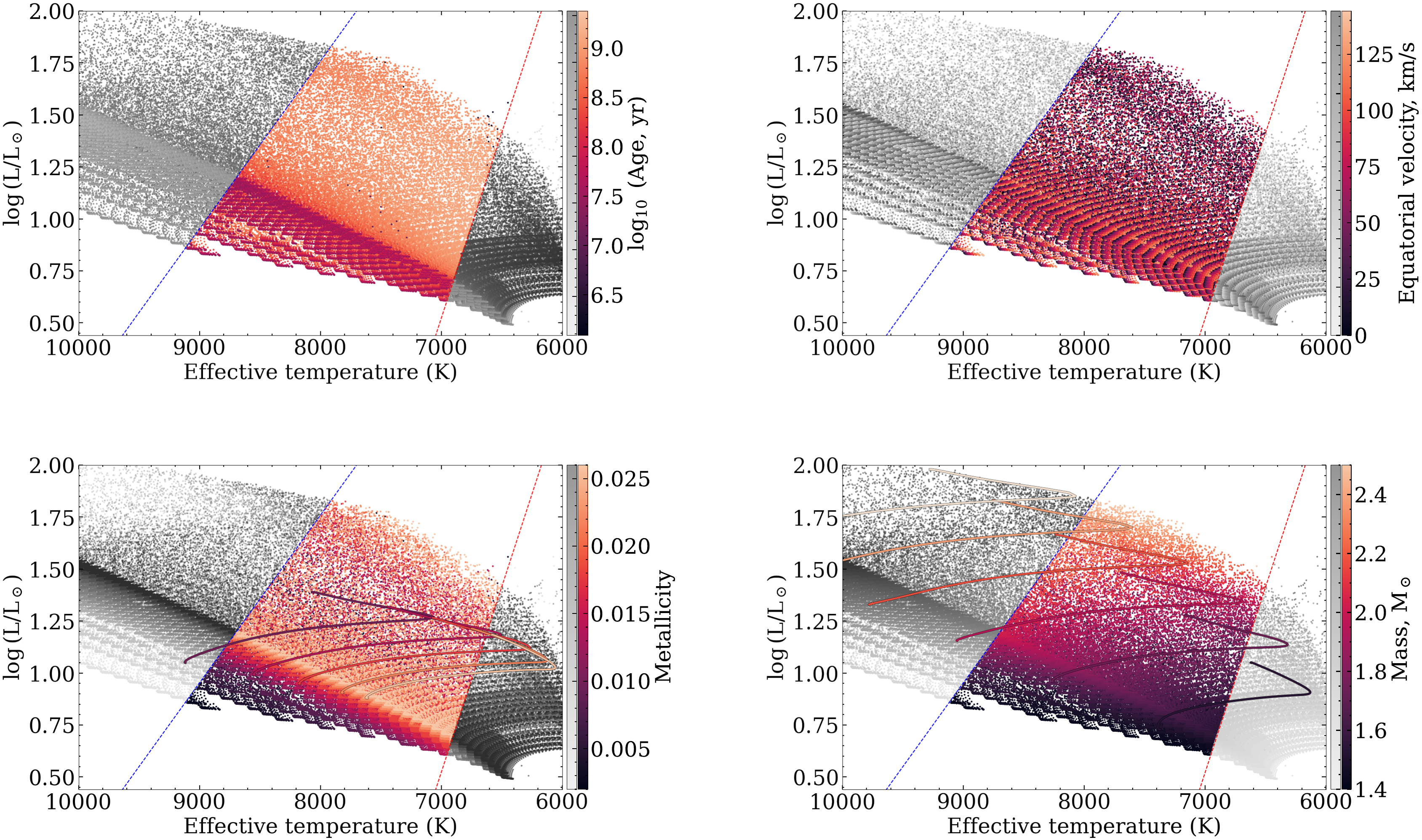}
                \caption{MS and post-MS evolution}
                \label{fig:HR_MS}
            \end{subfigure}
        
            \caption{Distribution of models across the HR diagram, separated into two panels: pre-MS models (top) and MS models (bottom). Each panel contains four subplots displaying the same models but coloured according to different stellar parameters: $\log(\mathrm{Age/Myr})$, equatorial rotation velocity ($v_{\rm eq} / {\rm km}\ {\rm s}^{-1}$), metallicity ($Z$), and stellar mass ($M/M_\odot$). For visual clarity, only every 200th pre-MS model and every 10th MS+post-MS model is plotted, with further thinning applied in metallicity. Models falling within the theoretical instability strip are shown in colour. In metallicity panels, we overlay evolutionary tracks for a fixed stellar mass of 1.7 M$_\odot$ at different metallicities, while the mass panels show tracks of different masses at fixed solar metallicity.}
            \label{fig:HR_plots}
        \end{figure*}

        The location and morphology of stellar evolutionary tracks on the Hertzsprung-Russell (HR) diagram are influenced by a star's mass, metallicity, and rotation. These dependencies shape the distribution of models across evolutionary phases and provide context for interpreting where pulsations occur in \dsct stars. We expect stellar mass to be the primary driver of HR diagram position, with metallicity introducing systematic temperature shifts, and rotation providing more subtle modifications to evolutionary tracks. Figure\:\ref{fig:HR_plots} shows how stellar parameters influence the distribution of models across the HR diagram through pre-MS, MS and post-MS evolution. 

        Mass is indeed the dominant factor setting a star’s position on the HR diagram across all evolutionary phases. As seen in the bottom-right panels, higher-mass stars occupy progressively hotter and more luminous regions, with this stratification already evident on the pre-MS and persisting through the MS and post-MS. The evolutionary pace is likewise mass-dependent: higher-mass models traverse the HR diagram more rapidly, resulting in a sparser distribution of evolved high-mass models.

        Metallicity systematically shifts evolutionary tracks in temperature, with metal-rich models appearing cooler than their metal-poor counterparts at equivalent mass and evolutionary stage. This temperature offset is due to increased opacity in high-metallicity stars, which increases the photospheric radius while maintaining similar luminosity. The metallicity effect on temperatures remains consistent across evolutionary phases. The overplotted 1.7\,M$_\odot$ tracks at various metallicities highlight this behaviour.

        Rotation has a more subtle effect on HR diagram morphology within the moderate rotation regime sampled by our grid. In the pre-MS phase (top-right panel), rotational velocities are relatively uniform, indicative of the imposed initial conditions. As stars evolve, centrifugal support from rotation leads to systematic reductions in $T_{\rm eff}$ at fixed luminosity \citep[e.g.,][]{perez_hernandez_photometric_1999, FoxMachado_pleiades_2006, Espinosa_gravdark_2011, Bedding_pleiades_2023}. For example, in 1.7-M$_\odot$ solar-metallicity models at a given age, those initialized with 4\,km\,s$^{-1}$ surface rotation are approximately 20\,K cooler than non-rotating models, while those starting at 8\,km\,s$^{-1}$ exhibit a temperature reduction of approximately 100\,K. At the same age, the corresponding decrease in luminosity is about 0.01\,L$_\odot$. These offsets scale non-linearly with rotation rate, such that higher rotation velocities produce progressively larger reductions in $T_{\rm eff}$. The effect persists throughout the MS and post-MS, shifting evolutionary tracks to cooler temperatures and lower luminosities as rotation increases.

        The instability strip intersects evolutionary tracks over a wide range of ages and masses. As indicated by the coloured models, stars within the strip span from early pre-MS contraction to post-MS expansion phases. Lower-mass stars ($M \lesssim 2\,\text{M}_\odot$) can cross the strip during both pre-MS and MS evolution, while higher-mass stars ($M \gtrsim 2,\text{M}_\odot$) tend to exhibit instability primarily during the pre-MS or near the TAMS. This pattern suggests that asteroseismic analyses of \dsct stars should account for the possibility of pre-MS as well as late-MS pulsators, particularly when studying young stellar populations or clusters with ongoing star formation. The pulsation properties readily distinguish between the two phases, as we discuss in the next subsection.

    \subsection{Evolution of p, g, and f modes}
    \label{sec:pulsation_modes_evolution}

\begin{figure*}
            \begin{center}
            \includegraphics[width=1.9\columnwidth]{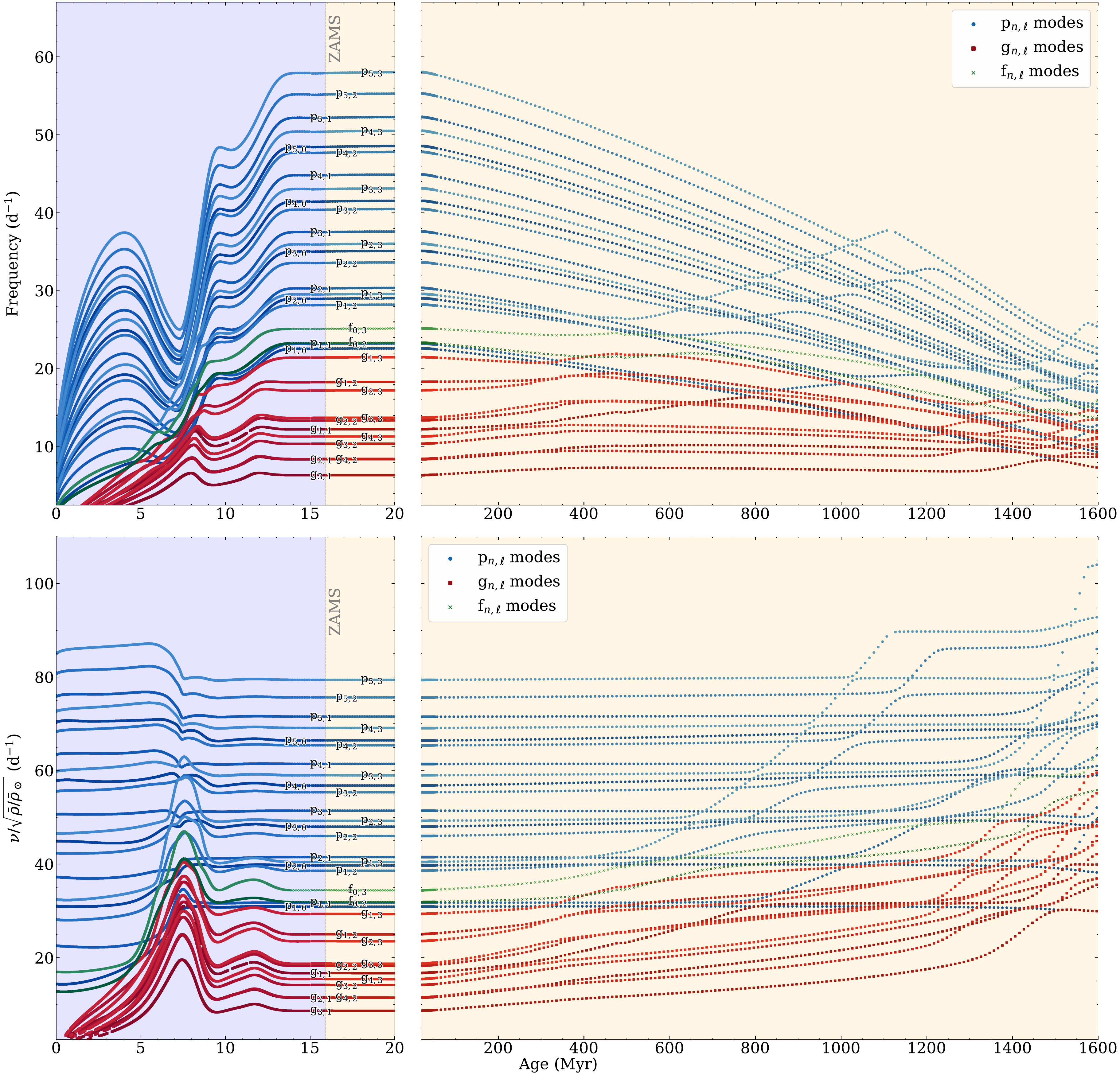}
            \caption{Evolution of stellar pulsation modes as a function of age for a 1.7-M$_\odot$, solar-metallicity star. The top panels show mode frequency evolution, while the bottom panels display frequencies scaled by mean density (normalized to solar values), which makes avoided crossings more readily visible. The left sub-panel shows the early evolution including the pre-MS and early MS phases, while the right sub-panel displays the MS evolution. The vertical dashed line marks the ZAMS. Different mode types are colour-coded: p\:modes in blue, g\:modes in red, and f\:modes in green, with lighter shades for higher degree modes. The subscripts $n,\ell$ denote the radial order and spherical degree of each mode. In addition to f\:modes for $\ell = 2$ and $\ell = 3$, only a subset of modes are shown: p\:modes with radial orders $1\le n_{pg} \le 5$ and g\:modes with radial orders $-4 \le n_{pg} \le -1$. Rotational splittings are omitted (only modes with $m=0$ are plotted). Note that the x-axis is intentionally split into two ranges (0 to 20 Myr and 20 to 1700 Myr) to highlight the rapid pre-MS evolution relative to the slower MS evolution.}
            \label{fig:mode_evolution}
            \end{center}
        \end{figure*}

\begin{figure*}
            \begin{center}
            \includegraphics[width=1.9\columnwidth]{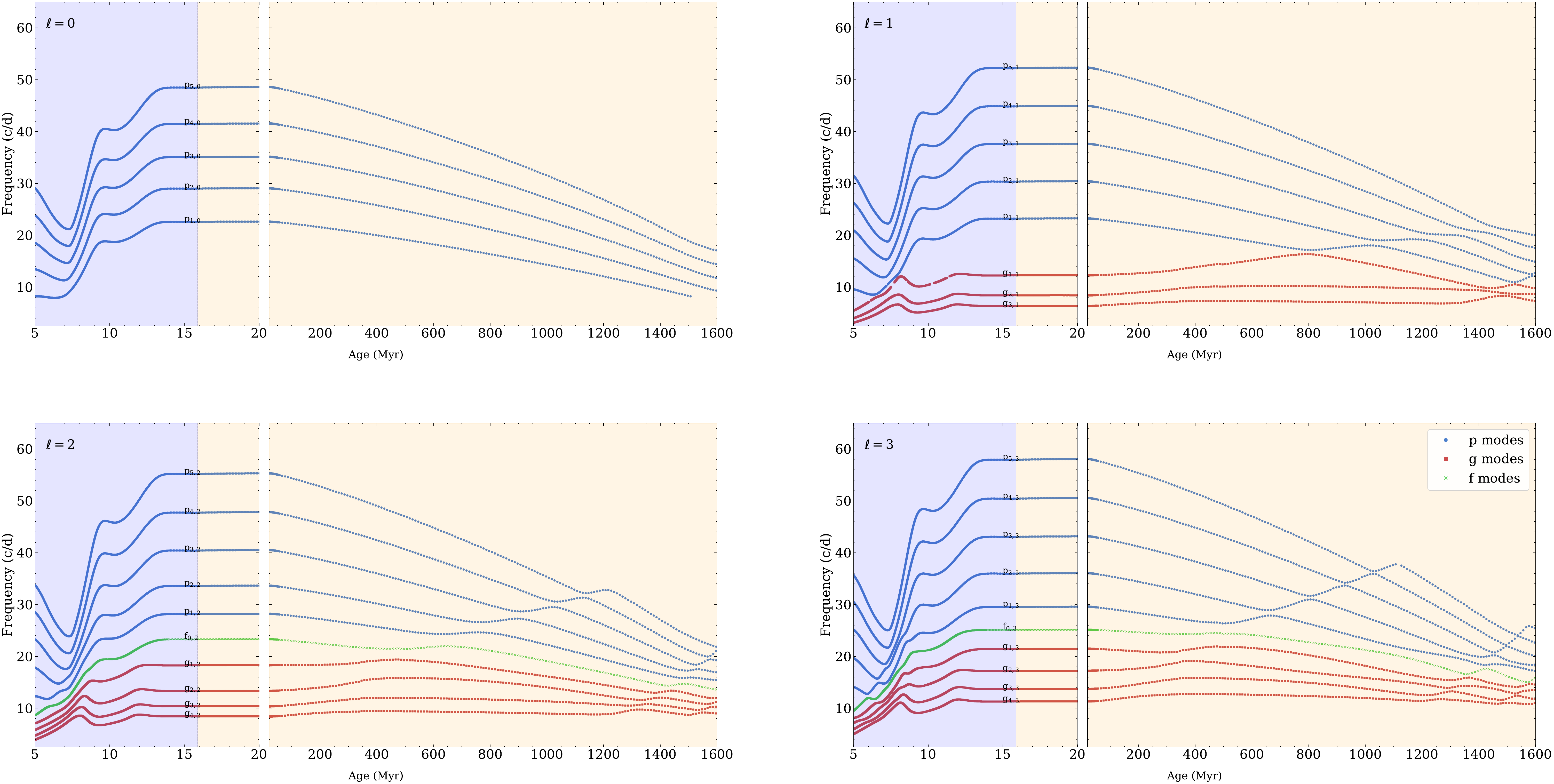}
            \caption{Evolution of stellar pulsation modes as a function of age for a 1.7-M$_\odot$, solar-metallicity star, organized by spherical degree $\ell$. Each panel displays modes of a specific degree: $\ell = 0$ (radial modes), $\ell = 1$ (dipole modes), $\ell = 2$ (quadrupole modes), and $\ell = 3$ (octupole modes).}
            \label{fig:mode_evolution_l}
            \end{center}
        \end{figure*}

        Beyond understanding the overall stellar evolution, examining the temporal behaviour of individual pulsation modes provides deeper insights into the changing internal structure. Figure\:\ref{fig:mode_evolution} illustrates the evolution of mode frequencies with stellar age for a representative 1.7~M$_\odot$ star with solar metallicity. Figure\:\ref{fig:mode_evolution_l} presents the same mode evolution but organized by spherical degree $\ell$, which more clearly reveals avoided crossing patterns. \change{Note that the classification of modes into p, f, or g\:mode types follows Takata/ESO scheme as implemented in \gyre. Consequently, the assigned mode labels correspond to the classification at the plotted age, rather than preserving a continuous mode identity along the track. As a result, neighbouring modes may exchange their physical character during avoided crossings, even though their discrete $(n,\ell)$ identifiers remain unchanged (see Sec.~\ref{sec:eigenfunctions_mode_inertia} for more discussion).}

        \subsubsection{Temporal trends in mode frequencies}
        \label{sec:mode_frequency_trends}
            
            Different pulsation modes in \dsct stars exhibit distinct evolutionary behaviours that reflect their sensitivity to changes in specific stellar regions. For the radial and non-radial p\:modes most commonly observed in \dsct stars, frequencies generally increase during the pre-MS phase until the ZAMS, after which they steadily decrease as the star ages. The most rapid frequency changes occur during the pre-MS evolution, reflecting the rapid change in the mean density of the star, with which these pulsation modes scale.

            The g\:modes, in contrast, follow a more complex evolutionary path. During early pre-MS evolution, g-mode frequencies initially increase as the radiative core develops and the Brunt-V\"ais\"al\"a  frequency (buoyancy frequency) profile becomes more pronounced in the stellar interior. These frequencies reach a local maximum near the ZAMS, followed by a slight decline. As evolution continues along the MS and the core gradually contracts, g-mode frequencies begin to increase again.
            
            This late MS rise in g-mode frequencies coincides with a steep gradient in the Brunt-V\"ais\"al\"a  frequency near the stellar core, driven by both core contraction and the buildup of a composition gradient at the core boundary. These structural changes mean the g-mode frequencies overlap with non-radial p\:modes, leading to interaction between p and g\:modes of the same degree. This gives rise to mode bumping and avoided crossings of mixed modes with hybrid p- and g\:mode characteristics. 
            
            Importantly, while the absolute frequencies of p\:modes and the large frequency separation both change with stellar evolution, the regular spacing pattern between consecutive overtones of the same angular degree is maintained throughout most of the MS evolution. For higher-order ($n=5$ to 9) radial ($\ell = 0$) p\:modes, this regular spacing roughly persists regardless of evolutionary stage since they cannot undergo avoided crossings because there are no radial g\:modes. This stability makes $\Delta\nu$ a reliable diagnostic of age for \dsct stars, as discussed in Section\,\ref{sec:asteroseismic_parameters}. In contrast, non-radial p\:modes ($\ell \geq 1$) eventually interact with g\:modes of similar frequency, causing their frequency spacings to deviate from regularity in more evolved models.
            
            Our models also include f\:modes ($n_{pg} = 0$) for degrees $\ell = 2$ and $\ell = 3$. Their frequencies evolve in a manner similar to g\:modes. In evolved models, f\:modes can also participate in avoided crossings with non-radial p\:modes of comparable frequency, further complicating the frequency spectrum. This interaction becomes particularly pronounced in post-MS models, where structural changes in the star's envelope cause f-mode frequencies to converge with those of p and g\:modes, leading to mode coupling and characteristic frequency perturbations.

            The occurrence of avoided crossings follows clear patterns across our grid. For a given mass, their onset occurs earlier at higher metallicity due to the earlier development of sharp internal chemical gradients that steepen the Brunt-V\"ais\"al\"a  frequency profile. While turning on diffusive mixing in the models can smooth these gradients locally, the overall effect on g-mode frequencies and the timing of avoided crossings remains largely unchanged, with fractional frequency differences typically below 1.5\%. We will characterise the effects of diffusive mixing more thoroughly in future work. The onset of avoided crossings first affects the higher-degree, low-order p\:modes, beginning with $\ell=3$ modes, followed by $\ell=2$, and then $\ell=1$ modes. For a given degree $\ell$, the lowest radial order p\:mode ($n=1$) is the first to interact directly with g\:modes of similar frequency. This interaction triggers a cascading effect: the perturbed $n=1$ mode bumps up the $n=2$ mode at a later evolutionary stage, which affects the $n=3$ mode at an even later stage, and so forth. Higher radial orders thus experience avoided crossings at progressively older ages. Among modes of similar radial order but different degrees, those with higher $\ell$ are affected first. 
              
        \subsubsection{Potential implications for mode identification}
        \label{sec:mode_identification_implications}
        
            One common challenge in modelling \dsct pulsations is the mismatch between observed and theoretical frequencies for the fundamental radial mode ($p_{1,0}$), even when higher-order p\:modes in the same star are well-modelled \citep{murphy_precise_2021, murphy_grid_2023}. The mode calculations in our grid demonstrate that what might appear to be a poorly modelled fundamental mode could plausibly be a g\:mode. 

           \begin{figure*}
                \centering
                \begin{subfigure}[t]{0.48\textwidth} 
                    \centering
                    \includegraphics[width=\linewidth]{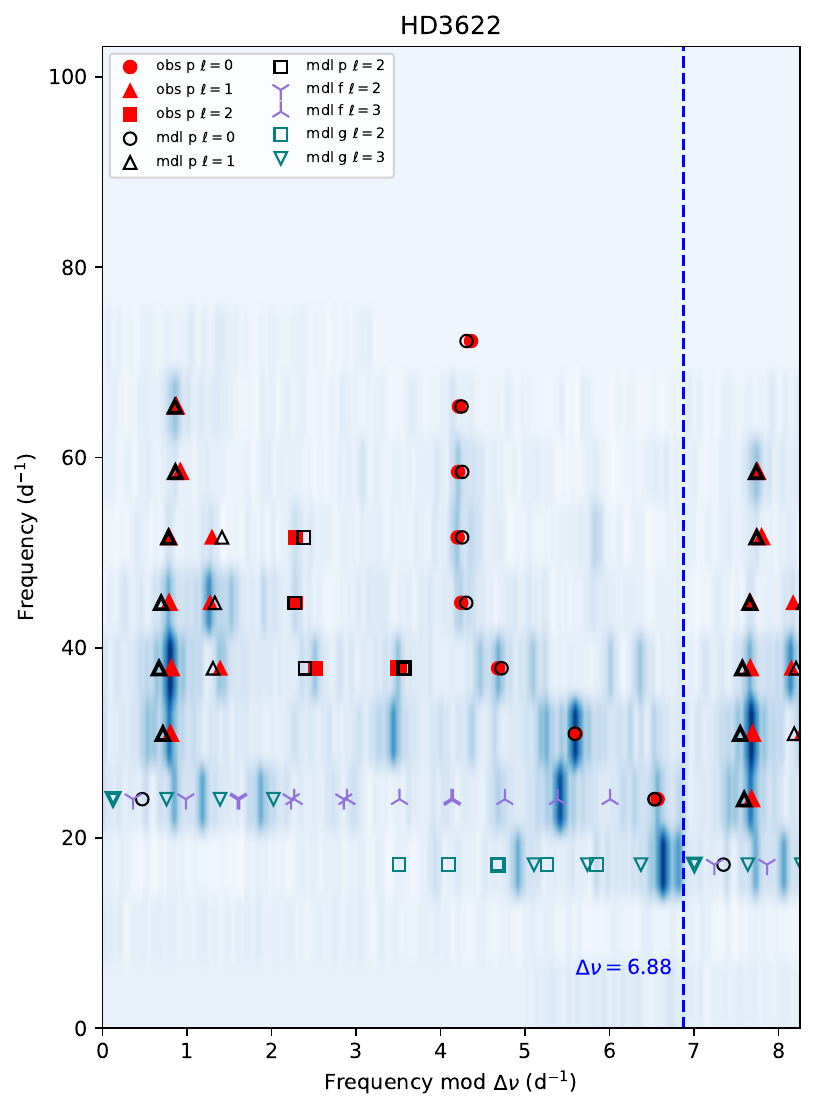}
                    \caption{HD 3622}
                \end{subfigure}
                \hfill 
                \begin{subfigure}[t]{0.48\textwidth} 
                    \centering
                    \includegraphics[width=\linewidth]{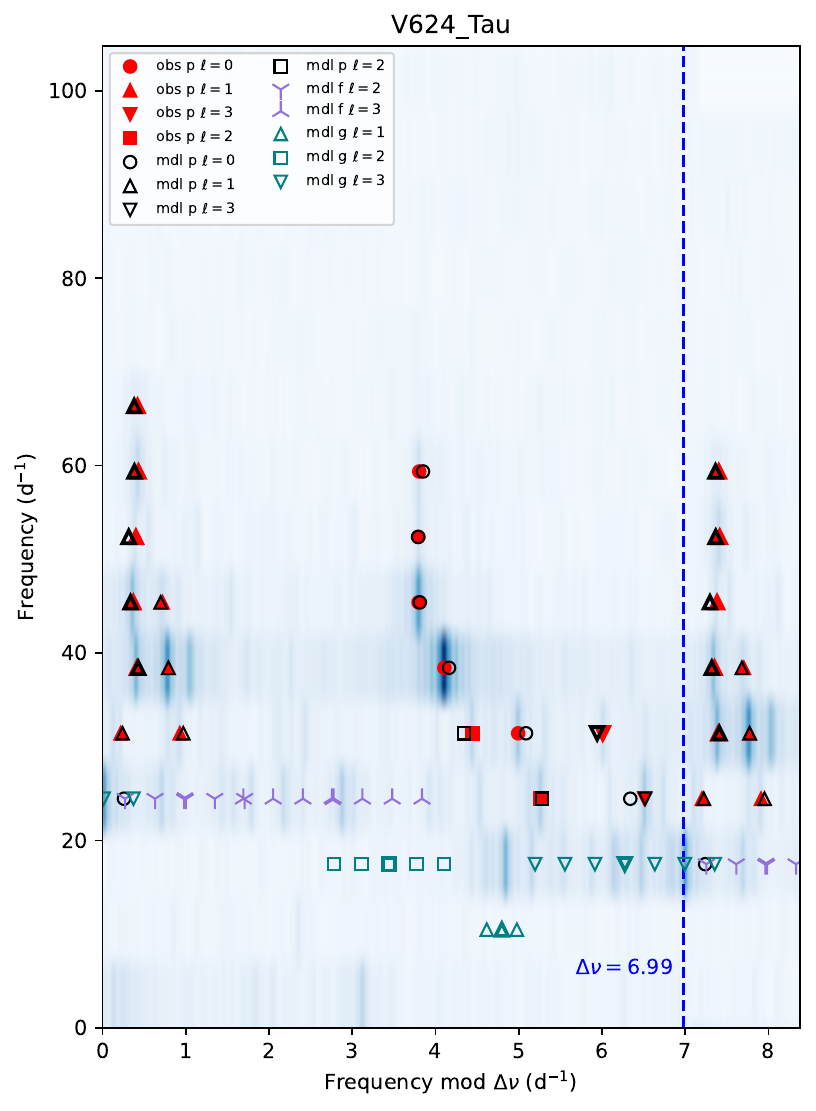}
                    \caption{V624 Tau}
                \end{subfigure}
                
                \caption{\'Echelle diagrams for HD\,3622 and the Pleiades star V624\,Tau. The red filled markers indicate the identified modes and the overlaid unfilled markers represent modes from the least squares best-fit model.}
                \label{fig:echelle_plots}
            \end{figure*}

            In young \dsct  stars, the frequencies of low-order g and f modes  can fall near the expected frequencies of the fundamental radial mode, as seen in stars like HD\,3622 and V624\,Tau (see Figure~\ref{fig:echelle_plots}). Peaks in the \'echelle diagram of frequencies slightly lower than the fundamental mode can only be explained by g\:modes with $n = 1$, $\ell = 2$ and $n = 1$, $\ell = 3$ or their rotational splittings. The f\:modes with $n = 0$, $\ell = 2$ and $n = 0$, $\ell = 3$ also lie nearby, but their frequencies are typically higher than the fundamental radial mode. The relatively high frequencies of these low-order g\:modes in \dsct stars make them distinct from the higher-order g\:modes typically found at lower frequencies in $\gamma$\,Dor stars and \dsct--$\gamma$\,Dor hybrids. 

            Model predictions indicate that these are the only modes with frequencies falling within this narrow window. As a result, the peaks observed near the fundamental radial mode and the first dipole mode in the \'echelle diagram are likely to be g and f modes. While g\:modes and f\:modes do not couple with radial modes in moderately rotating stars due to their different angular degrees, the proximity in frequency could lead to observational complications through mode misidentification or contribute to amplitude variations and mode visibility through non-adiabatic effects in the excitation region.

            Pulsational mode amplitudes for \dsct stars cannot be predicted from linear adiabatic theory, but one way to compare relative mode amplitudes is to investigate the mode inertia. Modes with higher inertia should have lower amplitudes, all else being equal, as they require more energy to sustain the same displacement amplitude. We computed an amplitude ratio of 0.77 between the fundamental radial mode ($n=1$, $\ell=0$) and the $\ell=3$ g\:mode in our least squares best-fit model for HD\,3622, based on the relative inertias of these modes. We expand on this topic in Sec~\ref{sec:eigenfunctions_mode_inertia}.

            This frequency coincidence between inherently different mode types stresses the importance of computing the full theoretical mode spectrum by including p, g and f modes along with their rotational splittings when modelling \dsct stars. By producing all possible theoretical mode frequencies across different spherical degrees, our grid takes a much-needed step toward resolving these identification challenges. 
            
    \subsection{Mode eigenfunctions and inertias}
    \label{sec:eigenfunctions_mode_inertia}
\begin{figure*}
            \centering
            \begin{subfigure}{0.342\textwidth}
                \centering
                \includegraphics[width=\linewidth]{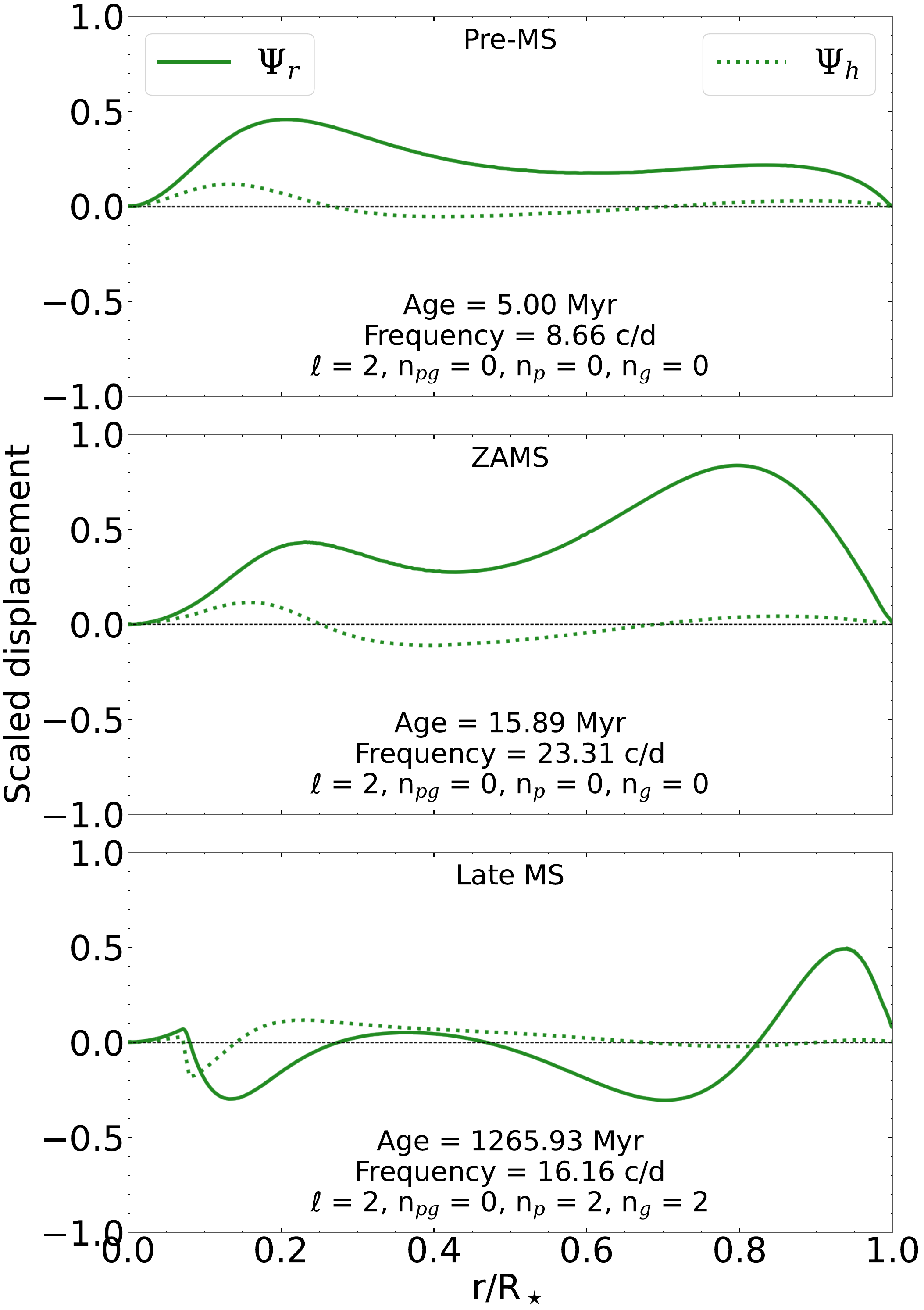}
                \caption{$\ell=2$ f\:mode}
                \label{fig:eigenfunction_f}
            \end{subfigure}
            \begin{subfigure}{0.3055\textwidth}
                \centering
                \includegraphics[width=\linewidth]{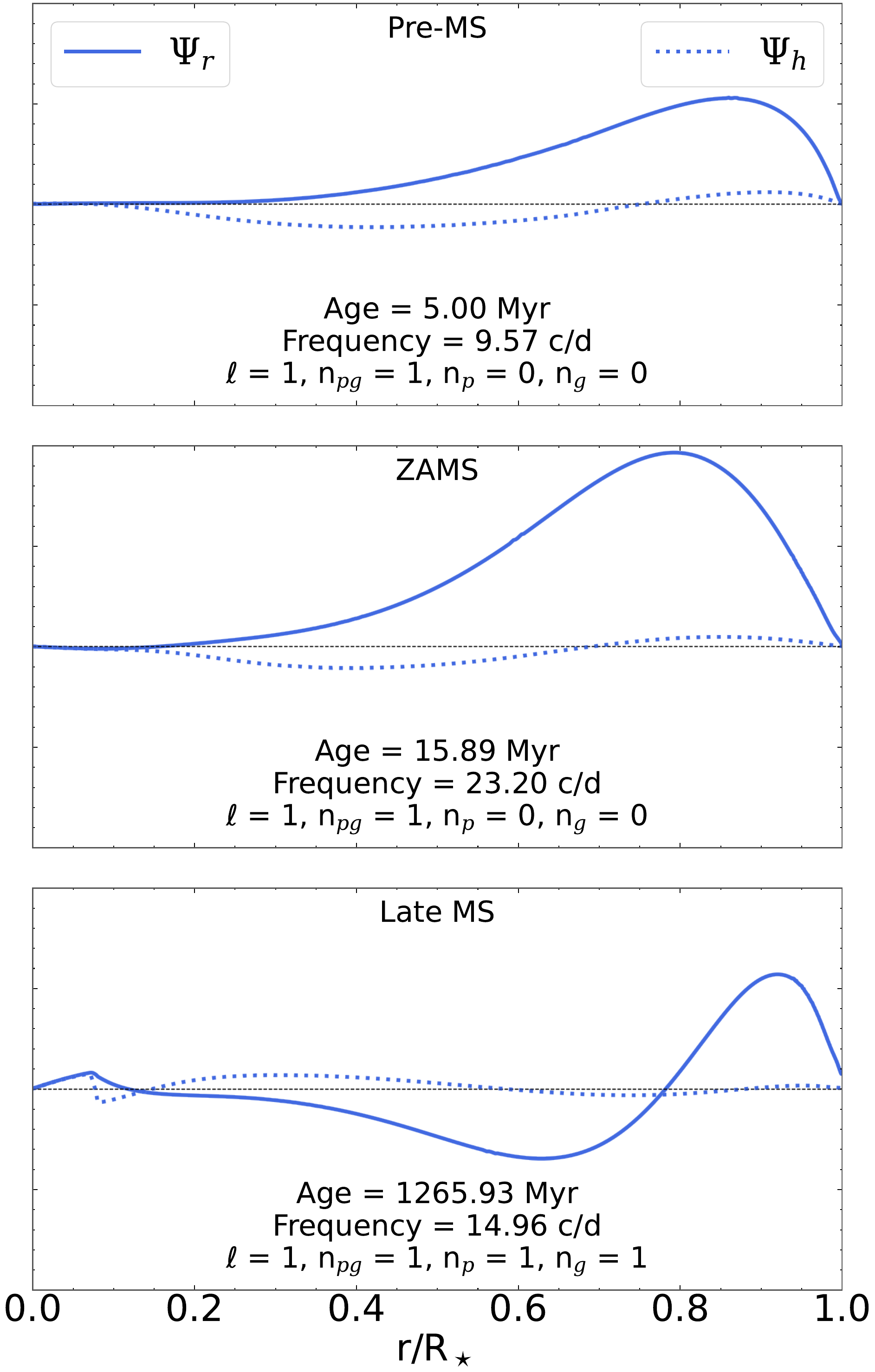}
                \caption{$\ell=1$, $n=1$ p\:mode}
                \label{fig:eigenfunction_p}
            \end{subfigure}
                \begin{subfigure}{0.335\textwidth}
                \centering
                \includegraphics[width=\linewidth]{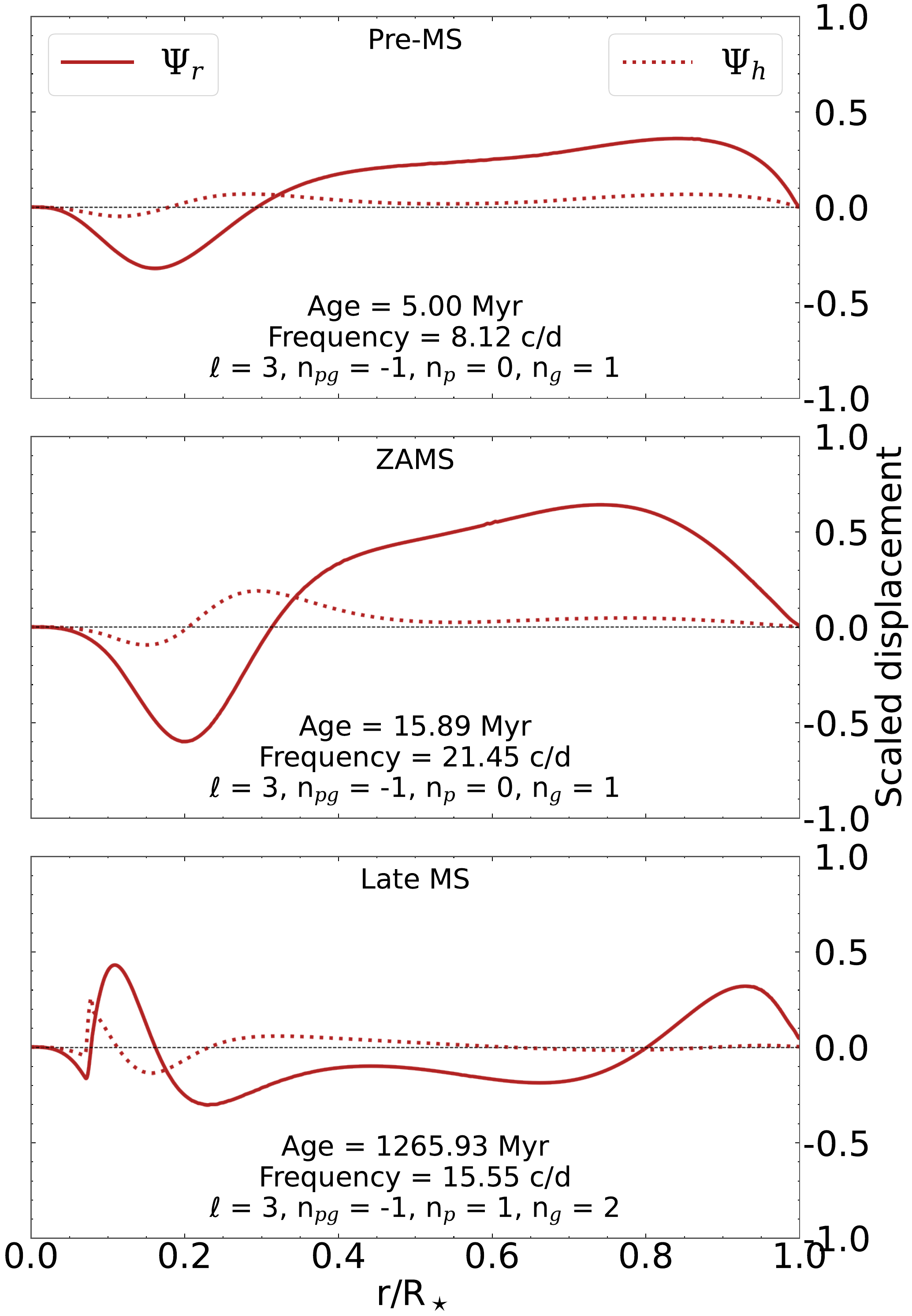}
                \caption{$\ell=3$, $n=1$ g\:mode}
                \label{fig:eigenfunction_g}
            \end{subfigure}
        
            \caption{\change{Scaled eigenfunctions for representative f, p, and g modes across evolution. Each subfigure shows the radial ($\Psi_r$, solid line) and horizontal ($\Psi_h$, dotted line) components of the scaled eigenfunctions at three evolutionary stages (top to bottom: pre-MS, ZAMS, MS). (a) $\ell=2$ f mode ($n=0$, $m=0$); (b) $\ell=1$ p$_1$ mode ($n=1$, $m=0$); (c) $\ell=3$ g$_1$ mode ($n=1$, $m=0$).}}
            \label{fig:eigenfunctions}
        \end{figure*}

        \change{To understand the relative importance of g and f modes with respect to p modes at different ages, it is useful to compare their mode eigenfunctions and inertias. We plot the scaled eigenfunctions $\Psi_r$ and $\Psi_h$ for selected p, g, and f modes in Figure \ref{fig:eigenfunctions}. Here, $\Psi_r = r\,\sqrt{\rho\,c_s}\,\xi_r$, where $\rho$ is the local mass density, $r$ is the radial coordinate measured from the stellar centre, $c_s$ is the adiabatic sound speed, and $\xi_r$ is the radial component of the Lagrangian displacement eigenfunction (analogously, $\Psi_h = r\,\sqrt{\rho\,c_s}\,\xi_h$ with $\xi_h$ being the horizontal component). We show these eigenfunctions for three representative evolutionary stages: a pre-MS model, a ZAMS mode, and a late-MS model when avoided crossings occur. The corresponding propagation diagrams for these models are provided in Appendix~\ref{app:propagation_diagrams}. The curves of $\Psi_r$ in Figure \ref{fig:eigenfunction_f} show that f\:modes have significant displacement throughout the bulk and are not just confined to the surface. Their eigenfunctions are very similar to the $n=1, \ell=1$ p\:mode (as well as to the fundamental radial mode) for the outer 50\% of the stellar radius, and they achieve similar radial displacements.}

\begin{figure*}
            \begin{center}
            \includegraphics[width=1.9\columnwidth]{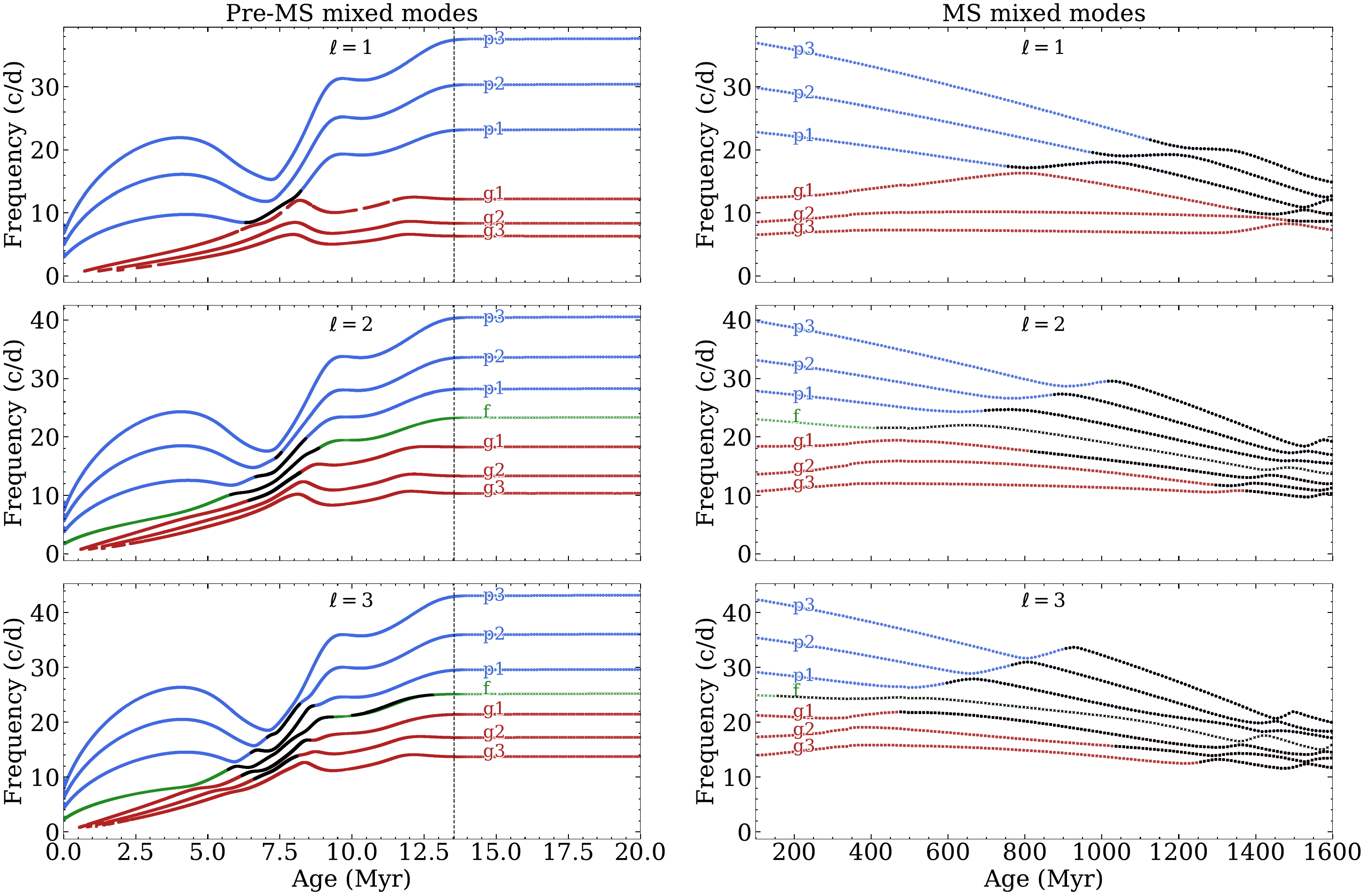}
            \caption{\change{Occurrence of mixed modes throughout the evolution from pre-MS to late MS. Mode evolution for $\ell=1,2,3$ is shown in vertically stacked panels. Left-hand panels focus on the pre-MS, where mixed character appears for a short while in certain modes prior to the establishment of a stable convective core; right-hand panels show the late-MS phase showing avoided crossings of mixed modes. Evolution of selected p, g and f modes are shown in blue, red and green colours, respectively. Mixed modes are overlaid in black.}}
            \label{fig:mixed_modes}
            \end{center}
        \end{figure*}

        \change{We note here that mixed modes have more radial nodes than their $n_{pg}$ values indicate. This is because their acoustic ($n_p$) and gravity ($n_g$) wave winding numbers are both non-zero, since these modes have mixed character. Notably, mixed modes are present not only during the late MS phase, where avoided crossings are common, but also during the pre-MS phase prior to the establishment of a stable convective core (see Fig.\,\ref{fig:mixed_modes}). Some modes change character several times on the pre-MS. This is because the radial–displacement eigenfunction, $\Psi_r$, develops a turning point that drifts toward and occasionally crosses the $\Psi_r=0$ line. When this extremum grazes or crosses zero, one or two additional nodes are created, respectively, each of which increments both the acoustic and gravity winding numbers $(n_p,n_g) \to (n_p + 1,n_g + 1)$. When the extremum recedes from zero, the nodes are removed, giving $(n_p,n_g) \to (n_p - 1,n_g - 1)$. Importantly, the $n_{pg}$ identification of the mode remains unchanged. This repeats throughout phases of mixed mode behaviour over short evolutionary intervals. A clear example is the $\ell=3$ $f$ mode, whose mixed characteristics come and go over $\lesssim 5$\,Myr during the pre-MS phase.}
    
\begin{figure*}
            \begin{center}
            \includegraphics[width=2\columnwidth]{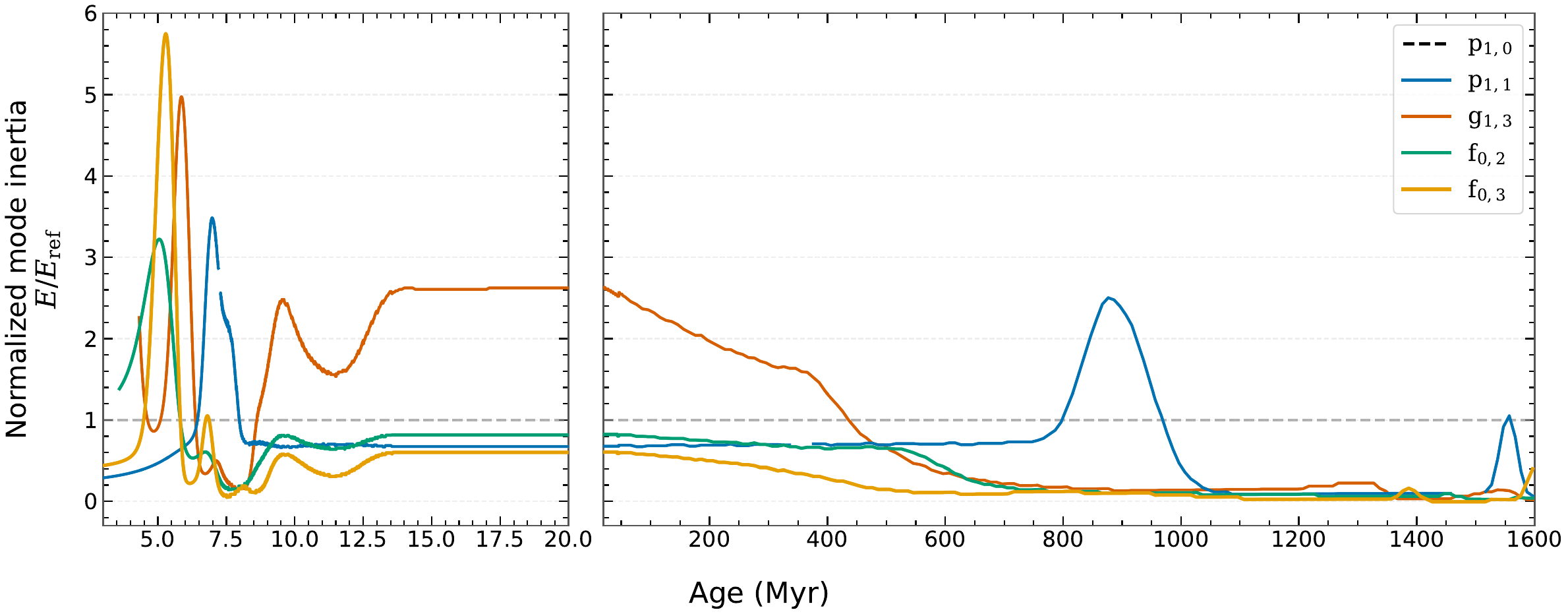}
            \caption{\change{Normalized mode inertia throughout the evolution of a 1.7-M$_\odot$, solar-metallicity star. For each mode we plot the fraction $E/E_{\rm ref}$, where $E_{\rm ref}$ is the value for the fundamental radial mode (p$_{1,0}$).}}
            \label{fig:mode_inertia}
            \end{center}
        \end{figure*}

        \change{We should presumably expect to observe these f modes in \dsct stars. To quantify their relative amplitudes, we computed their normalized mode inertias using the (dimensionless) radius $x=r/R_\star$ and the structure profile for $\rho(x)$. For each mode we formed the inertia kernel \citep[Eq. 3.139]{aerts_book_2010}
        \begin{equation}
            K(x) = \rho(x)\,x^{2}\,\Big(|\tilde\xi_r|^{2}\;+\;\lambda\,|\tilde\xi_h|^{2}\Big),
        \end{equation}
        where $\lambda$ is the eigenvalue of Laplace's tidal equation (under the traditional approximation of rotation) returned by \textsc{gyre}. In the non-rotating limit, we have $\lambda \rightarrow \ell(\ell + 1)$ \citep{townsend_asymptotic_2003, aerts_book_2010}. We integrated $E=\int_{0}^{1} K(x)\,\mathrm{d}x$ with a trapezoidal rule and normalized by the inertia of a reference mode, $E_{\rm ref}$, taken to be the fundamental radial mode. As can be seen in Fig. \ref{fig:mode_inertia}, f\:modes have lower mode inertias than the fundamental radial mode. Since $A_{\rm surf}\ \propto\ \sqrt{\eta/E}$, with $\eta$ as the (per-cycle) normalized growth rate (or stability parameter), lower inertia implies higher photometric amplitudes, assuming they have similar driving and damping. The comparable mode inertia and similar radial displacement supports the detectability of these f\:modes and low-order g\:modes in HD\,3622 and V624\,Tau, even with geometric cancellation for $\ell>1$. The modest decrease in relative f- and g-mode inertias during the first few hundred Myr of MS evolution suggest that these modes are detectable in many young \dsct stars.}
    
\section{$\Delta\nu$ scaling relations and asteroseismic diagnostics}
    Scaling relations for \dsct stars offer a tool for quick stellar characterization without the need for complex matching of individual frequencies. Particularly relevant are the connections between large frequency separation ($\Delta\nu$), the fundamental mode frequency ($\mathrm{p}_{n_1\ell_0}$), and mean stellar density ($\bar{\rho}$) \citep{garcia_hernandez_observational_2015, bedding_highf_2020, barac_revisiting_2022, murphy_grid_2023}. Here, we investigate how these relations manifest in our grid of models, and explore how they vary with stellar age, mass, metallicity, and rotation. \change{Throughout this section we restrict our analyses to models within the classical \dsct instability strip of \citet{dupret_theoretical_2004}, focusing on models representative of the observed \dsct population.}

    \subsection{The $\Delta\nu$--$\bar{\rho}$ scaling relation}
        The large frequency separation inherits the scaling behaviour of individual p-mode frequencies with square root of mean stellar density \citep{ulrich_determination_1986, kjeldsen_amplitudes_1995}:
        \begin{eqnarray}
            \Delta\nu \propto \sqrt{\bar{\rho}} \ \ .
        \end{eqnarray}
        \begin{figure}
            \begin{center}
            \includegraphics[width=\linewidth]{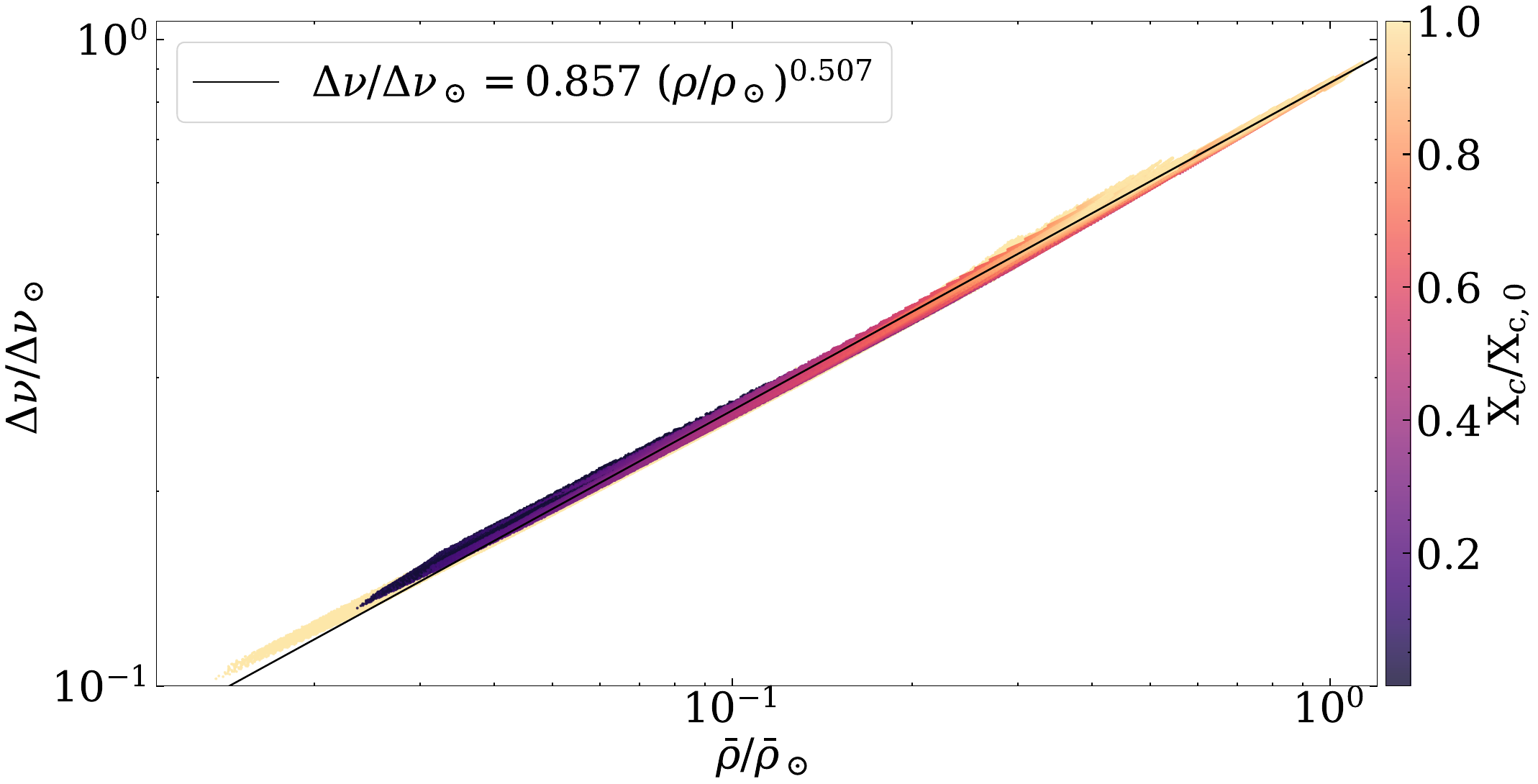}
            \caption{Power-law fit to the scaling relation between large frequency separation ($\Delta\nu$) and mean stellar density ($\bar{\rho}$) for models within the classical instability strip. Points are coloured by the central hydrogen fraction normalized to its initial value ($X_\mathrm{c}/X_\mathrm{c,0}$), providing a proxy for stellar age and evolutionary phase. Since our models are non-uniformly sampled in age, we display every 20th model for ages below 50\,Myr to reduce visual clutter, while all MS and post-MS models are shown in full. The best-fit relation, shown in black, was derived using the full set of instability strip models.}
            \label{fig:Dnu_rho_scaling}
            \end{center}
        \end{figure} 
        Figure\:\ref{fig:Dnu_rho_scaling} shows this relation for our grid. From a linear regression (in the log--log scale) of models within the instability strip, we find:
        \begin{eqnarray}
            \Delta\nu/\Delta\nu_\odot = 0.857 \times (\bar{\rho}/\bar{\rho}_\odot)^{0.507}.
        \end{eqnarray}
        To facilitate comparison, we adopt $\Delta\nu_\odot = 11.655$\,d$^{-1}$ or $134.9 \ \mu \text{Hz}$ \citep{kjeldsen_amplitudes_1995}. The best-fit power-law exponent of 0.5070 is in excellent agreement with the theoretical expectation of $0.5$, and the fitted scaling factor $f_{\Delta\nu} = 0.8574$ is broadly consistent with previous model-based and observational calibrations \citep{garcia_hernandez_observational_2015, murphy_grid_2023, bedding_highf_2020}. The standard deviation of the residuals is 0.007\,dex, which is a scatter of 1.017\% in linear space, and the fit achieves an R$^2$ score of 0.998. 

\begin{figure*}
            \centering
            \begin{subfigure}[t]{1.97\columnwidth} 
                \centering
                \includegraphics[width=1\linewidth]{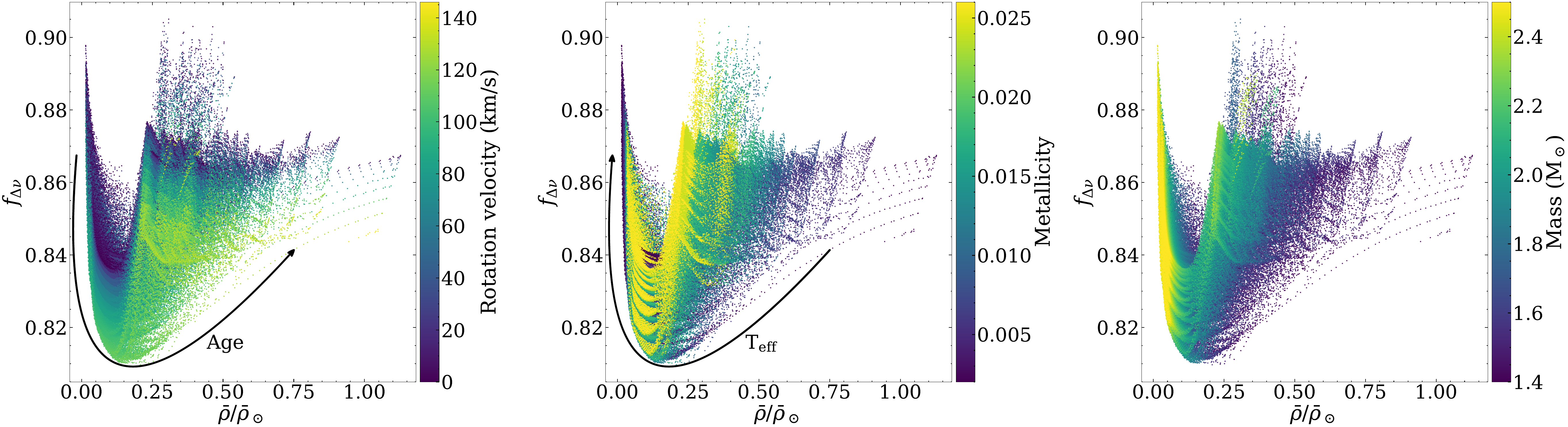}
                \caption{Pre-MS models}
                \label{fig:f_dnu_rho_spread_prems}
            \end{subfigure}
            
            \vspace{0.6cm}

            \begin{subfigure}[t]{1.97\columnwidth}
                \centering
                \includegraphics[width=1\linewidth]{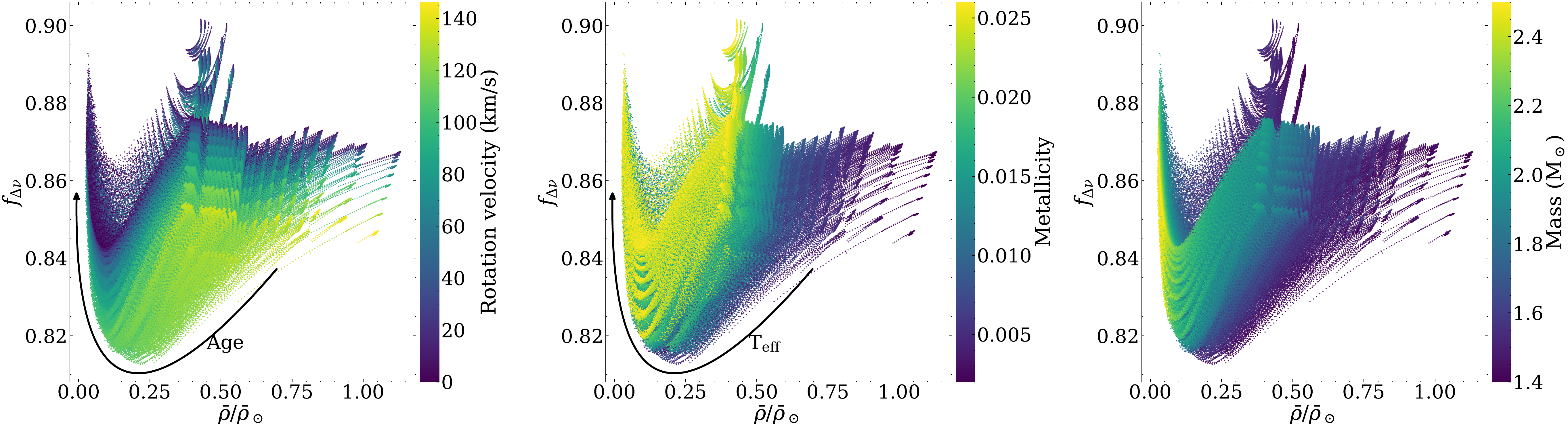}
                \caption{MS and post-MS contraction models}
                \label{fig:f_dnu_rho_spread_ms}
            \end{subfigure}
        
            \caption{Scaling behaviour of the $f_{\Delta\nu}$ relative to the mean stellar density. Each panel shows the dimensionless scaling factor $f_{\Delta\nu} = (\Delta\nu / \Delta\nu_\odot) \, (\bar{\rho}/\bar{\rho}_\odot)^{-0.5}$ as a function of the scaled mean density $\bar{\rho}/\bar{\rho}_\odot$, for models located within the classical instability strip of \citet{dupret_theoretical_2004}. Panels (a) and (b) correspond to the pre-MS and MS phases, respectively. Within each panel, colour denotes (i) the equatorial rotation velocity ($v_\mathrm{eq}$, km\,s$^{-1}$), (ii) the metallicity ($Z$), and (iii) the stellar mass ($M/M_\odot$). Here we plot every 20th pre-MS model and every 2nd MS model. The pre-MS models are also thinned in metallicity.}
            \label{fig:f1_vs_Dnu_rho_spread}
        \end{figure*}

        To allow for direct comparison with the theoretical scaling relation, we also derived a relation with the power-law exponent fixed at the theoretical value of 0.5. We find:
        \begin{eqnarray}
            f_{\Delta\nu} = \frac{\Delta\nu}{\Delta\nu_\odot} \left(\frac{\bar{\rho}}{\bar{\rho}_\odot}\right)^{-0.5} = 0.847 \pm 0.015,
        \end{eqnarray}
        where the quoted uncertainty represents the standard deviation of $f_{\Delta\nu}$ across all models in the instability strip. This fixed-exponent formulation yields a scaling factor of $f_{\Delta\nu} = 0.8472$. This scaling factor is consistent with the value of $f_{\Delta\nu} = 0.85$ reported by \citet{bedding_highf_2020} based on their non-rotating models. The majority of the TESS \dsct sample analysed by \citet{bedding_highf_2020} lies between the theoretical $f_{\Delta\nu}=1$ relation and their $f_{\Delta\nu}\simeq0.85$ model fit, with the observed spread plausibly arising from metallicity differences and rotational effects. Rotational oblateness lowers the mean density, while inclination shifts a star’s position in the HR diagram, which in turn biases the inferred radius, mass, and density. We show how $f_{\Delta\nu}$ scales with mean stellar density across mass, metallicity, rotation, and age in Figure\:\ref{fig:f1_vs_Dnu_rho_spread}.

        Previous studies have reported modest departures from the solar-calibrated scaling both in normalization and in exponent, though methodological differences in measuring $\Delta\nu$ complicate direct comparison. \citet{Suarez_rhoDnu_2014} derived a relation of the form $\Delta\nu/\Delta\nu_\odot = 0.776\,(\bar{\rho}/\bar{\rho}_\odot)^{0.46}$ using a grid of non-rotating MS models, where $\Delta\nu$ was calculated as an average over consecutive mode differences for each degree $\ell$ up to 3, then averaged across all degrees. \citet{garcia_hernandez_observational_2015} derived a mean density-frequency scaling relation from a sample of eclipsing binaries with \dsct components, computing $\Delta\nu = \mathrm{p}_{n+1,\ell} - \mathrm{p}_{n+1,\ell}$ for radial and non-radial modes ($\ell = 0$ to 3) within the observed frequency range. Their fit yielded $\Delta\nu/\Delta\nu_\odot = 0.806\,(\bar{\rho}/\bar{\rho}_\odot)^{0.491}$. \citet{rodriguez-martin_study_2020} calculated $\Delta\nu_\ell = \mathrm{p}_{n+1,\ell} - \mathrm{p}_{n+1,\ell}$ for radial orders $n = 2$--8 and took the median of individual $\Delta\nu$ values across all degrees, reporting $\Delta\nu/\Delta\nu_\odot = 0.974\,(\bar{\rho}/\bar{\rho}_\odot)^{0.559}$ for solid-body rotating models and $0.792\,(\bar{\rho}/\bar{\rho}_\odot)^{0.495}$ for mass-averaged fits. However, their models did not include core overshooting, which may have influenced the inferred relation. \citet{murphy_grid_2023} obtained a scaling factor of 0.87 (using the theoretical power-law exponent of 0.50) from a non-rotating model grid, measuring $\Delta\nu$ from a linear fit to radial modes with $n = 5$--9, the same approach adopted in this work. Our slightly lower coefficient is a direct consequence of rotational effects, which reduce $\bar{\rho}$ and $\Delta\nu$ in our models and thereby results in a lower scaling factor relative to non-rotating grids.
        
        Since \dsct stars pulsate in radial modes with low to intermediate radial orders ($n \sim 1$--9), where asymptotic approximations such as $\Delta\nu \propto \sqrt{\bar{\rho}}$ are not strictly valid, features such as acoustic cavity structure, mode trapping, and sharp sound-speed gradients can introduce mild but systematic departures from the scaling behaviour in lower-order modes. Nevertheless, the scaling relation remains approximately correct, suggesting that it continues to capture the dominant physical dependence on mean density even outside the asymptotic regime. In addition, the fitted scaling parameters appear to be influenced by the underlying mass distribution, the extent of evolutionary phase coverage, the treatment of rotation in each model grid, and the specific methodology used to compute $\Delta\nu$, including the mode range selected for the fit.

        While the small residuals indicate that the scaling relation remains approximately correct, the small but systematic deviations from the canonical relation may encode additional information about stellar parameters. However, the diagnostic power of individual frequency analysis makes detailed pulsation modelling the preferred approach for \dsct stars. Rather than extracting subtle information from scaling relation residuals, the clear mode patterns observed in young \dsct stars allow for direct asteroseismic inference from comprehensive frequency matching \citep[e.g.][]{murphy_grid_2023, scutt_asteroseismology_2023}.

    \subsection{The relation between $\mathrm{p}_{n_1\ell_0}$ and $\Delta\nu$}
         Observations by \citet{bedding_highf_2020} suggested a relationship for $\delta$ Sct stars between the frequency of the fundamental radial mode and the large separation, such that $\mathrm{p}_{n_1\ell_0}$ is approximately $3 \Delta\nu$. This has been validated as a tool for mode identification, alongside period ratios and the period--luminosity relation \citep{murphyetal2020b}.

        \begin{figure}
            \begin{center}
            \includegraphics[width=1\linewidth]{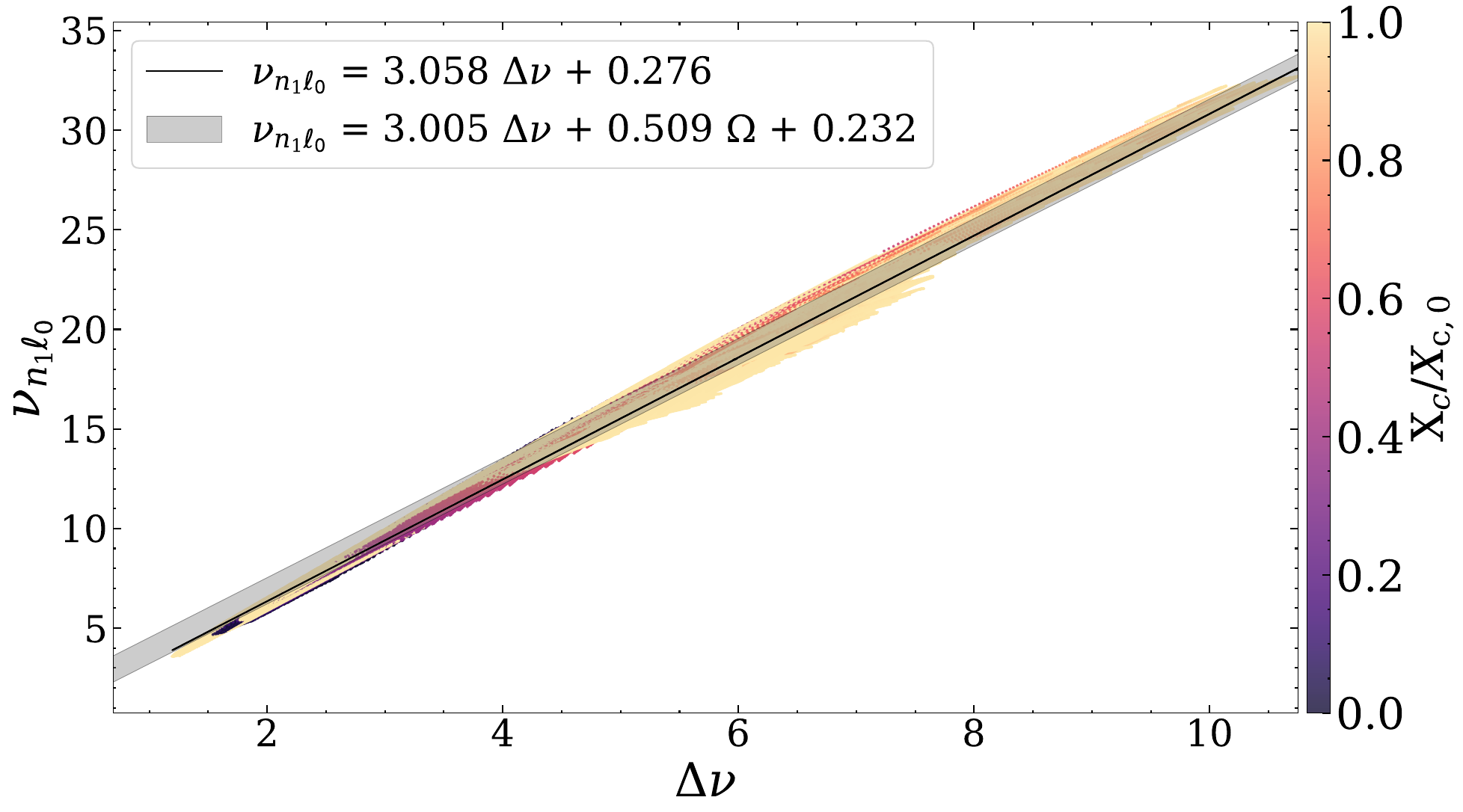}
            \caption{Relationship between the fundamental radial mode frequency ($\mathrm{p}_{n_1\ell_0}$) and the large frequency separation ($\Delta\nu$) across our grid of models that lie within the instability strip. The points represent individual models spanning pre-MS, MS, and post-MS contraction phases.}
            \label{fig:f1_vs_Dnu_fit}
            \end{center}
        \end{figure} 

        Figure\:\ref{fig:f1_vs_Dnu_fit} shows the relation between the fundamental mode frequency and large frequency separation across our grid for stars with masses between 1.5 and 2.5\,M$_\odot$, spanning pre-MS, MS, and post-MS contraction phases. We used only those models that lie within the classical instability strip, where $\delta$ Sct pulsations are theoretically expected to be driven. A remarkably consistent linear trend emerges across all evolutionary stages, which we quantify as:
        \begin{eqnarray}
            \mathrm{p}_{n_1\ell_0} = 3.058\ \Delta\nu + 0.276\,{\rm d}^{-1}.
        \end{eqnarray}
        This linear fit has a standard deviation of residuals of 0.357 d$^{-1}$ and an R$^2$ score of 0.996, confirming the tight correlation persists throughout stellar evolution. 
        
        When we incorporate a rotation term, the relation becomes:
        \begin{eqnarray}
            \mathrm{p}_{n_1\ell_0} = 3.005 \ \Delta\nu + 0.509 \ \frac{\Omega}{2\pi} + 0.232\,{\rm d}^{-1},
        \end{eqnarray}
        where $\Omega$ is the angular rotation frequency (with units of rad\,d$^{-1}$). The inclusion of rotation further improves the fit, yielding a reduced standard deviation of residuals of 0.253 d$^{-1}$ and a slightly increased R$^2$ score of 0.998.
        
        It is noteworthy that this relationship holds across vastly different evolutionary stages with such precision, suggesting a fundamental connection between the radial fundamental mode frequency and the sound-travel time across the star that persists despite evolutionary changes in stellar structure. When the fundamental radial mode ($\mathrm{p}_{n_1\ell_0}$) can be confidently identified in an observed frequency spectrum, it offers an immediate estimate of the large frequency separation ($\Delta\nu$), and hence of the stellar density. This initial $\Delta\nu$ estimate allows for the efficient construction of \'echelle diagrams with the appropriate frequency modulus, rather than testing numerous values to achieve vertical alignment of mode ridges. Once the correct $\Delta\nu$ is established, mode identification becomes substantially more straightforward.
        
        The weak dependence of the $\mathrm{p}_{n_1\ell_0}$--$\Delta\nu$ relation on rotation frequency ($\Omega$) shows that rotation has only a modest effect on this scaling relation, with faster rotators (within our moderate rotation regime) showing slightly higher $\mathrm{p}_{n_1\ell_0}$ values at a given $\Delta\nu$. This relative insensitivity to rotation makes the $\mathrm{p}_{n_1\ell_0}$--$\Delta\nu$ relation particularly useful for mode identification in $\delta$ Sct stars with unclear mode patterns and can serve as a discriminator between slow and rapid rotators  \citep{murphy_cepher_2024}.

        \begin{figure*}
            \centering
            \begin{subfigure}[t]{1.97\columnwidth} 
                \centering
                \includegraphics[width=1\linewidth]{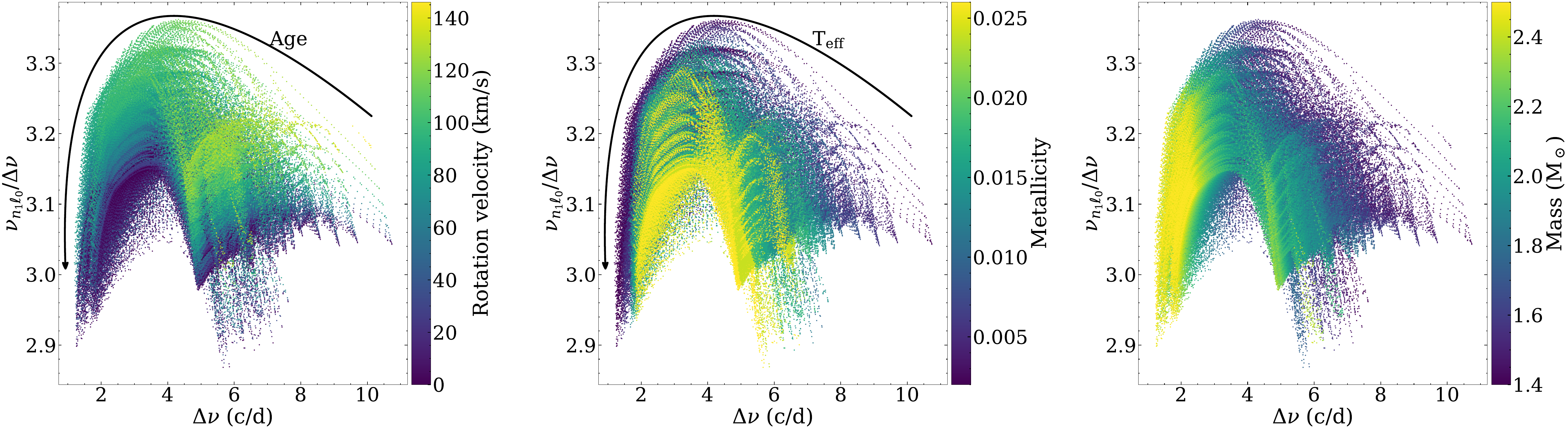}
                \caption{Pre-MS models}
                \label{fig:f1_vs_Dnu_spread_preMS}
            \end{subfigure}
            
            \vspace{0.6cm}

            \begin{subfigure}[t]{1.97\columnwidth}
                \centering
                \includegraphics[width=1\linewidth]{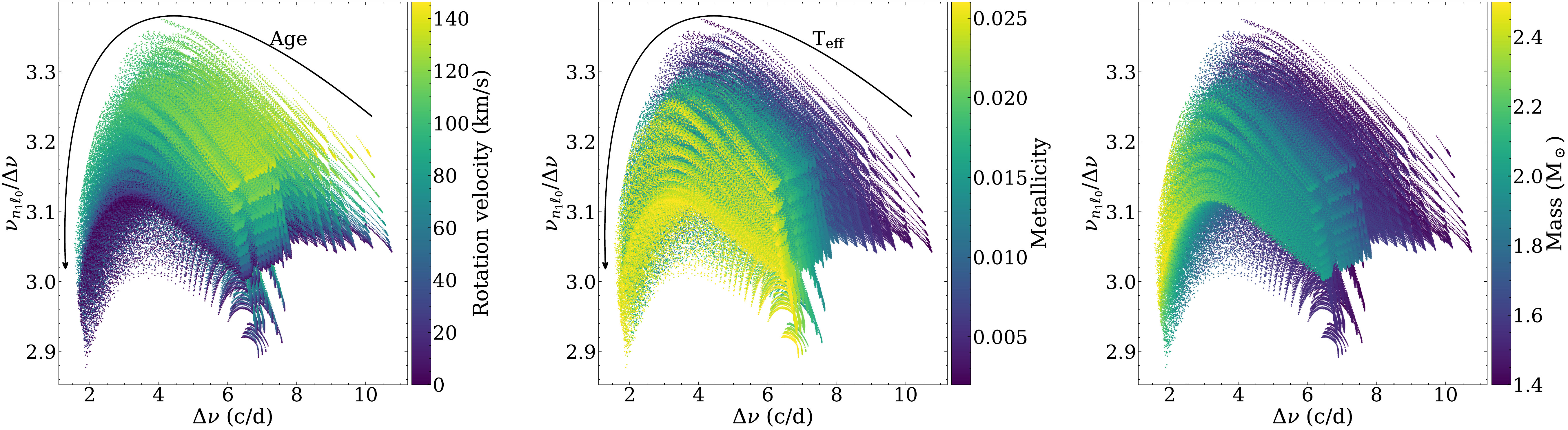}
                \caption{MS and post-MS contraction models}
                \label{fig:f1_vs_Dnu_spread_MS}
            \end{subfigure}
        
            \caption{Visualization of the ratio between $\mathrm{p}_{n_1\ell_0}$ and large frequency separation $\Delta\nu$, for models within the classical instability strip \citep{dupret_theoretical_2004}. The three plots in each panel display trends with stellar parameters shown by colour: (a) equatorial rotation velocity ($v_\mathrm{eq}$, km\,s$^{-1}$), (b) metallicity ($Z$), and (c) stellar mass ($M/M_\odot$). To aid readability, every 20th pre-MS model and every 2nd MS model is plotted. To further reduce visual clutter, a sparse sampling in metallicity is shown for the pre-MS models.}
            \label{fig:f1_vs_Dnu_spread}
        \end{figure*}

        Figure\:\ref{fig:f1_vs_Dnu_spread} examines how the ratio $\mathrm{p}_{n_1\ell_0}/\Delta\nu$ varies with stellar parameters across our grid. Overall, the ratio clusters tightly around a value of $\sim$3.1, but systematic deviations reveal secondary dependencies. Age effects are most pronounced at the extremes: very young ($<$10 Myr) and more evolved ($\gtrsim$1 Gyr) models show a downward trend in $\mathrm{p}_{n_1\ell_0}/\Delta\nu$, due to structural changes in the acoustic cavity during pre- and post-MS evolution. These outliers predominantly correspond to models with low $\Delta\nu$ values. Metallicity also introduces a subtle trend, with metal-rich stars exhibiting slightly lower $\mathrm{p}_{n_1\ell_0}/\Delta\nu$ ratios than metal-poor counterparts. Rotation, by contrast, appears to have only a weak influence on this relation, consistent with the shallow rotational dependence inferred from the global fit. Notably, low-mass models ($M < 1.6,M_\odot$) exhibit the largest deviations from the mean relation, with $\mathrm{p}_{n_1\ell_0}/\Delta\nu$ ratios dropping as low as $\sim$2.9 in some cases.

        Beyond variations in the ratio, these panels also illustrate how $\Delta\nu$ itself is distributed across different regions of parameter space. The spread in $\Delta\nu$ is strongly modulated by metallicity and mass \citep{murphy_grid_2023}. High-metallicity models are tightly confined to $\Delta\nu \sim 2$-5 d$^{-1}$, while low-metallicity stars span a broader range from $\sim$2 to 8 d$^{-1}$. In terms of age, stars in the 10--500 Myr range exhibit relatively stable and narrow $\Delta\nu$ distributions, whereas the youngest (early pre-MS) and oldest (near- and post-TAMS) stars tend toward lower values, reflecting the lower mean densities characteristic of these evolutionary phases. Rotation has some impact on the $\Delta\nu$ spread, but this effect is smaller than those of age or metallicity. Finally, a clear mass dependence is observed: lower-mass stars, being typically denser, populate the high-$\Delta\nu$ regime and show significant scatter, while higher-mass models cluster tightly at lower $\Delta\nu$ values, though the $\mathrm{p}_{n_1\ell_0}/\Delta\nu$ ratio shows little mass dependence. Irrespective of evolutionary phase, models with masses $M \lesssim 2.2\,\mathrm{M}_\odot$ show a systematic increase in $\Delta\nu$ as mass decreases. 

    \subsection{The $\varepsilon$ parameter across the grid}
        The phase parameter, $\varepsilon$, in the asymptotic relation \citep{tassoul_1980, aerts_book_2010}  
        \begin{eqnarray}
            \nu = \Delta\nu(n + \ell/2 + \varepsilon),
        \end{eqnarray}
        contains additional diagnostic potential \citep{white_asteroseismic_2011, white_calculating_2011}. This is because $\varepsilon$ encodes information about the upper turning point of the acoustic cavity \citep{gough_1986, aerts_book_2010}.

        \begin{figure*}
            \centering
            \begin{subfigure}[t]{1.97\columnwidth} 
                \centering
                \includegraphics[width=1\linewidth]{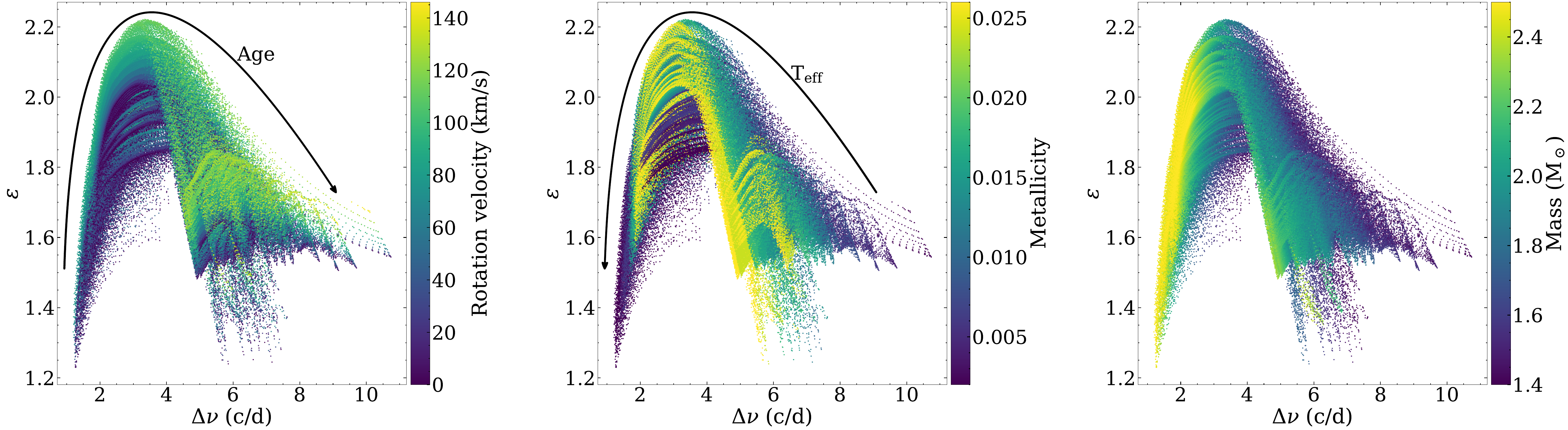}
                \caption{Pre-MS models}
                \label{fig:eps_vs_Dnu_spread_preMS}
            \end{subfigure}
            
            \vspace{0.6cm}

            \begin{subfigure}[t]{1.97\columnwidth}
                \centering
                \includegraphics[width=1\linewidth]{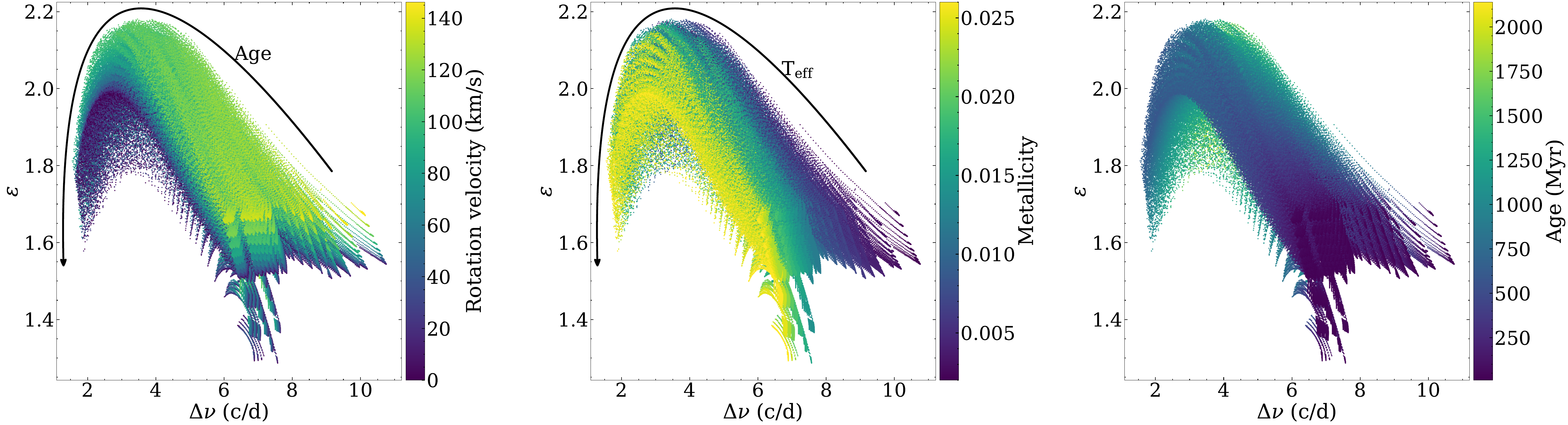}
                \caption{MS and post-MS contraction models}
                \label{fig:eps_vs_Dnu_spread_MS}
            \end{subfigure}
        
            \caption{The two panels show $\varepsilon$ as a function of $\Delta\nu$, across models that lie within the classical instability strip. The top and the bottom panels present pre-MS and MS models, respectively. Each plot within a panel shows the dependence of $\varepsilon$ with a different global stellar parameter encoded via colour: (a) equatorial rotation velocity ($v_\mathrm{eq}$, in km\,s$^{-1}$), (b) metallicity ($Z$), and (c) stellar mass ($M/M_\odot$). Here we show every 20th  pre-MS model, and every 2nd MS model. The pre-MS models are thinned in metallicity, while the MS panels show all models.}
            \label{fig:eps_vs_Dnu_spread}
        \end{figure*}

        Figure \ref{fig:eps_vs_Dnu_spread} presents how $\varepsilon$ varies across our model grid as a function of $\Delta\nu$, with different panels showing its dependence on age, rotation, metallicity, and mass. Several clear trends emerge from this analysis.

        The age behaviour of $\varepsilon$ reveals dependence on evolutionary phases. Pre-MS models generally exhibit a wide spread in $\varepsilon$ (1.2--2.2), indicative of rapid structural changes during these phases. As stars approach and settle onto the MS (by $\sim$50 Myr), this spread narrows considerably (1.3--1.7). Throughout most of the MS, models occupy a well-defined band along which $\varepsilon$ slowly increases with increasing age. This band begins with $\varepsilon$ values between 1.3--1.7 in the early MS and broadens to approximately 1.7--2.15 by 600--700 Myr. Towards the TAMS and in the post-MS contraction phase, the $\varepsilon$ range expands again (1.6--2.2), indicating renewed structural adjustments following core hydrogen exhaustion. This three-phase pattern in $\varepsilon$ is similar to the behaviour observed in $\mathrm{p}_{n_1\ell_0}/\Delta\nu$.

        A particular feature across all parameters is the non-monotonic relationship between $\Delta\nu$ and $\varepsilon$. Models with high $\Delta\nu$ values (5--9\,d$^{-1}$) consistently show low $\varepsilon$ values (around 1.6), while models with intermediate $\Delta\nu$ values (3--5\,d$^{-1}$) exhibit the highest $\varepsilon$ values. Models with the lowest $\Delta\nu$ values ($<$3\,d$^{-1}$) show intermediate $\varepsilon$ values, higher than in the high-$\Delta\nu$ regime but lower than in the intermediate-$\Delta\nu$ regime. This pattern creates a characteristic arc in the $\varepsilon$--$\Delta\nu$ plane that appears consistently across variations in mass, metallicity, and rotation. The arc corresponds to an underlying temperature sequence, with cooler models occupying the lower-$\Delta\nu$ region on the left and progressively hotter models extending toward higher $\Delta\nu$ values on the right.

        Rotation shows subtle effects on $\varepsilon$ values, particularly at lower $\Delta\nu$ values where rapid rotators (with equatorial velocities $\sim$120 km\,s$^{-1}$) predominantly exhibit higher $\varepsilon$ values (1.4--2.2) compared to slower rotators ($\sim$30 km\,s$^{-1}$), which show lower values (1.2--2). Non-rotating models display the widest overall spread in $\varepsilon$ (1.2--2). This trend is consistent with rotationally induced shifts in pulsation frequencies and phase relations, especially for low-order p\:modes \citep{ouazzani_rotational_2012, guo_oscillation_2024}. 

        Metallicity produces a negligible effect on $\varepsilon$ during the MS and the post-MS phases. Increasing metallicity systematically decreases $\Delta\nu$, while $\varepsilon$ is largely independent of it. The increase in $\Delta \nu$ is primarily driven by an increase in opacity which leads to a larger radius and a lower density. This trend is reflected in the pre-MS phase as well, albeit more loosely because of rapid structural changes.

        Stellar mass exhibits a direct relationship with both $\varepsilon$ and $\Delta\nu$. During the pre-MS phase, higher-mass models ($M > 2.2\,\mathrm{M}_\odot$) correspond to lower $\Delta\nu$ values with a large spread in $\varepsilon$ ($\sim$1.2 to 2.2). Throughout the MS and post-MS contraction phases, higher-mass models exhibit a smaller spread (1.75--2.05) in $\varepsilon$ and low $\Delta\nu$ values. 

\section{Conclusions}
\label{sec:conclusions}

We have developed an extensive grid of stellar pulsation models for \dsct stars spanning masses from 1.4 to 2.5 M$_\odot$, metallicities from $Z = 0.001$ to 0.026, and rotation rates from zero up to $\Omega/\Omega_{\rm crit} = 0.3$. This work extends prior efforts by incorporating rotation, including higher-degree modes ($\ell \leq 3$), including g and f modes, and tracking stars from the pre-MS through the MS to post-MS contraction. 

We have revisited the systematic relationships between pulsation properties and fundamental stellar parameters in \dsct stars. The large frequency separation ($\Delta\nu$) follows a tight power-law relation with mean stellar density, with $\Delta\nu/\Delta\nu_\odot = 0.857 \times (\bar{\rho}/\bar{\rho}_\odot)^{0.507}$. This scaling factor of 0.857 reflects the non-asymptotic nature of the observed p\:modes in \dsct stars, while the power-law exponent closely matches the theoretical expectation of 0.5. We have also established a linear relationship between the fundamental radial mode frequency and the large frequency separation, $\mathrm{p}_{n_1\ell_0} = 3.058\ \Delta\nu + 0.276\,{\rm d}^{-1}$, which holds across our entire parameter space with remarkably low scatter. When rotation is incorporated, the relation becomes $\mathrm{p}_{n_1\ell_0} = 3.005 \ \Delta\nu + 0.509 \ \frac{\Omega}{2\pi} + 0.232\,{\rm d}^{-1}$.
We also analysed the behaviour of the phase offset parameter $\varepsilon$ and large frequency separation $\Delta\nu$ across our model grid to see how these vary with fundamental stellar parameters including mass, metallicity, age, and rotation.

The different mode types in our grid exhibit varying penetration depths within stellar interiors, providing complementary sensitivity to the internal density profile at different radial locations. Pressure modes primarily sample the outer envelope and are sensitive to mean density variations, while gravity modes probe deeper regions. Both are sensitive to evolutionary changes. The inclusion of rotational effects allows quantification of rotation rates through the analysis of mode splitting patterns, breaking the degeneracy between rotational and evolutionary effects on mode frequencies that has historically limited asteroseismic inference for \dsct stars.

\change{We presented two \dsct stars (HD\,3622 and V624\,Tau) whose observed p\:mode spectra are reproduced well by our models. Both stars show additional frequency peaks near to and on either side of the fundamental radial mode that p\:modes cannot explain. Our grid places the f\:modes and low-order g\:modes in the observed frequency ranges of these peaks. To assess their plausibility and expected visibility we computed mode inertias along a representative stellar track and found that the f and low-order g\:modes have inertias comparable to or lower than the fundamental radial mode during the late pre-MS, near the ZAMS, and through most of the MS before the first avoided crossings occur. This suggests that these modes reach observable amplitudes in general, and supports our observations and inference of their presence in HD\,3622 and V624\,Tau. We anticipate that f\:modes and low-order g\:modes will be observable in a large fraction of young \dsct stars, providing a complementary diagnostic to p\:mode asteroseismology for constraining rotation and stellar age.}

In a forthcoming paper, we will quantify systematic uncertainties in the inferred stellar parameters that arise from variations in the input physics of our models. This analysis will encompass variations in several physical factors, including convective overshoot, nuclear reaction networks, angular momentum diffusion, and rotational oblateness, among others. This analysis will inform a neural network inference framework for determining stellar parameters from observed \dsct frequencies (Gautam et al. in preparation), building on previous machine learning approaches \citep{scutt_asteroseismology_2023, murphy_grid_2023}. We also plan to use our models to refine the age of the Pleiades cluster, a benchmark open cluster whose precise age is useful for calibrating stellar evolution models. By doing so, we aim to test and validate our method and models. 

\section*{Acknowledgements}
AG was supported by an Australian Government Research Training Program (RTP) Scholarship and University of Southern Queensland (UniSQ) International Fees Research Scholarship. SJM was supported by the Australian Research Council (ARC) through Future Fellowship FT210100485. TRB acknowledges support from the ARC through Laureate Fellowship FL220100117. We thank Joel Ong and the USyd--UniSQ \dsct Research Group for helpful comments on this paper.

\subsection*{Computational details}
\label{sec:computing}

The grid comprises approximately 20\,000 stellar evolution tracks. For each track, the combined evolution and pulsation-frequency calculations required on average $\sim$50~CPU~hours, corresponding to a total computational investment exceeding 1~million~CPU~hours for the complete grid.

The grid was developed through successive computational grants that allowed for a staged approach where the initial grants supported the computation of a sparse set of models used to refine the parameter space and modelling strategy, while subsequent larger allocations facilitated the construction of the full, densely sampled grid. 

This work utilised the UniSQ Fawkes High Performance Computing (HPC) facility and the Gadi supercomputer at the National Computational Infrastructure (NCI), supported by the Australian Government. Computational resources were provided through the NCI Adapter Scheme (200\,kSU in Q3\,2023 and 200\,kSU in Q4\,2023), the Australian Research Data Commons (ARDC) Nectar Research Cloud (20\,kSU for 12\,months in 2023 and 2024), the Queensland Cyber Infrastructure Foundation (QCIF) NCI Share (75\,kSU per quarter throughout 2024), and the Astronomy Supercomputer Time Allocation Committee (ASTAC) under the NCI Astronomy Program (2,500\,kSU in Q3--Q4\,2024).

\section*{Data availability}
We provide the complete model grid at \href{https://zenodo.org/records/15839323}{Zenodo}. All supporting materials, including the \mesa and \gyre inlists, and usage documentation, are available in a public \href{https://github.com/gautam-404/dSct-model-grid}{GitHub repository}.



\bibliographystyle{mnras}
\bibliography{bibliography} 



\appendix



\change{\section{Typical propagation diagrams}}
\label{app:propagation_diagrams}

    \change{In this appendix we show propagation diagrams corresponding to ages as those in Fig. \ref{fig:eigenfunctions} (main text). The examples are computed with \gyre in the adiabatic approximation using the same numerical settings as in the main text, and typical frequencies are overplotted on the propagation diagrams for context.}
    
    \change{In the Cowling and adiabatic approximations, the local radial wave number of non-radial waves satisfy \citep{lopes_nonradial_2001}
    \begin{equation}
        k_r^2 \simeq \frac{\omega^2}{c_s^2}\left(1-\frac{S_\ell^2}{\omega^2}\right)\left(1-\frac{N^2}{\omega^2}\right),
    \end{equation}
    so turning points occur where $\omega^2=N^2$ or $\omega^2=S_\ell^2$. Turning points are radii where $k_r^2=0$. They separate oscillatory (propagating, $k_r^2>0$) and evanescent (decaying, $k_r^2<0$) regions. Thus, we get:
    \begin{itemize}
      \item \textit{Acoustic (p\:mode) cavity:} $\omega^2 > \max(N^2,S_\ell^2)$; waves propagate where the mode frequency lies above both $N$ and $S_\ell$.
      \item \textit{Gravity (g\:mode) cavity:} $\omega^2 < \min(N^2,S_\ell^2)$; waves propagate where the mode frequency lies below both $N$ and $S_\ell$.
      \item \textit{Evanescent zones:} between the $N$ and $S_\ell$ curves where the two factors in the $k_r^2$ equation have opposite signs. This corresponds to the radii where $N^2<\omega^2<S_\ell^2$ or $S_\ell^2<\omega^2<N^2$, such that $k_r^2<0$ and solutions decay.
      \item \textit{Low-degree f\:modes:} eigenfrequencies that place the mode on the p-g interface with two turning points, an inner one at $\omega=N$ and an outer one at $\omega=S_\ell$. The interior ($\omega<N$) is g-like, the exterior ($\omega>S_\ell$) is p-like, and these are separated by an evanescent layer where $N<\omega<S_\ell$. For $\ell \lesssim 15$ the evanescent barrier is thin (Lamb-like with appreciable interior amplitude), whereas, at higher $\ell$ the barrier thickens and the mode becomes surface-trapped \citep{perrot_topological_2019,leclerc_topological_2022,saux_core-sensitive_2025}.
    \end{itemize}}

\begin{figure*}
        \centering
        \begin{subfigure}[t]{1.5\columnwidth} 
            \centering
            \includegraphics[width=1\linewidth]{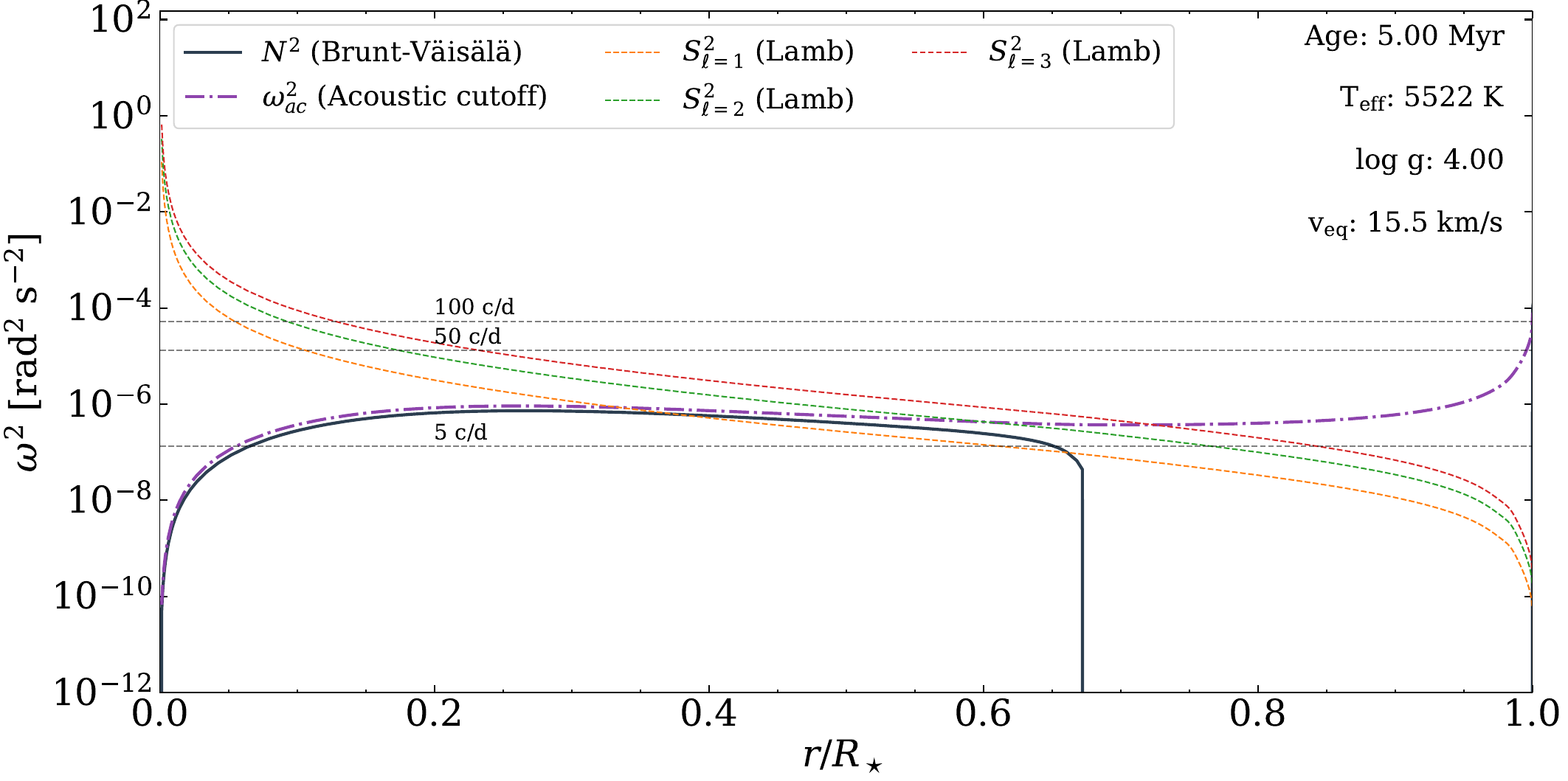}
            \caption{Pre-MS model}
            \label{fig:prop_diag1}
        \end{subfigure}
        
        \vspace{0.6cm}
    
        \begin{subfigure}[t]{1.5\columnwidth}
            \centering
            \includegraphics[width=1\linewidth]{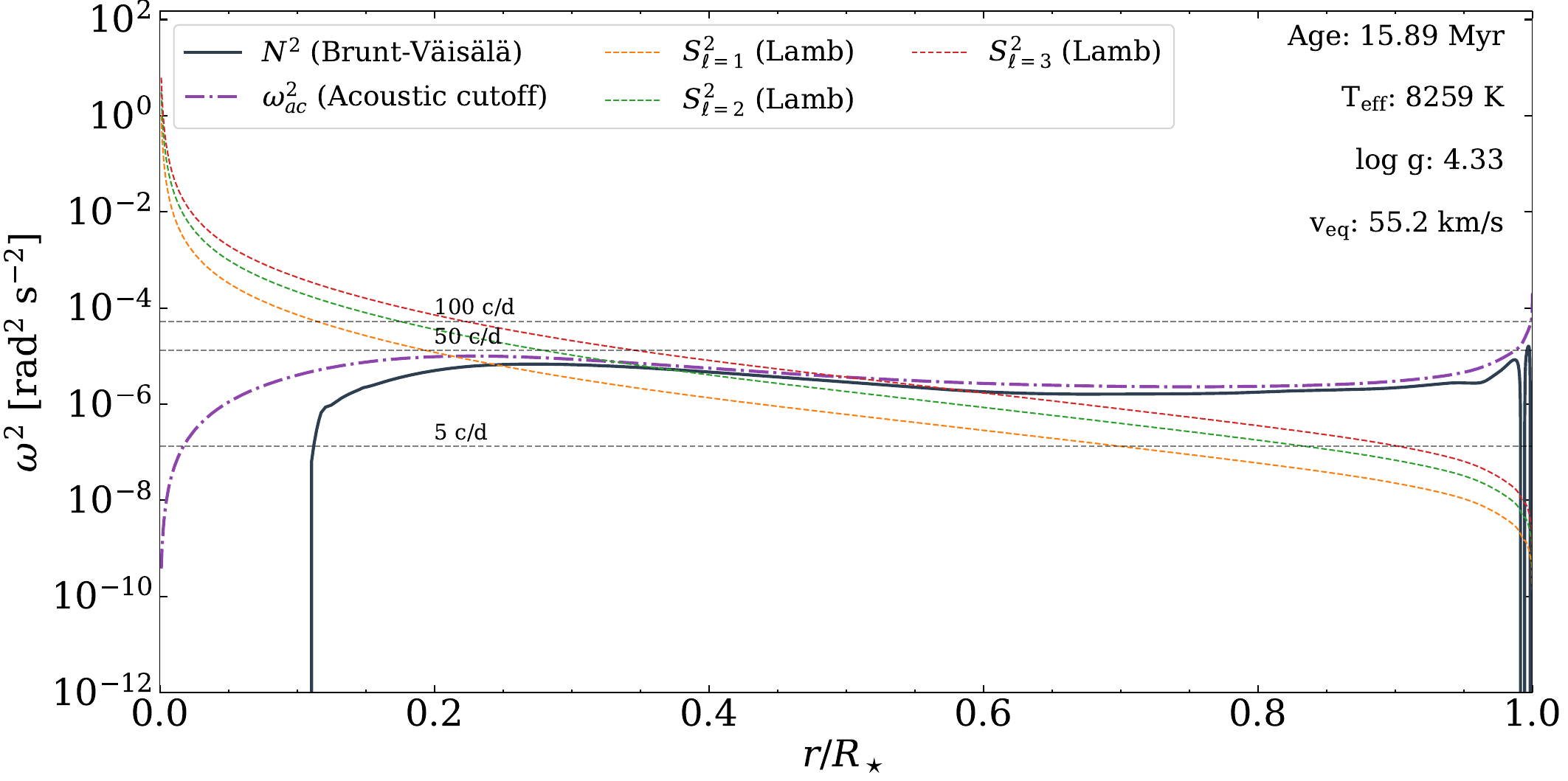}
            \caption{ZAMS model}
            \label{fig:prop_diag2}
        \end{subfigure}

        \vspace{0.6cm}

        \begin{subfigure}[t]{1.5\columnwidth}
            \centering
            \includegraphics[width=1\linewidth]{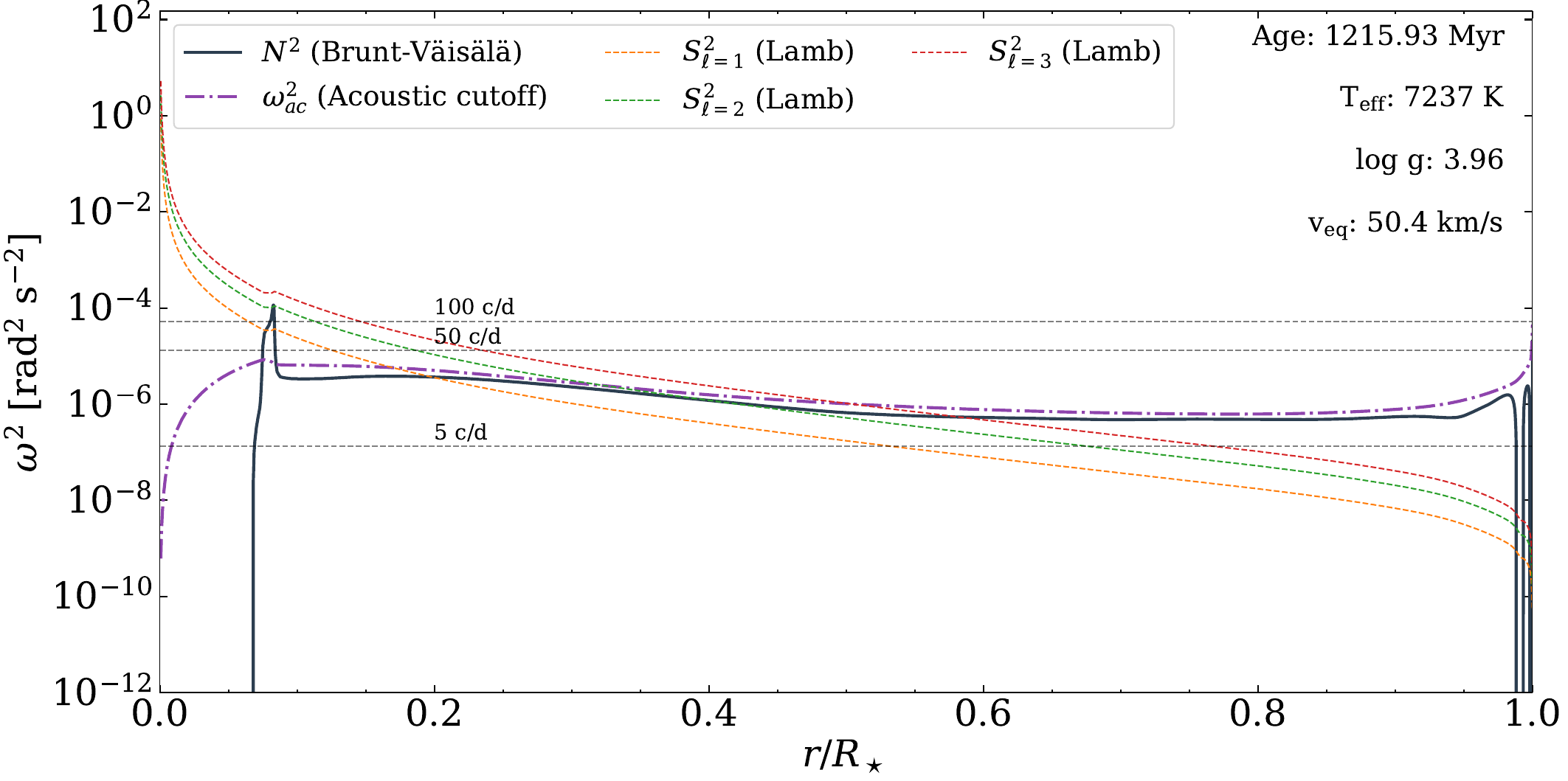}
            \caption{MS avoided crossing model}
            \label{fig:prop_diag3}
        \end{subfigure}
    
        \caption{\change{Propagation diagrams showing how wave cavities evolve from the pre-MS (panel a) through the ZAMS (panel b) to the late-MS phase where avoided crossings occur (panel c). The models shown here are picked from the evolutionary models for a solar-metallicity model ($Z=0.015$) of 1.7\,M$_\odot$, as in Fig.\,\ref{fig:mode_evolution}. Each panel shows the Brunt-V\"ais\"al\"a  frequency $N^2$ (black), the acoustic cut-off $\omega_c^2$ (purple), and the Lamb frequencies $S_\ell^2$ for $\ell=1,2,3$ (orange, green and red, respectively, as labelled).}}
        \label{fig:prop_diags}
    \end{figure*}

    \change{Figure \ref{fig:prop_diags} shows how propagation cavities evolve for $\ell=1$--$3$ modes from the pre-MS to the ZAMS and toward the late MS stage, where avoided crossings occur. Key changes in these cavities between panels a and b include the establishment of a convective core and the disappearance of a thick surface convection zone, which both modify the local Brunt-V\"ais\"al\"a frequency. Between panels b and c, the convective core has retreated somewhat, and there is a bump in the Brunt-V\"ais\"al\"a frequency at the edge of the convective core because of the sharp increase in mean molecular weight there.}
    
\bsp	
\label{lastpage}
\end{document}